\documentclass[12pt,a4paper]{article}
\pdfoutput=1

\usepackage{graphicx}
\usepackage{caption}
\usepackage{subcaption}
\usepackage{jheppub}
\usepackage{enumitem}
\usepackage{framed,float}
\usepackage{booktabs}
\usepackage{subcaption}
\usepackage{tikz}
\usetikzlibrary{calc,arrows,decorations.markings}
\usepackage{mathrsfs}
\usepackage[utf8]{inputenc}

\usepackage{color}

\title{Higgsing and T\!wisting of 6d $D_N$ gauge theories.} 

\author[a]{Hee-Cheol Kim,\,}
\author[b]{Sung-Soo Kim,\,}
\author[c]{and Kimyeong Lee\,}

\affiliation[a]{Department of Physics, POSTECH, Pohang 37673, Korea}
\affiliation[b]{School of Physics, University of Electronic Science and Technology of China,\\
North Jianshe Road, Chengdu, Sichuan 610054, China}
\affiliation[c]{School of Physics, Korea Institute for Advanced Study,\\
85 Hoegi-ro Dongdaemun-gu, Seoul 02455, Korea}

\abstract
{We propose Type IIB 5-brane configurations that engineer the 6d $\mathcal{N}=(1,0)$ SCFTs with $SO(N)$ gauge symmetry coupled to a single tensor multiplet on a circle, by considering RG flows on Higgs branches of $D$-type conformal matter theories. We test the brane systems against known  Calabi-Yau threefolds for the  6d SCFTs on a circle. In addition we study a new RG flow involving Higgs vevs of scalar operators with Kaluza-Klein momentum along the circle. The new RG flow results in the 5-brane webs for the 6d SCFTs of $D_N$ gauge symmetry compactified on a circle with $Z_2$ outer-automorphism twist.}

\begin{document}
\begin{flushright}
\tt 
KIAS-P19021\\
\end{flushright}

\maketitle

\section{Introduction}\label{sec:intro}
Type IIB 5-brane web \cite{Aharony:1997ju, Aharony:1997bh} is a powerful tool to study 5d superconformal theories (SCFTs)\cite{Seiberg:1996bd,Morrison:1996xf,Douglas:1996xp,Intriligator:1997pq}. 
Perturbative and non-perturbative aspects of many SCFTs be understood from 5-brane webs. As an effective description of the SCFT at low energy, one can readily read off the prepotential of 5d $\mathcal{N}=1$ gauge theory from a given web diagram of the theory. Also, novel non-perturbative aspects of the SCFTs can be computed or seen through 5-brane configurations. For instance, for a given 5-brane configuration, one can  compute the instanton partition functions using topological vertex formulation \cite{Aganagic:2003db,Iqbal:2007ii}. 
Global symmetry enhancement \cite{DeWolfe:1999hj,Gaberdiel:1997ud,Gaberdiel:1998mv} at UV fixed points and various dualities 
 between low energy gauge theories, which are hard to explain in the gauge theories, often have simpler explanations in the 5-brane webs.

Five-dimensional $\mathcal{N}=1$ supersymmetric gauge theories can 
also be engineered via M-theory compactification on a local Calabi-Yau threefolds (CY$_3$) \cite{Aharony:1997ju, Aharony:1997bh}. BPS particles and magnetic monopole strings in 5d theory correspond to M2 branes wrapping holomorphic 2-cycles and magnetic dual M5 branes wrapping holomorphic 4-cycles, whose volumes are given as masses and tensions of the 5d BPS objects, respectively. 
The prepotential of the 5d theory is realized as the triple intersection number in geometry.  Geometry itself provides another powerful tool to study 5d SCFTs as well.   
In particular, recent progress on geometric classifications of 5d SCFTs \cite{Jefferson:2017ahm, DelZotto:2017pti,Jefferson:2018irk,Bhardwaj:2018yhy,Bhardwaj:2018vuu,Apruzzi:2019vpe,Apruzzi:2019opn} enlarged lists of possible SCFTs and also proposed intriguing new dualities and RG flows.

Equivalence between 5-brane web and geometry for toric Calabi-Yau threefolds is well established \cite{Leung:1997tw}. 
Given a toric geometry, it is straightforward for one to find 
the corresponding 5-brane web diagram. Volumes of 2-cycles and 4-cycles in geometry correspond to the lengths of edges and the areas of compact faces in the 5-brane web, respectively. 

Non-toric web diagrams are 5-brane configurations of orientifold planes such as O7$^\pm$-, O5- and ON$^0$-planes  and those obtained via non-perturbative Higgsings \cite{Benini:2009gi, Hayashi:2015fsa, Bergman:2015dpa, Hayashi:2015zka, Hayashi:2015vhy, Zafrir:2015ftn}. 
These non-toric 5-brane webs, along with F-theory classification of 6d SCFTs, has provided a new perspective on 6d $\mathcal{N}=(1,0) $ SCFTs. 
In particular, some of 6d $\mathcal{N}=1$ SCFTs with $SO(N)$ gauge symmetry coupled to a single tensor multiplet, when compactified on a circle, have natural Type IIB 5-brane configurations with two O5-planes, as its (T-dual) Type IIA brane configuration contains an $O6$-plane \cite{Hanany:1997gh, Hayashi:2015vhy}.
Though such large class of non-toric 5-branes are known, unlike toric case,  equivalence between non-toric 5-brane webs and geometry is still limited and hence worth further study. 

The aim of this paper is to complete Type IIB 5-brane engineering of some of 6d $\mathcal{N}=1$ SCFTs which involve non-toric realizations of 5-branes with O5-planes. In particular, we consider 5-brane realizations of a family of 6d $\mathcal{N}=(1,0)$ SCFTs with $SO(N)$ gauge symmetry 
on a $-2$ or $-3$ curve in the base of F-theory \cite{Heckman:2015bfa} when compactified on a circle. 
  We construct 5-brane diagrams of those and compute their prepotentials (or monopole string tensions) which are consistent with the triple intersection numbers from the corresponding Calabi-Yau geometries \cite{Bhardwaj:2018yhy}. More specifically, by considering RG flows on Higgs branches of $D$-type conformal matter, we construct 5-brane web configurations with two O5-planes describing 6d SCFTs with an $SO(N)$ gauge group and then study possible Higgsings and Hanany-Witten transitions to generate Higgs branch flows of 6d SCFTs with $SO(N)$ gauge symmetry on a circle.
   Moreover, by tuning non-zero holonomies such that a charged scalar mode carrying non-zero Kaluza-Klein momentum along the 6d circle becomes light and then by Higgsing the theory with a vev of the light mode, we propose new RG flows leading to 5-brane web configurations for twisted compactifications of 6d theories.

The organization of the paper is as follows: 
In section \ref{sec:2}, we review equivalence between 5-brane web and toric CY geometry and extend its equivalence to non-toric cases, and then discuss how to obtain the prepotential from 5-brane and also from geometry.  In section \ref{sec:SON03}, we further develop non-toric geometry with O5-planes and discuss Type IIB 5-brane configurations for 6d SCFTs with $SO(N)$ gauge symmetry, which are realized as elliptically fibered (non-toric) Calabi-Yau threefold. Especially we propose 5-brane webs realizing $SO(N)$ theories on $-3$ curve. In section \ref{sec:SON02}, we propose 5-brane webs for $SO(N)$ theories on $-2$ curve. In section \ref{sec:SON04}, we discuss twisting of 6d $SO(N)$ gauge theories on $-4$ curve. We then summarize and conclude with future directions in section \ref{sec:conclusion}.

\section{5-Brane webs and Calabi-Yau geometries}\label{sec:2}
We start with a short summary of the relationship between Type IIB $(p,q)$ 5-brane webs and non-compact Calabi-Yau threefolds in M-theory compactification, 
with the notations used throughout the subsequent sections.

Let us review the salient features of M-theory compactified on a non-compact Calabi-Yau three (CY$_3$) manifold $X$ to five dimensions \cite{Witten:1996qb}. This compactification gives rise to a 5d field theory at low energy. We can consider a smooth threefold $X$ with K\"ahler moduli $\phi_i$ ($i=1,\cdots,n$) associated to K\"ahler 4-cycles $S_i$ where $S_i$ is a divisor basis of $h^{1,1}(X)=n$. The K\"ahler moduli $\phi_i$  for compact 4-cycles $S_i$ with $i=1,\cdots,r$ are identified with the Coulomb branch moduli in the 5d field theory, whereas $\phi_i$ for non-compact surfaces $S_i$ with $i=r+1,\cdots,n$ are associated to the mass parameters $m_i$. So when the 5d field theory admits a gauge theory description, the rank of the gauge group agrees with the number of compact K\"ahler surfaces $r$ and $\phi_i$ becomes the vacuum expectation value of the scalar field in the charged vector multiplet parameterizing the Coulomb branch.

In the singular limit, when all the K\"ahler parameters $\phi_i$ are turned off so that all 4-cycles collapse to a point, the smooth threefold $X$ will reduce to a singular Calabi-Yau threefold $Y$. This singular threefold $Y$ realizes either a 6d SCFT if the CY$_3$ admits an elliptic fibration or, otherwise, a 5d SCFT. In the former case, the smooth threefold $X$ describes a circle compactification of the associated 6d SCFT where the circle size is implemented as an additional mass parameter in the geometry. So $X$ provides a resolution of the singular elliptic CY$_3$ when elliptic fiber class has finite volume. 

In M-theory compactification, BPS states in the 5d field theory come from M2- and M5-branes. M2- and M5-branes wrapped on 2- or 4-cycles in the CY$_3$ provide charged vector multiplets and magnetic monopole strings respectively in the low energy field theory. The masses of vector multiplets and the tensions of the monopole strings are volumes of 2- and 4-cycles and they are controlled by K\"ahler parameters $\phi_i$.

A non-compact threefold $X$ can be modeled by a union of connected compact surfaces $\displaystyle S=\cup_{i} S_i$ and each K\"ahler surface $S_i$ is either 
$\mathbb{P}^2$ or a ruled surface of degree $n$ over genus $g$ curve blown up at $p$ points, which we denote by $\mathbb{F}^p_{n,g}$ \cite{Jefferson:2018irk,Bhardwaj:2018yhy,Bhardwaj:2018vuu}. 
We will often omit super(or sub)-scripts $p,n,g$ if those numbers are zero. Let $S_1$ and $S_2$ be two surfaces in $S$. These two surfaces are glued along a holomorphic curve $C=S_1\cap S_2$ of genus $g$. In order to have a consistent Calabi-Yau threefold, the gluing curves should intersect transversally which imposes the Calabi-Yau condition 
\begin{equation}
	C^2|_{S_1} + C^2|_{S_2} = 2g-2 \ ,
\end{equation}
where the subscripts denote the projections of $C$ on $S_{1,2}$ respectively. Also, a curve class in each surface must satisfy the adjunction formula
\begin{equation}
	(K\cdot C)|_{S_i}+C^2|_{S_i} = 2g-2 \ ,
\end{equation}
where $K|_{S_i}$ is the canonical class of the surface $S_i$. There can be more than one curve at the intersection of two surfaces satisfying above conditions.

\begin{figure}
  \centering
  \includegraphics[width=.4\linewidth]{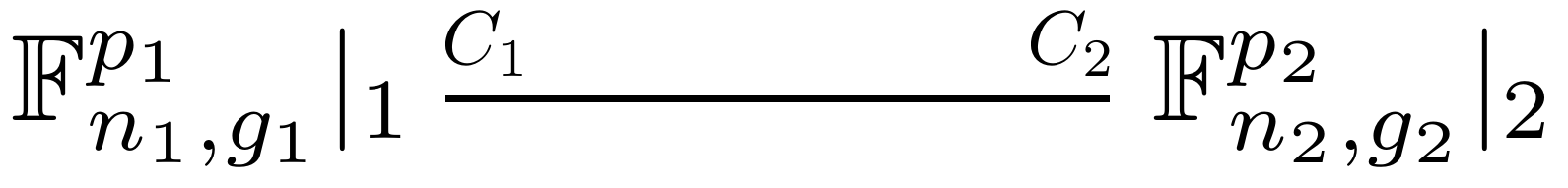}
  \caption{Two surfaces (or 4-cycles) $S_1=\mathbb{F}_{n_1,g_1}^{p_1}$ and $S_2=\mathbb{F}_{n_2,g_2}^{p_2}$ glued along a curve $C$ whose projection to the surface $S_i$ is denoted by $C_i$.  The subscript of $|_i$ denotes $i$-th surface.}
  \label{fig:gluing}
\end{figure}

Diagrammatically, two surfaces $S_1=\mathbb{F}_{n_1,g_1}^{p_1}$ and $S_2=\mathbb{F}_{n_2,g_2}^{p_2}$ glued together by a curve $C$ will be drawn as the diagram in Figure \ref{fig:gluing}. Here, the gluing curve is represented by a line connecting two surfaces and the symbols $C_1$ and $C_2$ on the line denote the projections of $C$ to the first and the second surfaces respectively. A diagram can have more than one line between two surfaces if two surfaces are glued along multiple curves. Note also that two holomorphic curves within a single surface can be glued each other. In this case, there is a line connecting the surface to itself in the diagram. We will call this 
{\it self-gluing} \cite{Jefferson:2018irk}.

The canonical class of a local $S=\mathbb{P}^2$ is $K|_S = -3\ell$ where $\ell$ is a base class with self-intersection $\ell^2=1$. The canonical class of a Hirzebruch surface of degree $n$ with $p$ blownup points $S=\mathbb{F}_n^p$ is given by
\begin{equation}
	K|_S = -2h +(n-2)f + \sum_{i=1}^p x_i \ .
\end{equation}
Here $h$ is a section class and $f$ is a fiber class and $x_i$'s are exceptional curve classes at the blowup points. There is an additional distinct section class $e$. They satisfy
\begin{eqnarray}\label{eq:canonicalclass}
	&&e^2 = -n \ , \quad h^2 = n \ , \quad f^2 = 0 \ , \quad x_i \cdot x_j= -\delta_{ij} \ , \nonumber \\
	&&e\cdot f = h \cdot f = 1 \ , \quad e\cdot h=e\cdot x_i = h\cdot x_i = f\cdot x_i = 0 \ .
\end{eqnarray}
A ruled surface over a genus $g$ curve can be obtained by $g$ {\it self-gluings} of  a rational Hirzebruch surface with a number of blow-up points. This geometric transition is introduced in \cite{Jefferson:2018irk}. For example, a surface $\mathbb{F}_{n,g}$ can be obtained from $\mathbb{F}_{n}^{2g+p}$ by $g$ self-gluings. For this, we glue $g$ pairs of exceptional curves $(x_i,y_i)$ with $i=1,\cdots ,g$. The canonical class $K_S'$ after self-gluings is then given by 
\begin{equation}\label{eq:K-self-gluing}
	K_S' = K_S + \sum_{i=1}^g (x_i+y_i) \ .
\end{equation}
See \cite{Jefferson:2018irk,Bhardwaj:2018vuu} for more details and other geometric transitions.

We are now interested in physics on the Coulomb branch of the low energy 5d theory. A smooth threefold $X$ is described by the K\"ahler class $J$ defined as
\begin{equation}
  J = \phi_i S_i\ , \quad i=1,\cdots ,n \ ,
\end{equation}
with compact and non-compact moduli $\phi_i$.
Then the triple intersection numbers of 4-cycles in $X$ can be computed as
\begin{equation}
  \mathcal{F} = \frac{1}{3!}\int_X J\cdot J \cdot J = \frac{1}{3!}\sum_{i,j,k}^n \phi_i\phi_j\phi_k\int_X S_i\cdot S_j \cdot S_j \ . 
\end{equation}
As one can see, $\mathcal{F}$ is a cubic polynomial of K\"ahler parameters $\phi_i$ and it computes the prepotential on the Coulomb branch in the 5d field theory. 
The metric $\tau_{ij}$ on the Coulomb branch and the volumes of 4-cycles $T_i$ can be extracted from the prepotential $\mathcal{F}$ as
\begin{equation}
  \tau_{ij} = \partial_i\partial_j \mathcal{F} = \int_X J\cdot S_i\cdot S_j \ , \quad T_i = \partial_i\mathcal{F}=\frac{1}{2}\int_X J \cdot J\cdot S_i \ ,
\end{equation}
with $i=1,\cdots, r$.

We can compute the triple intersection numbers $S_i\cdot S_j\cdot S_k$ among 4-cycles $S_i$ in $X$. First, if all surfaces are the same, the self-intersection number is determined by the canonical class as
\begin{equation}\label{eq:triple1}
	S_i \cdot S_i \cdot S_i = (K\cdot K)|_{S_i} \ .
\end{equation}
If a surface $S_i$ has {\it self-gluings}, we should replace $K$ in this equation by a modified divisor $K'$ given in \eqref{eq:K-self-gluing}. 
Second, if $i=j\neq k$, then the intersection number is given by
\begin{equation}\label{eq:triple2}
	S_i\cdot S_i \cdot S_k = \sum_a (C^a \cdot C^a)|_{S_k} \ ,
\end{equation}
where $C^a|_{S_k}$ stands for a gluing curve between two surfaces projected on $S_k$. Lastly, if all surfaces are distinct, i.e. $i\neq j \neq k$, the triple intersection number can be written as
\begin{equation}\label{eq:triple3}
  S_i \cdot S_j \cdot S_k = \sum_a (C^a_j \cdot C^a_k)|_{S_i} \ ,
\end{equation}
where $C_j^a$ are the gluing curve classes between $S_i$ and $S_j$. The prepotential of the 5d field theory is therefore fully determined by intersection data of compact surfaces in the associated Calabi-Yau threefold. We note that the
volume of an effective 2-cycle $C$ in a surface $S_i \in S$ can also be computed as an intersection product
\begin{equation} \label{eq:edgevol}
  vol(C) = - (J\cdot C) = -\sum_{j=1}^n\phi_j (K\cdot C)|_{S_j} - \sum_{i,k=1}^n \phi_i(C\cdot C_{ik})|_{S_i} \ ,
\end{equation}
where $C_{ik}$ denotes a collection of all gluing 2-cycles except for $C$ at the intersection $S_i\cap S_k$.

\begin{figure}
  \centering
  \includegraphics[width=.8\linewidth]{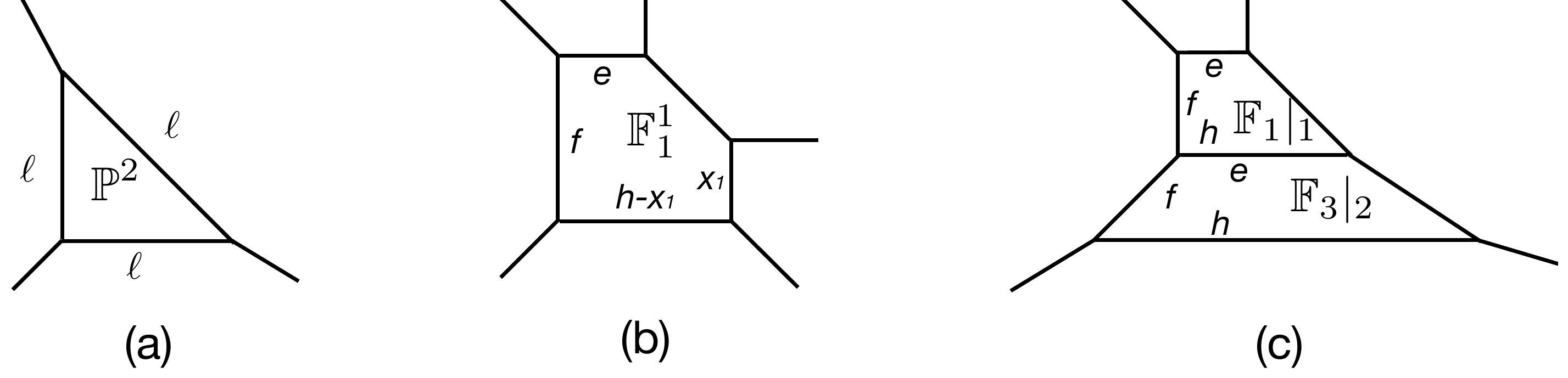}
  \caption{(a) 5-brane web for a local $\mathbb{P}^2$. 
  (b) A 5-brane web for $\mathbb{F}^1_1=\mathbb{F}_0^1$. (c) A 5-brane web for two surfaces $S_1=\mathbb{F}_1$ and $S_2 = \mathbb{F}_3$.}
  \label{fig:P2-F1-F3}
\end{figure}

Now we discuss 
the relation between Calabi-Yau geometries explained above and $(p,q)$ 5-brane webs in Type IIB string theory.
When $X$ is a toric Calabi-Yau threefold, there exists 
one-to-one map between $X$ and a 5-brane web via the duality between M-theory and Type IIB theory \cite{Leung:1997tw}. A 5-brane web consists of edges 
and vertices forming a number of closed (or internal) 
and open (or external) 
faces on a plane. Some examples of 5-brane webs are given in Figure \ref{fig:P2-F1-F3}. The diagrams (a) and (b) in Figure \ref{fig:P2-F1-F3} are 5-brane counterparts of the Calabi-Yau threefold of a local $\mathbb{P}^2$ and a del Pezzo surface ${\rm dP}_2=\mathbb{F}_1^1$, respectively. Note that Figure \ref{fig:P2-F1-F3}(b)  can be also thought of as an $\mathbb{F}_0^1$, since 
$\mathbb{F}_0^1$ is isomorphic to 
$\mathbb{F}_1^1$. 
Figure \ref{fig:P2-F1-F3}(c) 
realizes a threefold of a K\"ahler surface $S_1 \cup S_2$ where two compact surfaces $S_1=\mathbb{F}_1$ and $S_2=\mathbb{F}_3$ are connected along a curve $C=S_1\cap S_2$ whose projections on to each surface are $C|_{S_1}=h$ and $C|_{S_2}=e$. 

In a 5-brane web, closed faces correspond to compact surfaces of 
the associated Calabi-Yau geometry. So the closed face in Figure \ref{fig:P2-F1-F3}(a) implements a local $\mathbb{P}^2$ and the closed face in Figure \ref{fig:P2-F1-F3}(b) is an $\mathbb{F}_1^1$ (or $\mathbb{F}_0^1$). The volume of a compact K\"ahler surface in geometry is therefore the area of the compact face of the corresponding 5-brane web.

\begin{figure}
  \centering
  \includegraphics[width=1.\linewidth]{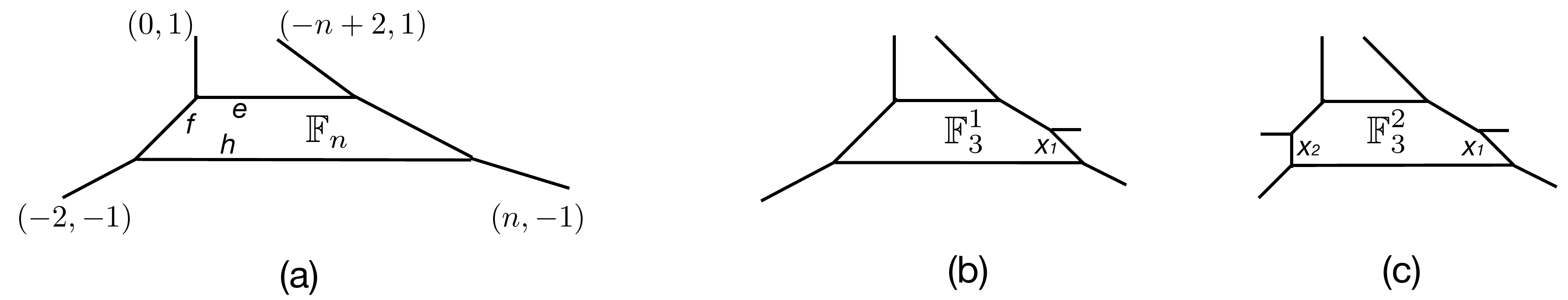}
  \caption{(a) A local 5-brane web for an $\mathbb{F}_n$. (b) A 5-brane web for $\mathbb{F}^1_3$. (c) A 5-brane web for $\mathbb{F}_3^2$.}
  \label{fig:Hirzebruch}
\end{figure}

A Hirzebruch surface $\mathbb{F}_n$ of degree $n$ can be described as a local 5-brane web in Figure \ref{fig:Hirzebruch}(a). 
A Calabi-Yau geometry consisting of a single surface $\mathbb{F}_n$ with $n>3$ is non-toric. It can be however understood as a component surface in a bigger toric CY$_3$. For example, a single surface $\mathbb{F}_3$ can be a component surface of the dual threefold of the brane web in Figure \ref{fig:P2-F1-F3}(c). 
Blowing up a point in a Hirzebruch surface is equivalent to adding an additional external 5-brane as in Figure \ref{fig:Hirzebruch}(b) for one blowup and also as in Figure \ref{fig:Hirzebruch}(c) for two blowups. 
Bigger brane webs for toric threefolds can be constructed by gluing such closed faces following the intersection structure of compact surfaces in the geometry.

{A collection of connected edges on 
a face is mapped to an effective curve class in the geometry. The volumes of each curve are the lengths of the corresponding edges in a given 5-brane web.}
 Gluing curves between two surfaces in geometry correspond to 
 internal edges between two closed faces of a web diagram. For example, the base curve $\ell$ in a $\mathbb{P}^2$ is one of three edges of 
 the closed face in Figure \ref{fig:P2-F1-F3}(a), and a fiber class $f$ in an $\mathbb{F}_1^1$ is the edge connecting two parallel internal edges as drawn in Figure \ref{fig:P2-F1-F3}(b). 
 Also the gluing curve between an $\mathbb{F}_1$ and an $\mathbb{F}_3$ is the edge between two internal faces in Figure \ref{fig:P2-F1-F3}(c). 

One non-trivial test for the above maps is to compare the prepotential from geometry with that from 5-brane web. 
In geometry, using the formulae given 
in \eqref{eq:triple1}-\eqref{eq:triple3}, we can compute the prepotentials of threefolds involving only a single surface $S=\mathbb{P}^2$ and $S=\mathbb{F}_1^1$ as
\begin{eqnarray}\label{eq:FofP2F1}
  6\mathcal{F}_{\mathbb{P}^2} &=& \phi^3 (K\cdot K) =  \phi^3(-3\ell)^2 = 9\, \phi^3 \ , \nonumber \\
  6\mathcal{F}_{\mathbb{F}_1^1} &=& \phi^3 (K\cdot K) =  \phi^3(-2h-f+x_1)^2  = 7 \phi^3 \ ,
\end{eqnarray}
where $\phi$ is the K\"ahler parameter and we used $\ell^2=1$ and \eqref{eq:canonicalclass}. Here we 
turned off all the non-compact K\"ahler parameters\footnote{When computing or comparing the prepotentials, we will always turn off all the non-compact K\"ahler parameters.
}. 
In 5-brane web, we first compute the volumes of internal faces and then integrate them 
to obtain a cubic prepotential. In order to compute the volumes of faces in a brane web, we need to fix the lengths of internal edges against K\"ahler parameters in the corresponding geometry. For this, we can use the above map for 2-cycles. 
For example, the volume of the base class $\ell$ in an isolated $\mathbb{P}^2$ is $vol(\ell) = 3\,\phi$. This implies the length of an internal edge in Figure \ref{fig:P2-F1-F3}(a) is $3\,\phi$. 
It follows then that the area of the internal face in the web diagram is $T_{\mathbb{P}^2} = \frac{9}{2}\phi^2$. 
By integrating this area, one finds the cubic prepotential that indeed agrees with \eqref{eq:FofP2F1}. Similarly, one can compute the prepotential for Figure \ref{fig:P2-F1-F3}(b). The volumes of 2-cycles in an $\mathbb{F}_1^1$ are $vol(e) = vol(x_1)=\phi$, $vol(f)=2\phi$, and $vol(h)=3\phi$. They correspond to the lengths of the edges $e,x_1,f,h$ of the 5-brane web in Figure \ref{fig:P2-F1-F3}(b). Using these lengths, one readily finds the area of internal face $T_{\mathbb{F}_1^1}=\frac{7}{2}\phi^2$. This leads to the same prepotential $\mathcal{F}_{\mathbb{F}_1^1}$ as \eqref{eq:FofP2F1}, as expected. 

The prepotential computation for Figure \ref{fig:P2-F1-F3}(c) is more involved, but straightforward. 
Let us first compute the prepotential in geometry. The intersection numbers of $S=S_1\cup S_2$ with $S_1 = \mathbb{F}_1,\, S_2 = \mathbb{F}_3$ are
\begin{align}
&&&S_1^3 = (K\cdot K)|_{\mathbb{F}_1} = 8 \ , &&S_2^3 = (K\cdot K)|_{\mathbb{F}_3} = 8 \ , &\nonumber \\
&&&S_1^2 S_2 =  e^2|_{S_2} = -3 \ , && S_1 S_2^2 = h^2|_{S_1} = 1 \ .	&
\end{align}
One then finds
\begin{equation}\label{eq:PrePF1F3}
  6\mathcal{F}_{\mathbb{F}_1\cup \mathbb{F}_3} = 8\phi_1^3 + 8\phi_2^3 -9\phi_1^2\phi_2 +3\phi_1\phi_2^2 \ .
\end{equation}

The length of the internal edges in the brane web can be identified as the volumes of curves \eqref{eq:edgevol}
\begin{eqnarray}
  &&vol(e|_{S_1}) = \phi_1 \ , \qquad vol(f|_{S_1}) = 2\phi_1-\phi_2 \ , \qquad vol(f|_{S_2}) = 2\phi_2-\phi_1 \ , \nonumber \\
  &&vol(h|_{S_1}) = vol(e|_{S_2}) = 3\phi_1-\phi_2 \ , \qquad vol(h|_{S_2}) = 5\phi_2 \ .
\end{eqnarray}
Note that the fiber classes $f|_{S_{1,2}}$ amount to the strings connecting two D5-branes and their volumes form the Cartan matrix of $SU(3)$ gauge algebra as 
\begin{equation}
  (vol(f|_{S_1}),vol(f|_{S_2})) = \left(\begin{array}{cc}2 & -1 \\ -1 & 2\end{array}\right)\cdot \left(\begin{array}{c}\phi_1 \\ \phi_2\end{array}\right)  \ ,
\end{equation}
which is expected as 
 the low energy field theory of the brane web is the $SU(3)$ gauge theory with 
the Chern-Simons level $\kappa=-2$.
Now the area of each internal face can be computed as
\begin{equation}
  T_{1} = \frac{1}{2}(2\phi_1-\phi_2)(4\phi_1-\phi_2) \ , \quad T_{2} = \frac{1}{2}(2\phi_2-\phi_1)(3\phi_1+4\phi_2) \ .
\end{equation}
This gives rise to the prepotential \eqref{eq:PrePF1F3} of $\mathcal{F}_{\mathbb{F}_1\cup \mathbb{F}_3}$ above.

5-brane webs for non-toric threefolds are not well-understood. Only case-by-case studies have been carried out and there is so far no systematic construction of brane webs for non-toric geometry. All known non-toric examples are quite non-trivially realized in brane systems with 
other additional ingredients such as 7-branes and orientifold planes. 
Let us list some known examples of non-toric brane webs. 

\begin{figure}
  \centering
  \includegraphics[width=.8\linewidth]{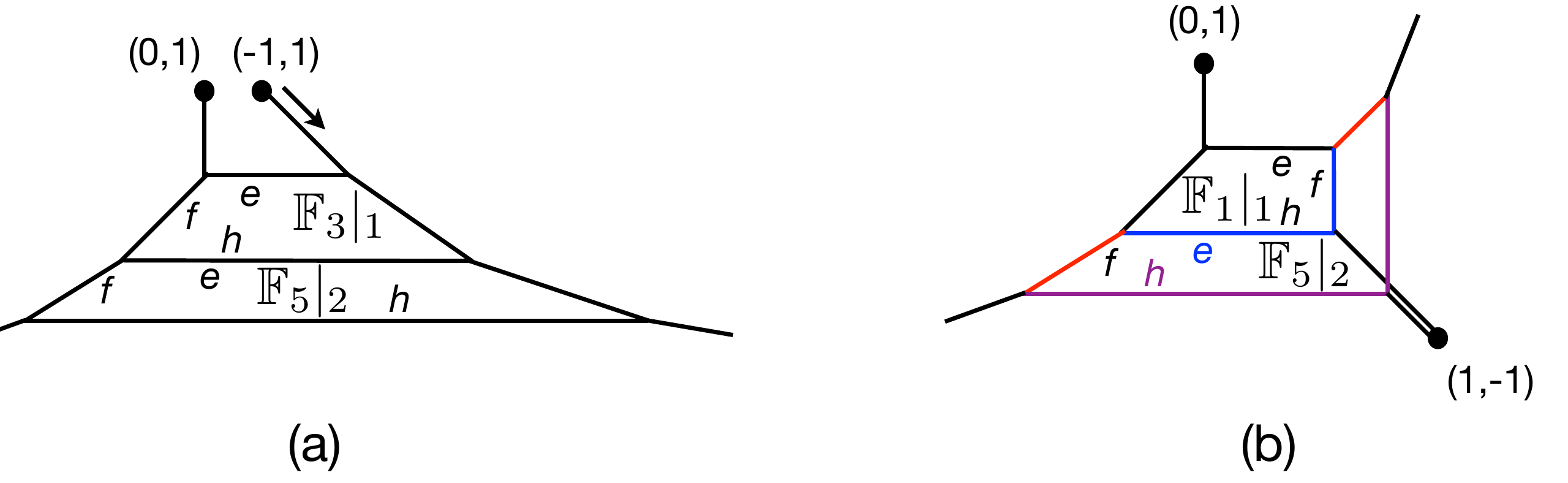}
  \caption{(a) The 5-brane web for a non-toric threefold of two intersecting surfaces $S_1=\mathbb{F}_3$ and $S_2=\mathbb{F}_5$. (b) The 5-brane web for a non-toric threefold of $S_1=\mathbb{F}_1$ and $S_2=\mathbb{F}_5$ which is the resulting 5-brane web after the Hanany-Witten transition is implemented on the 5-brane web (a) with $(-1,1)$ 7-brane.}
  \label{fig:F3-F5}
\end{figure}

One simple example is the web diagram in Figure \ref{fig:F3-F5}(a). The corresponding geometry is a non-toric CY$_3$ of two surfaces $S_1=\mathbb{F}_3$ and $S_2 =\mathbb{F}_5$ intersecting along a curve $C|_{S_1}=h, C|_{S_2}=e$. In order to draw a web diagram for this geometry, we introduced two 7-branes denoted by black dots in Figure \ref{fig:F3-F5}(a). 
 This non-toric threefold hence has a 5-brane web realization.
This geometry however leads to an inconsistent 5d field theory according to the criteria suggested in \cite{Jefferson:2018irk}. The associated field theory violates unitarity on Coulomb branch when all masses are switched off. So it has no consistent Coulomb branch unless we turn on certain mass parameters for the matter fields, which signals that the 5d theory has no consistent UV completion.
However, we can deform 
this geometry to another geometry by applying 
complex structure deformation \cite{Jefferson:2018irk}. The deformed geometry is a CY$_3$ of $S_1=\mathbb{F}_1,\,S_2=\mathbb{F}_5$ glued along a curve $C|_{S_1}=h+f,\, C|_{S_2}=e$. This geometry then leads to a consistent 5d field theory with non-trivial Coulomb branch while showing the same physics as the original threefold when masses are turned on, up to some decoupled sectors. {So it makes sense to study physics of M-theory compactification on the deformed geometry. }
This complex structure deformation in geometry can be implemented in brane web as a Hanany-Witten transition \cite{Hanany:1996ie}. In the case at hand, the complex structure deformation is a Hanany-Witten transition moving the $(-1,1)$ 7-brane along with the $(1,-1)$ direction as shown with the arrow in Figure \ref{fig:F3-F5}(a). 
As a result of the transition, we have the diagram 
in Figure \ref{fig:F3-F5}(b) with the $(1,-1)$ 7-branes on which two $(1,-1)$ 5-branes end. 
This deformed brane web in Figure \ref{fig:F3-F5}(b) 
realizes the threefold of $\mathbb{F}_1 \cup \mathbb{F}_5$. Here the Hirzebruch surface $\mathbb{F}_1$ is manifest, while 
the second surface $\mathbb{F}_5$ is non-trivially realized. The surface $\mathbb{F}_5$ is a combination of two other faces enclosed by colored edges in the web diagram in Figure \ref{fig:F3-F5}(b). The gluing curve $C$ is a union of the blue edges at the intersection $\mathbb{F}_1\cap \mathbb{F}_5$. The union of the purple edges in $\mathbb{F}_5$ is the section class $h$. Such non-trivial realization of $\mathbb{F}_5$ should be understood as an effect of the $(1,-1)$ 5-brane crossing over $\mathbb{F}_5$ and ending on the $(1,-1)$ 7-brane.


One can compare the brane web against the threefold of $S=S_1\cup S_2$ with $S_1=\mathbb{F}_1, S_2=\mathbb{F}_5$.  One readily finds the prepotential of the geometry 
\begin{equation}\label{eq:PrepF1F5}
  6\mathcal{F}_{\mathbb{F}_1\cup \mathbb{F}_5} = 8\phi_1^3 + 8\phi_2^3 -15\phi_1^2\phi_2 +9\phi_1\phi_2^2 \ .
\end{equation}
The lengths of the internal edges  in Figure \ref{fig:F3-F5}(b) 
are fixed from the volumes of geometric 2-cycles given as
\begin{align}
vol(e|_{S_1}) &= \phi_1 - \phi_2 \ ,  & vol(f|_{S_1}) &=2\phi_1-\phi_2 \ , \cr  vol(h|_{S_1}) &= 3\phi_1-2\phi_2 \ , \ & vol(f|_{S_2}) &= 2\phi_2-\phi_1  \ .	
\end{align}
With these parameters in the brane web, one obtains 
 the areas of the closed faces as
\begin{equation}
  T_{1} = \frac{1}{2}(2\phi_1-\phi_2)(4\phi_1-3\phi_2) \ , \qquad T_{2} = \frac{1}{2}(2\phi_2-\phi_1)(5\phi_1+4\phi_2) \ .
\end{equation}
The integration of these areas yields the same  prepotential as \eqref{eq:PrepF1F5}, as expected. 
It implies in turn that the tensions of magnetic monopole strings corresponding to 
the volumes of 4-cycles in the geometry are same as the areas of the compact faces in the corresponding 5-brane web in Figure \ref{fig:F3-F5}(b). This hence supports the brane realization of the non-toric threefold given by $S=\mathbb{F}_1\cup \mathbb{F}_5$.

\begin{figure}
  \centering
  \includegraphics[width=.85\linewidth]{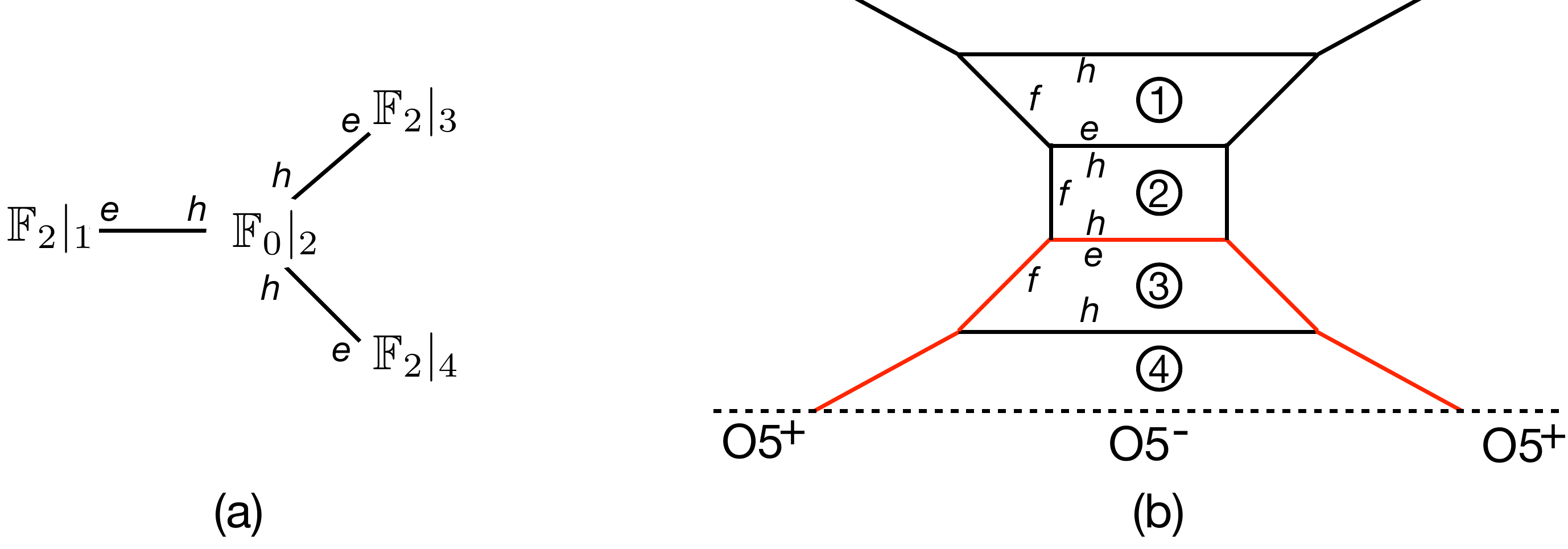}
  \caption{(a) A non-toric threefold of four surfaces $\mathbb{F}_2\cup \mathbb{F}_0 \cup \mathbb{F}_2 \cup \mathbb{F}_2$. (b) The brane web for the threefold (a). This is also a brane web for the 5d $SU(8)$ gauge theory at low energy.}
  \label{fig:SO8}
\end{figure}

Another non-toric example
 is given in Figure \ref{fig:SO8}. The diagram in Figure \ref{fig:SO8}(b) is a typical 5-brane web for the 5d $SO(8)$ gauge theory. The associated geometry is a non-toric Calabi-Yau threefold formed by four Hirzebruch surfaces $S_1 = \mathbb{F}_2, S_2 = \mathbb{F}_0, S_3 = \mathbb{F}_2 ,$ and $ S_4=\mathbb{F}_2$ which is illustrated as the diagram in Figure \ref{fig:SO8}(a). 
 An interesting yet simple structure of these surfaces is that there are three gluing curves $C_i$ with $i=1,2,3$ at the intersections $S_1\cap S_2$, $S_3\cap S_2$, and $S_4\cap S_2$, respectively. More precisely, the gluing curves are
\begin{eqnarray}
  C_1|_{S_1} = e|_{S_1} , \quad C_2|_{S_3} = e|_{S_3} , \quad C_3|_{S_4} = e|_{S_4} , \quad C_1|_{S_2} = C_2|_{S_2} =C_3|_{S_2} =h|_{S_2}. \ \
\end{eqnarray}
One can identify the compact faces labeled by circled numbers in Figure \ref{fig:SO8}(b) 
as 
the Hirzebruch surfaces in CY$_3$. The map is given as follows:
\begin{equation}
  \mathbb{F}_2|_1 = \small\textcircled{\scriptsize 1}  \ , \qquad \mathbb{F}_0|_2 = \small\textcircled{\scriptsize 2}  \ , \qquad \mathbb{F}_2|_3 = \small\textcircled{\scriptsize 3}  \ , 
\end{equation}
and
\begin{equation}
  \mathbb{F}_2|_4 = \small\textcircled{\scriptsize 3}  + 2\cdot \small\textcircled{\scriptsize 4} \ .
\end{equation}
As one may notice, the fourth surface $S_4=\mathbb{F}_2$ is not a single face but rather non-trivially realized as 
a combination of multiple faces surrounded with the red edges. This happens due to an O5-plane 
 at the bottom. So one way of constructing brane webs for non-toric threefolds is to use orientifold planes 
in a 5-brane configuration.

One can again check the brane web realization by comparing the prepotential. In geometry, the prepotential computation is straightforward. The intersection structure above yields 
\begin{equation}\label{eq:PrepF2F0F2F2}
  6\mathcal{F}_{\mathbb{F}_2\cup\mathbb{F}_0\cup\mathbb{F}_2\cup\mathbb{F}_2} = 8\sum_{i=1}^4\phi_i^3 -6\phi_2^2(\phi_1+\phi_3+\phi_4) \ .
\end{equation}
The volumes of the 2-cycles in the threefold are
\begin{eqnarray}\label{eq:SO8-vol}
  &&vol(h|_{S_2})=vol(e|_{S_1}) =vol(e|_{S_3}) =vol(e|_{S_4}) = 2\phi_2 \ , \nonumber \\
  && vol(h|_{S_1}) = 4\phi_1 \ ,  \quad vol(h|_{S_3}) = 4\phi_2 \ , \quad vol(h|_{S_1}) = 4\phi_3 \ ,
\end{eqnarray}
and
\begin{equation}
  (vol(f|_{S_1}),vol(f|_{S_2}),vol(f|_{S_3}),vol(f|_{S_4})) = \left(\begin{array}{cccc}2 & -1 & 0 &0\\ -1 & 2 & -1 & -1 \\ 0 & -1 & 2 & 0 \\ 0 & -1 & 0 & 2\end{array}\right)\cdot\left(\begin{array}{c}\phi_1\\\phi_2 \\ \phi_3 \\ \phi_4\end{array}\right) \ .
\end{equation}
So the fiber classes $f_{S_i}$ form the Cartan matrix of $SO(8)$ algebra.

The volumes of curves in 
\eqref{eq:SO8-vol} 
fix the parameters of the brane web in Figure \ref{fig:SO8}(b).
In terms of the geometric parameters, the areas of the internal faces read 
\begin{eqnarray}
  &&T_{\small\textcircled{\scriptsize 1}} = (2\phi_1-\phi_2)(2\phi_1+\phi_2) \ , \qquad T_{\small\textcircled{\scriptsize 2}} = 2\phi_2(2\phi_2-\phi_1-\phi_3-\phi_4) \ , \nonumber \\
 && T_{\small\textcircled{\scriptsize 3}} = (2\phi_3-\phi_2)(2\phi_3+\phi_2) \ , \qquad T_{\small{\textcircled{\scriptsize 3}}+2\small\textcircled{\scriptsize 4}} = (2\phi_4-\phi_2)(2\phi_4+\phi_2) \ .
\end{eqnarray}
Indeed, 
the prepotential obtained by integrating these areas agrees with the prepotential \eqref{eq:PrepF2F0F2F2} of the non-toric threefold above.




\begin{figure}
  \centering
  \includegraphics[width=.55\linewidth]{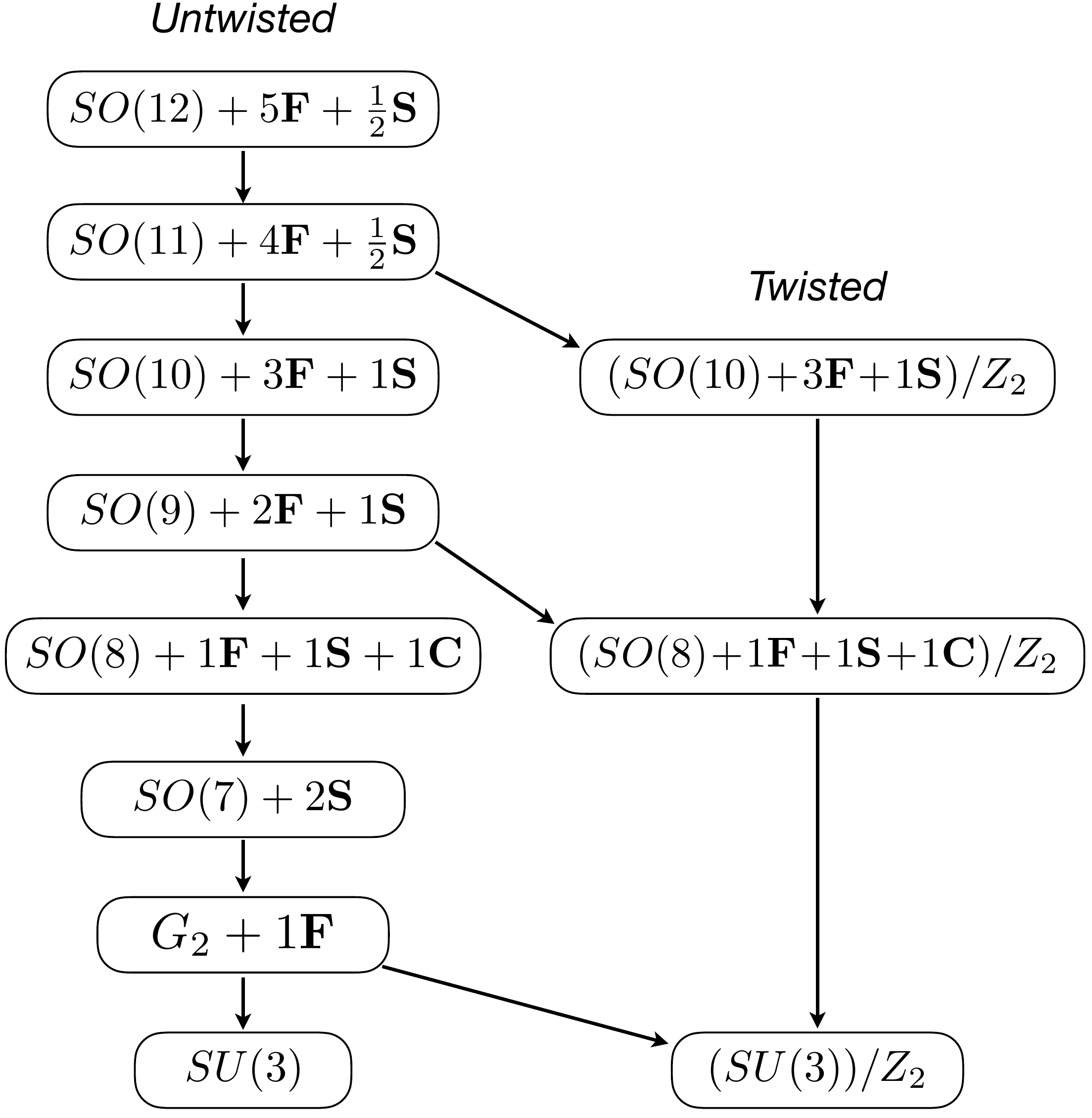}
  \caption{Higgsing hierarchy on a $-3$ curve. ${\bf F}, {\bf S},$ and ${\bf C}$ denote the fundamental, the spinor, and the conjugate spinor representation of gauge groups respectively.}
  \label{fig:HiggsChainO-3}
\end{figure}

\section{$SO(N)$ theories on $-3$ curve}\label{sec:SON03}
In this section we construct 5-brane webs for circle compactifications of 6d SCFTs with $SO(N)$ gauge symmetry supported on a $-3$ curve in the base of the elliptic Calabi-Yau threefold in F-theory compactification \cite{Heckman:2015bfa,Heckman:2013pva}. We shall first consider a Type IIB 5-brane web for the 6d SCFT with $SO(12)\times Sp(2)$ gauge symmetry when compactified on a circle and discuss how the brane web changes following RG flows in its Higgs branches. These RG flows in the brane system yield the $(p,q)$ 5-brane realizations for the theories on a $-3$ curve we are interested in.  We will also extend this construction to involve new Higgs branch flows giving rise to twisted compactifications of 6d SCFTs. The hierarchy of RG fixed points that we will discuss in this section is given in Figure \ref{fig:HiggsChainO-3}.

\subsection{Untwisted theories}
\begin{figure}
  \centering
  \includegraphics[width=1.\linewidth]{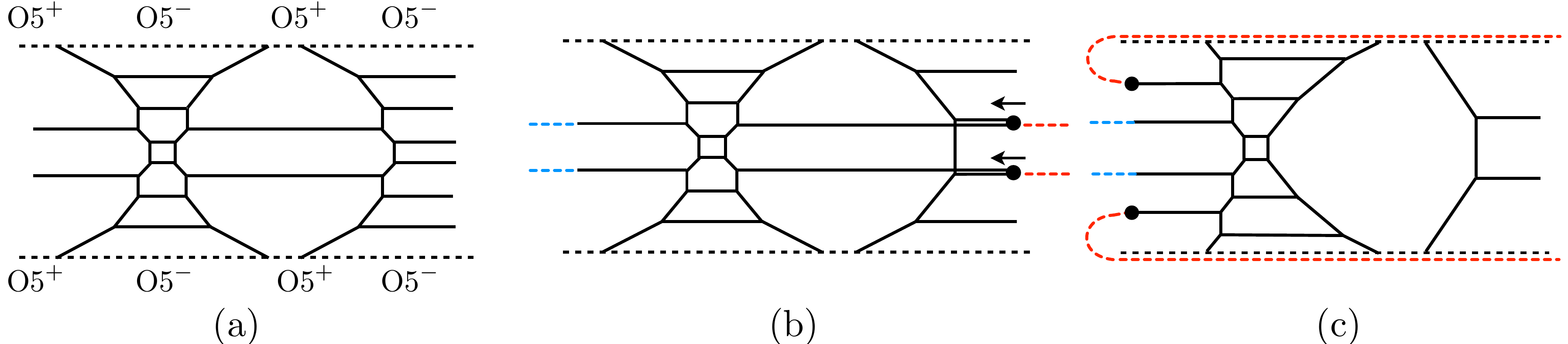}
  \caption{(a) 
  A 5-brane configuration for the 6d $[Sp(2)]  \stackrel{\mathfrak{so}_{12}}{\bf 4} \ \stackrel{\mathfrak{sp}_2}{\bf 1}  [SO(12)]$ theory. (b) Higgsing (a) by using a pair of two $Sp(2)$ fundamental hypermultiplets. The resulting 5-brane web is dual to the 6d $SO(12)$ gauge theory on a $-4$ curve intersecting with a $-1$ curve, $[Sp(4)]  \stackrel{\mathfrak{so}_{12}}{\bf 4} \ \stackrel{}{\bf 1}  [SO(4)]$. { The (blue) dotted lines denote the monodromy cuts of (1,0) 7-branes on the left, while the (red) dotted lines denote the monodromy cuts of Higgsed 7-branes on the right. (c) An equivalent diagram by sending two D7-branes in the digram (b) to the left. The (red) dotted lines are oriented such that they point to the right through the O5-planes.}}
  \label{fig:SO12-Sp2}
\end{figure}
 We start with  the 6d SCFT with a product gauge group $SO(12)\times Sp(2)$ where the $SO(12)$ gauge symmetry is on a $-4$ curve while the $Sp(2)$ gauge symmetry is over a $-1$ curve. In the convention in \cite{Heckman:2015bfa}, this theory can be depicted as
\begin{equation}
	[Sp(2)] \ \stackrel{\mathfrak{so}_{12}}{\bf4} \ \stackrel{\mathfrak{sp}_2}{\bf1} \ [SO(12)] \ .
\end{equation}
We shall consider various RG flows of this theory developed by vevs of Higgs scalar fields in the hypermultiplets.

This theory has a 5-brane web realization in Type IIB string theory as depicted in Figure \ref{fig:SO12-Sp2}(a). There are six and two internal D5-branes suspended between other $(p,q)$ 5-branes in the middle of the diagram. With two O5-planes on the top and the bottom of the diagram, string states stretched between them implement the $SO(12)\times Sp(2)$ gauge bosons and bi-fundamental matter. Other external D5-branes give rise to the global symmetry $Sp(2)\times SO(12)$. 

{We remark that in Figure \ref{fig:SO12-Sp2}, we have chosen the directions of 7-brane monodromy cuts such that they point still the same direction even after the Hanany-Witten transition. For instance, the monodromy cuts of two 7-branes on the right of Figure \ref{fig:SO12-Sp2}(b) are oriented along (1,0) direction (to the right), and the same 7-branes moved to the left by the Hanany-Witten transition are still assigned such that they point to the right along the O5-planes, as shown in Figure \ref{fig:SO12-Sp2}(b). It is then straightforward that the charges are conserved on the O5-planes. 
From here on, the orientation of 7-brane monodromy cuts should be understood such that the charges are conserved on 5-branes. See Appendix \ref{sec:appB} for more details.
}

Let us now discuss Higgs branches associated with 
the scalars in the $Sp(2)$ fundamental hypermultiplets charged under the $SO(12)$ flavor symmetry. We shall give a non-trivial vev to these Higgs scalars such that only the $SO(8)$ part in the $SO(12)$ flavor symmetry remains unbroken. The RG flow followed by this Higgs vev leads to a new IR SCFT which now has $SO(12)\times Sp(1)$ gauge symmetry. Given this IR theory, we can perform another Higgsing breaking the $SO(8)$ symmetry to $SO(4)$ flavor symmetry. Consequently we reach another IR theory that is the 6d SCFT with $SO(12)$ gauge symmetry. This sequence of RG flows can be summarized as \cite{Heckman:2016ssk}: 
\begin{equation}
	[Sp(2)]  \stackrel{\mathfrak{so}_{12}}{\bf 4} \ \stackrel{\mathfrak{sp}_2}{\bf 1}  [SO(12)] \  \rightarrow \ [Sp(3)]  \stackrel{\mathfrak{so}_{12}}{\bf 4} \ \stackrel{\mathfrak{sp}_1}{\bf1}  [SO(8)] \ \rightarrow  \ [Sp(4)]  \stackrel{\mathfrak{so}_{12}}{\bf 4}  {\bf 1} \  [SO(4)] \ .
\end{equation}


In the brane web, the first Higgsing of the $Sp(2)$ fundamental scalars amounts to aligning two external $SO(12)$ D5-branes and an internal $Sp(2)$ D5-brane together so that two external D5-branes can end on a single D7-brane while keeping the S-rule \cite{Hanany:1996ie}. The diagram in Figure \ref{fig:SO12-Sp2}(b) shows the RG flow in the brane web that Higgses the UV $SO(12)\times Sp(2)$ gauge symmetry to the $SO(12)$ symmetry in IR.
Two black dots in this diagram denote D7-branes and two external D5-branes are ending on each 7-brane. As one can see the $Sp(2)$ gauge symmetry on the right side is fully Higgsed. This brane web hence corresponds to the 6d SCFT with the $SO(12)$ gauge symmetry over a $-4$ curve intersecting with a $-1$ curve. This theory (or the brane web) has $Sp(4)\times SO(4)$ global symmetry. This will become manifest in an equivalent brane web given in the Figure \ref{fig:SO12-Sp2}(c) which we can obtain from the diagram (b) by the Hanany-Witten transitions after moving two 7-branes to the left-side.

\begin{figure}
  \centering
  \includegraphics[width=1.\linewidth]{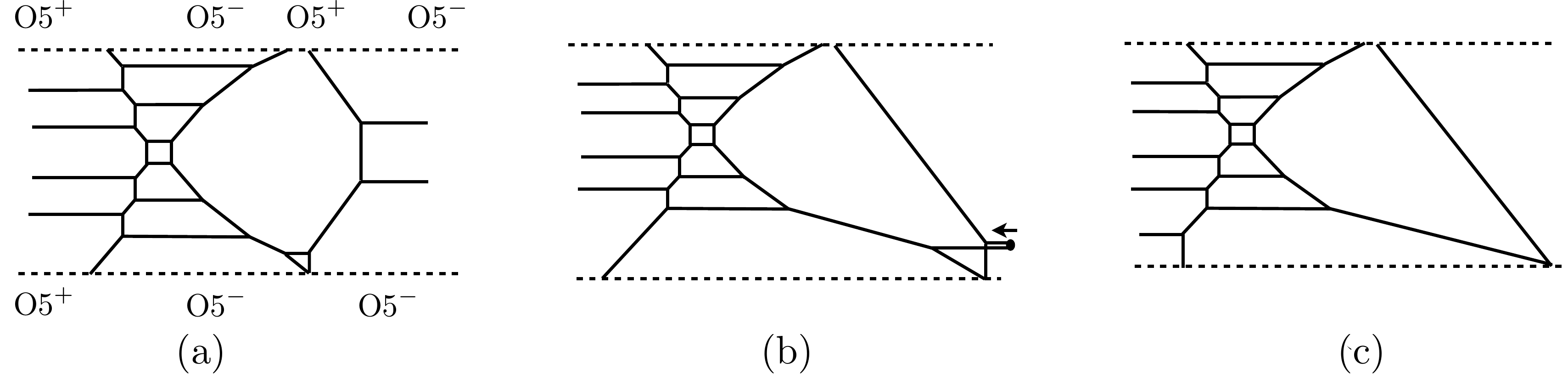}
  \caption{(a) A 5-brane web for the 6d $[Sp(4)]  \stackrel{\mathfrak{so}_{12}}{\bf4}  {\bf1} \  [SO(4)]$ theory. 
  (b) The brane web for the 6d $SO(12)$ gauge theory on a $-3$ curve, $[Sp(5)]  \stackrel{\mathfrak{so}_{12}}{\bf 3}$. (c) An equivalent web diagram after the Hanany-Witten transition moving the 7-brane on the right hand side to the left hand side on the diagram (b).}
  \label{fig:SO12-Sp0}
\end{figure}
\paragraph{\underline{$SO(12)$ gauge theory}} Now we will blow down the $-1$ curve in the base. Blowdown of an exceptional curve decreases the self-intersection numbers of adjacent curves by one. So, after doing this, we expect that our theory reduces to a 6d theory over a single curve of the self-intersection $-3$ in the base. This blowdown of $-1$ curve can also be done by a Higgsing.  In the brane diagrams in Figure \ref{fig:SO12-Sp0}, we first deform the diagram in Figure \ref{fig:SO12-Sp2}(c) to Figure \ref{fig:SO12-Sp0}(a). This deformation is a smooth transition by tuning the positions of $(p,q)$-branes near orientifold planes as discussed in \cite{Hayashi:2017btw}. Now a new Higgs branch can open up in the new diagram where the heights of two external D5-branes on the right hand side are aligned  with height of the adjacent internal D5-brane,  as depicted in Figure \ref{fig:SO12-Sp0}(b). 
We claim that this is the Higgsing blowing down the $-1$ curve in the base while leaving a single $-3$ curve with gauge group $SO(12)$. More precisely, the transition from the 5-brane configuration of Figure \ref{fig:SO12-Sp0}(a) to that of Figure \ref{fig:SO12-Sp0}(b) is equivalent to the Higgsing in the 6d field theory
\begin{equation}
 [Sp(4)]  \stackrel{\mathfrak{so}_{12}}{\bf4}  {\bf1} \  [SO(4)] \ \ \rightarrow \ \ [Sp(5)]  \stackrel{\mathfrak{so}_{12}}{\bf3} \ .
\end{equation}
So we propose that the brane diagram in Figure \ref{fig:SO12-Sp0}(b) (or Figure \ref{fig:SO12-Sp0}(c) after Hanany-Witten transition) is a Type IIB realization of the 6d SCFT of $SO(12)$ gauge symmetry on a $-3$ curve coupled to five hypermultiplets in the fundamental representation and an half-hypermultiplet in the spinor representation of the $SO(12)$. 

We can associate this brane web with the dual CY$_3$ geometry of the 6d $SO(12)$ SCFT on $-3$ curve when compactified on a circle. The corresponding CY$_3$ geometry is given in \cite{Bhardwaj:2018yhy}.
An elliptic Calabi-Yau threefold for a 6d theory after resolving all the singularities can be represented by a collection of compact K\"ahler surfaces (or four-cycles) with finite volumes, each of which 
is a ruled surface over a genus $g$ curve possibly with a number of blowups as we discussed above. 


\begin{figure}
  \centering
  \includegraphics[width=1\linewidth]{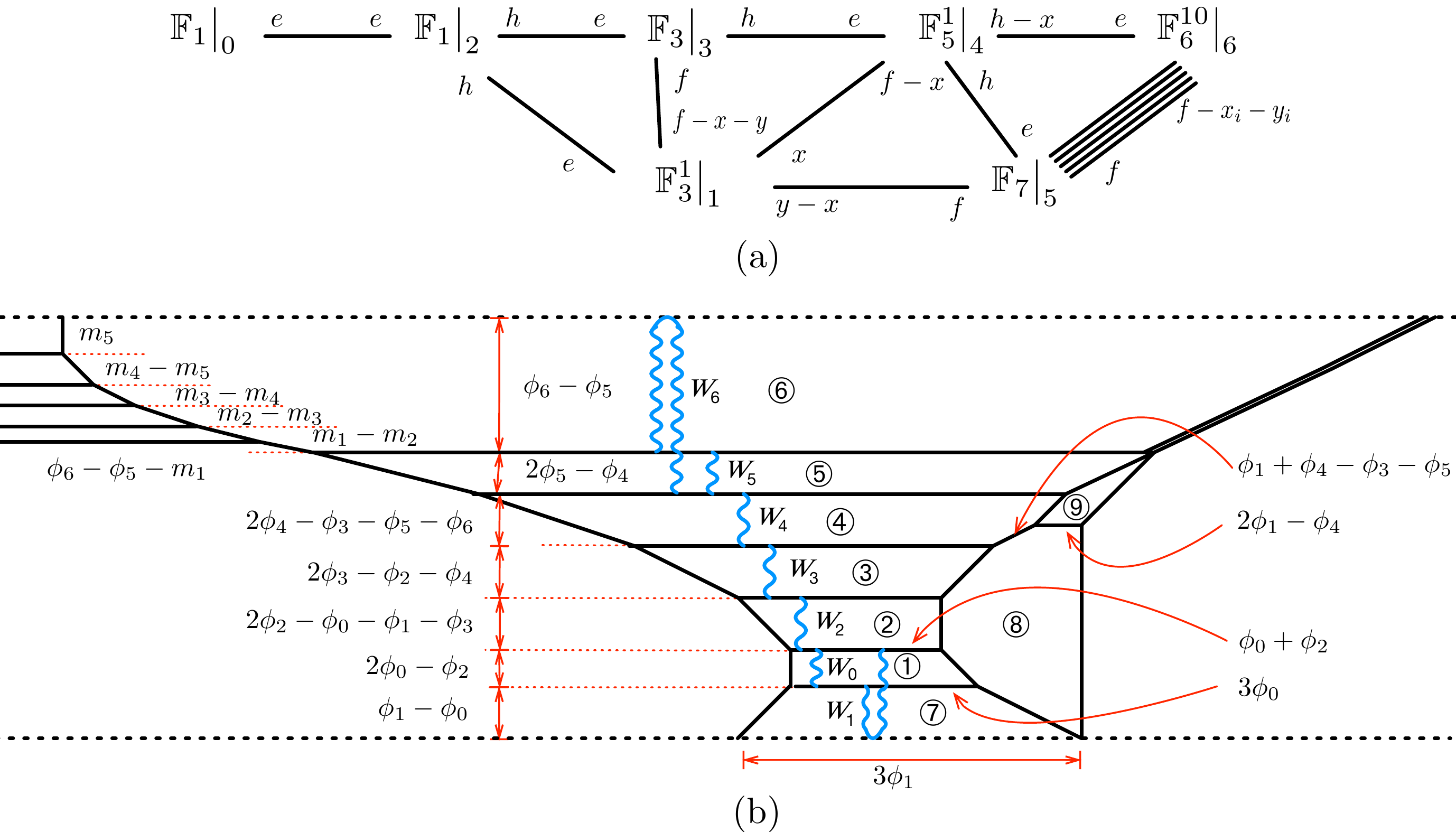}
  \caption{(a) The elliptic threefold for the 6d $SO(12)$ theory with 5 fundamentals and a $1/2$ spinor.
  (b) The 5-brane web dual to the geometry in (a).}
  \label{fig:SO12-5F}
\end{figure}

The geometry for the $SO(12)$ theory is drawn in Figure \ref{fig:SO12-5F}(a). 
This diagram represents the intersection structure of 7 complex ruled surfaces $S_i,$ $(i=0,1,\cdots 6)$ with K\"ahler parameters $\phi_i$ in an elliptically fibered Calabi-Yau threefold.
We can compute the cubic prepotential (or the triple intersection numbers) of this geometry by employing the computation algorithm given in the previous section. The result is
\begin{eqnarray}\label{eq:SO12-3}
  6\mathcal{F}_{SO(12)} &=& 8\phi_0^3 + 6\phi_1^3 + 8\phi_2^3 + 8\phi_3^3+7\phi_4^3+8\phi_5^3-2\phi_6^{3}-3\phi_0^2\phi_2-3\phi_0\phi_2^2-9\phi_1\phi_2^2 \nonumber \\
 && +3\phi_1^2\phi_2-6\phi_1\phi_3^2-3\phi_1^2\phi_4-3\phi_1\phi_4^2-6\phi_1\phi_5^2-9\phi_2^2\phi_3+3\phi_2\phi_3^2-15\phi_3^2\phi_4\nonumber \\
 &&+9\phi_3\phi_4^2 -21\phi_4^2\phi_5+15\phi_4\phi_5^2-18\phi_4^2\phi_6+12\phi_4\phi_6^2-30\phi_5^2\phi_6 \nonumber \\
 &&+6\phi_1\phi_2\phi_3+6\phi_1\phi_3\phi_4+6\phi_1\phi_4\phi_5+30\phi_4\phi_5\phi_6 \ .
\end{eqnarray}
We can compare this geometry against our brane web configuration. For this, we 
use the brane web in Figure \ref{fig:SO12-5F}(b) which can be obtained from the diagram in Figure \ref{fig:SO12-Sp0}(c) by a series of flop transitions as well as brane moves discussed in Appendix \ref{sec:appA}.

A collection of internal faces in a 5-brane web can be identified with compact K\"ahler surfaces in the dual geometry. In the case at hand, the internal faces of the 5-brane web in Figure \ref{fig:SO12-5F}(b) 
can be associated with the surfaces $S_i$ in the geometry as follows:
\begin{eqnarray}
  &&\mathbb{F}_1|_{0} = \textcircled{\scriptsize 1} \ , \quad 
  \mathbb{F}^2_3|_{1} = \textcircled{\scriptsize 1}+2\cdot\textcircled{\scriptsize 7}+\textcircled{\scriptsize 8}\ , \quad
  \mathbb{F}_1|_{2} = \textcircled{\scriptsize 2} \ , \quad \mathbb{F}_3|_{3} = \textcircled{\scriptsize 3} \ , \quad \nonumber \\
  && \mathbb{F}_5^1|_{4} = \textcircled{\scriptsize 4} \ , \quad \mathbb{F}_7|_{5} = \textcircled{\scriptsize 5} + \textcircled{\scriptsize 9} \ , \quad \mathbb{F}^{10}_6|_{6} = \textcircled{\scriptsize 5}+2\cdot\textcircled{\scriptsize 6} \  .
\end{eqnarray}
The 2-cycles in the geometry can also be identified with internal edges in a 5-brane web. For example, all the fiber classes $f|_{S_i}$ are identified with the edges denoted by $W_i$'s, which give rise to W-bosons in the low energy field theory. 
We note that the charges of the  7 fiber classes form the Cartan matrix of the affine $SO(12)$ gauge algebra as
\begin{equation}
  vol(f|_{S_0},f|_{S_1},f|_{S_2},f|_{S_3},f|_{S_4},f|_{S_5},f|_{S_6}) = 
  \left(\begin{array}{ccccccc}
  2 & 0 & -1 & 0 & 0 & 0 & 0 \\
  0 & 2 & -1 & 0 & 0 & 0 & 0 \\
  -1 & -1 & 2 & -1 & 0 & 0 & 0 \\
  0 & 0 & -1 & 2 & -1 & 0 & 0 \\
  0 & 0 & 0 & -1 & 2 & -1 & -1 \\
  0 & 0 & 0 & 0 & -1 & 2 & 0 \\
  0 & 0 & 0 & 0 & -1 & 0 & 2
  \end{array}\right)\cdot \left(\begin{array}{c}\phi_0 \\ \phi_1 \\ \phi_2 \\ \phi_3 \\ \phi_4 \\ \phi_5 \\ \phi_6 \end{array}\right) \ ,
\end{equation}
which is expected for $SO(12)$ gauge algebra.
The edge between compact faces $\footnotesize\textcircled{\scriptsize1}$ and $\footnotesize\textcircled{\scriptsize2}$ is identified with the gluing curve of $C|_{S_0}=e$ and $C|_{S_2}=e$ 
between $S_0$ and $S_2$ whose volume is $vol(c)=\phi_0+\phi_2$.
With this identification, lengths of internal edges  in the  brane web are fixed by the volumes of the associated 2-cycles in terms of the K\"ahler parameters $\phi_i$ (also $m_i$ for non-compact K\"ahler parameters) and the result is written on the brane diagram in Figure \ref{fig:SO12-5F}(b). The areas of the internal compact faces of the 5-brane web in Figure \ref{fig:SO12-5F}(b), $\frac{\partial\mathcal{F}_{SO(12)}}{\partial \phi_i}$  ($i=0, \cdots,6$), are then given by
\begin{align}
  &T_{\footnotesize\textcircled{\scriptsize1}} = \frac{1}{2}(4\phi_0\!+\!\phi_2)(2\phi_0\!-\!\phi_2)  
  \ , \nonumber \\
  &T_{\footnotesize\textcircled{\scriptsize 1}+2\footnotesize\textcircled{\scriptsize 7}+\footnotesize\textcircled{\scriptsize 8}} = \frac{1}{2}(6\phi_1^2\!-\!3\phi_2^2\!-\!2\phi_3^2\!-\!\phi_4^2\!-\!2\phi_5^2\!+\!2\phi_2(\phi_1\!+\!\phi_3)\!+\!2\phi_4(-\phi_1\!+\!\phi_3\!+\!\phi_5)) 
   \ , \nonumber \\
  &T_{\footnotesize\textcircled{\scriptsize 2}} = \frac{1}{2}(4\phi_2\!+\!\phi_0\!-\!\phi_1\!-\!\phi_3)(2\phi_2\!-\!\phi_0\!-\!\phi_1\!-\!\phi_3) \ , \nonumber \\
  &T_{\footnotesize\textcircled{\scriptsize 3}} = \frac{1}{2}(2\phi_3\!-\!\phi_2\!-\!\phi_4)(4\phi_3\!-\!2\phi_1\!+\!3\phi_2\!-\!3\phi_4) \ , \nonumber \\
  &T_{\footnotesize\textcircled{\scriptsize 4}} = \frac{1}{2}(-\phi_1^2\!-\!5\phi_3^2\!+\!7\phi_4^2\!+\!5\phi_5^2\!+\!4\phi_6^2\!+2\phi_1(\phi_3\!+\!\phi_5)\!+\!10\phi_5\phi_6\!+\!2\phi_4(3\phi_3\!-\!\phi_1\!-\!7\phi_5\!-\!6\phi_6)) , \nonumber \\
  &T_{\footnotesize\textcircled{\scriptsize 5}+\footnotesize\textcircled{\scriptsize 9}} =\frac{1}{2}(2\phi_5\!-\!\phi_4)(4\phi_5\!-\!2\phi_1\!+\!7\phi_4\!-\!10\phi_6) \ , \nonumber \\
  &T_{\footnotesize\textcircled{\scriptsize 5}+2\footnotesize\textcircled{\scriptsize 6}} = -3\phi_4^2\!-\!5\phi_5^2\!-\!\phi_6^{2}+\phi_4(5\phi_5+4\phi_6) \ .
\end{align}
The result perfectly agrees with the volumes of compact surfaces which come from first derivatives of the triple intersection numbers with respect to $\phi_i$ in the threefold of the $SO(12)$ theory given in equation (\ref{eq:SO12-3}).
This strongly supports our proposal of the 5-brane web for the 6d $SO(12)$ SCFT on a $-3$ curve, given in Figure \ref{fig:SO12-5F}(b).

\begin{figure}
  \centering
  \includegraphics[width=.9\linewidth]{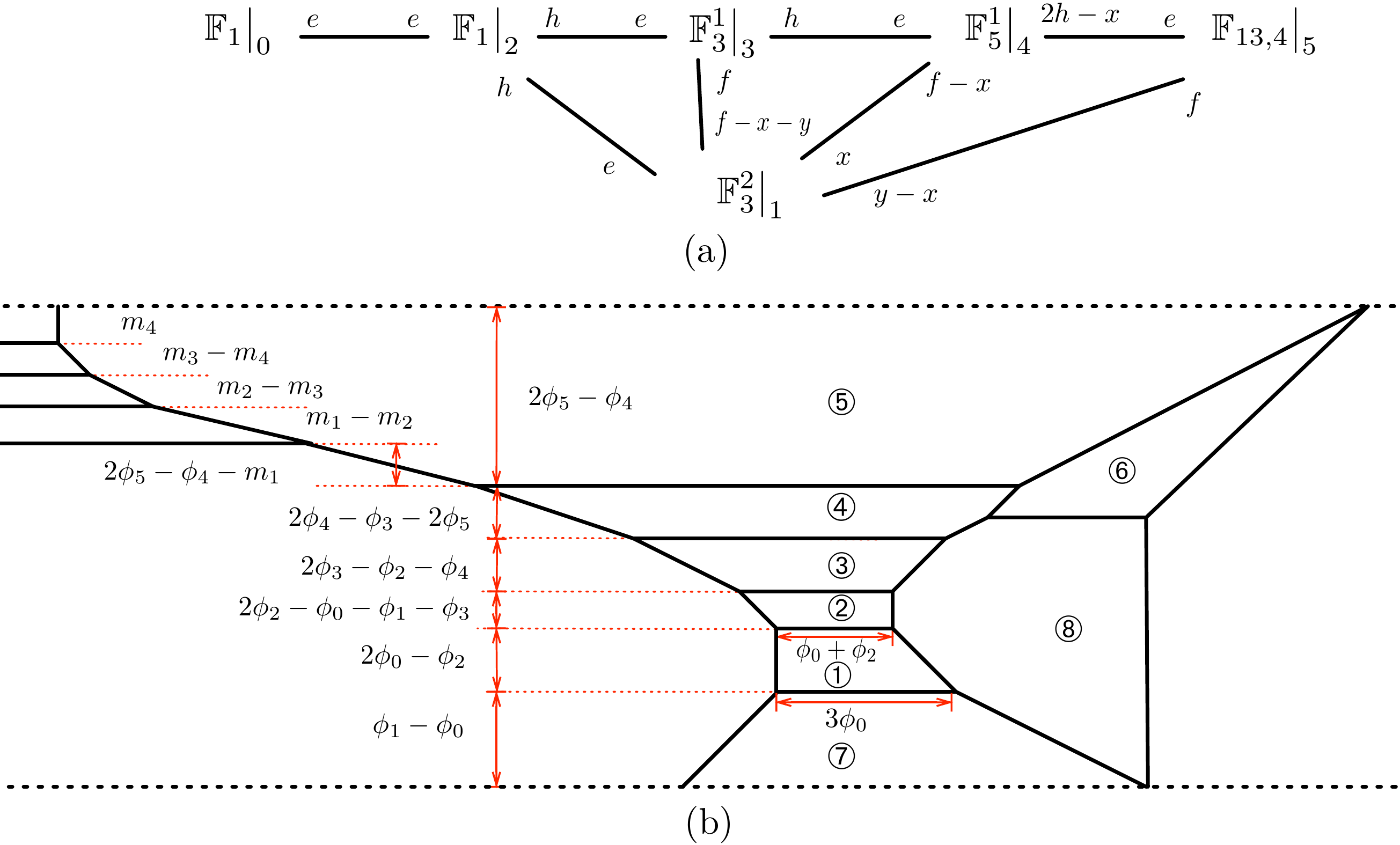}
  \caption{(a) The elliptic threefold for the 6d $SO(11)$ theory with 4 fundamentals and a $1/2$ spinor hyper.
  (b) A dual brane web for (a).}
  \label{fig:SO11-4F}
\end{figure}

\paragraph{\underline{$SO(11)$ gauge theory}}
We shall now Higgs the above brane diagram for the 6d $SO(12)$ theory to 
obtain the brane configurations for a family of 6d SCFTs on $-3$ curve.
Our first example is the Higgsing to the 6d SCFT of $SO(11)$ gauge symmetry with 4 fundamental hypermultiplets and a 1/2 spinor hypermultiplet. 
As the Higgs branch opens up in the massless limit at certain subspace in the Coulomb moduli, this Higgsing is achieved by bringing the top internal and the top external D5-branes together toward the top orientifold 5-brane in Figure \ref{fig:SO12-5F}(b).
This is a standard Higgsing procedure from $SO(2N)$ to $SO(2N-1)$ in a 5-brane web,  
where a hypermultiplet in the fundamental representation of $SO(2N)$ becomes massless. 
In other words, we set the masses (or heights of D5-branes) as
\begin{equation}
  \phi_6-\phi_5 = m_5 = 0 \ , 
\end{equation}
in 
the 5-brane web given in Figure \ref{fig:SO12-5F}(b).
This leads to the 5-brane configuration in Figure \ref{fig:SO11-4F}(b).
We claim that this brane configuration implements the 6d SCFT with $SO(11)$ gauge symmetry coupled to 4 fundamental hypers and a 1/2 spinor hyper.

The resolved CY$_3$ geometry for this theory is given in \cite{Bhardwaj:2018yhy}, which is summarized in Figure \ref{fig:SO11-4F}(a). From the intersection structure of 6 compact surfaces $S_i,\, i=0,1,\cdots,5$, we compute the cubic prepotential for this geometry
\begin{eqnarray}
  6\mathcal{F}_{SO(11)} &=& 8\phi_0^3+6\phi_1^3+8\phi_2^3+8\phi_3^3+7\phi_4^3-24\phi_5^3 -3\phi_0^2\phi_2-3\phi_0\phi_2^2-9\phi_1\phi_2^2+3\phi_1^2\phi_2 \nonumber \\
  &&-6\phi_1\phi_3^2-3\phi_1^2\phi_4-3\phi_1\phi_4^2-6\phi_1\phi_5^2-9\phi_2^2\phi_3+3\phi_2\phi_3^2-15\phi_3^2\phi_4+9\phi_3\phi_4^2 \nonumber \\
  &&-39\phi_4^2\phi_5+57\phi_4\phi_5^2+6\phi_1\phi_2\phi_3+6\phi_1\phi_3\phi_4+6\phi_1\phi_4\phi_5 \ .
\end{eqnarray}

We can identify our 5-brane configuration with this geometry. The relation between the compact K\"ahler surfaces in geometry and the internal faces in the brane diagram is given by
\begin{eqnarray}
  &&\mathbb{F}_1|_{0} = \textcircled{\scriptsize 1} \ , \qquad 
  \mathbb{F}_3^2|_{1} = \textcircled{\scriptsize 1}+2\cdot\textcircled{\scriptsize 7}+\textcircled{\scriptsize 8} \ , \qquad
  \mathbb{F}_1|_{2}=\textcircled{\scriptsize 2} \ , 
  \nonumber \\
  &&
 \mathbb{F}_3|_{3} = \textcircled{\scriptsize 3} \ , \qquad
  \mathbb{F}^1_5|_{4} = \textcircled{\scriptsize 4} \ , \qquad 
  \mathbb{F}_{13,4}|_{5} = 2\cdot \textcircled{\scriptsize 5}+\textcircled{\scriptsize 6} \ . 
\end{eqnarray}
This relation can be checked by comparing the monopole string tensions with respect to the K\"ahler parameters $\phi_i$. In the brane configuration, the monopole string tensions (or the areas of the internal faces) are
\begin{align}
  &T_{\footnotesize\textcircled{\scriptsize 1}} = \frac{1}{2}(4\phi_0+\phi_2)(2\phi_0-\phi_2) \ , \cr 
  &T_{\footnotesize\textcircled{\scriptsize 1}+2\footnotesize\textcircled{\scriptsize 7}+\footnotesize\textcircled{\scriptsize 8}} = \frac{1}{2}(6\phi_1^2\!-\!3\phi_2^2\!-\!2\phi_3^2\!-\!\phi_4^2\!-\!2\phi_5^2\!+\!2\phi_2(\phi_1\!+\!\phi_3)\!-\!2\phi_4(\phi_1\!-\!\phi_3\!-\!\phi_5)) \ , \cr   
  &T_{\footnotesize\textcircled{\scriptsize 2}} = \frac{1}{2}(4\phi_2\!+\!\phi_0\!-\!\phi_1\!-\!\phi_3)(2\phi_2\!-\!\phi_0\!-\!\phi_1\!-\!\phi_3) \ , \nonumber \\
  &T_{\footnotesize\textcircled{\scriptsize 3}} = \frac{1}{2}(2\phi_3\!-\!\phi_2\!-\!\phi_4)(4\phi_3\!-\!2\phi_1\!+\!3\phi_2\!-\!3\phi_4) \ , \\ 
  &T_{\footnotesize\textcircled{\scriptsize 4}} = \frac{1}{2}(-\phi_1^2\!-\!5\phi_3^2 \!+\!7\phi_4^2\!+\!19\phi_5^2 \!+\! 2\phi_1(\phi_3\!+\!\phi_5)+2\phi_4(-\phi_1\!+\!3\phi_3\!-\!13\phi_5))\ , \nonumber \\
  &T_{2\footnotesize\textcircled{\scriptsize 5}+\footnotesize\textcircled{\scriptsize 6}} = \frac{1}{2}(\phi_4\!-\!2\phi_5)(2\phi_1\!-\!13\phi_4\!+\!12\phi_5) 
 .\quad \nonumber
\end{align}
As expected, these tensions agree with the volumes of K\"ahler surfaces in the geometry.

\begin{figure}
  \centering
  \includegraphics[width=.9\linewidth]{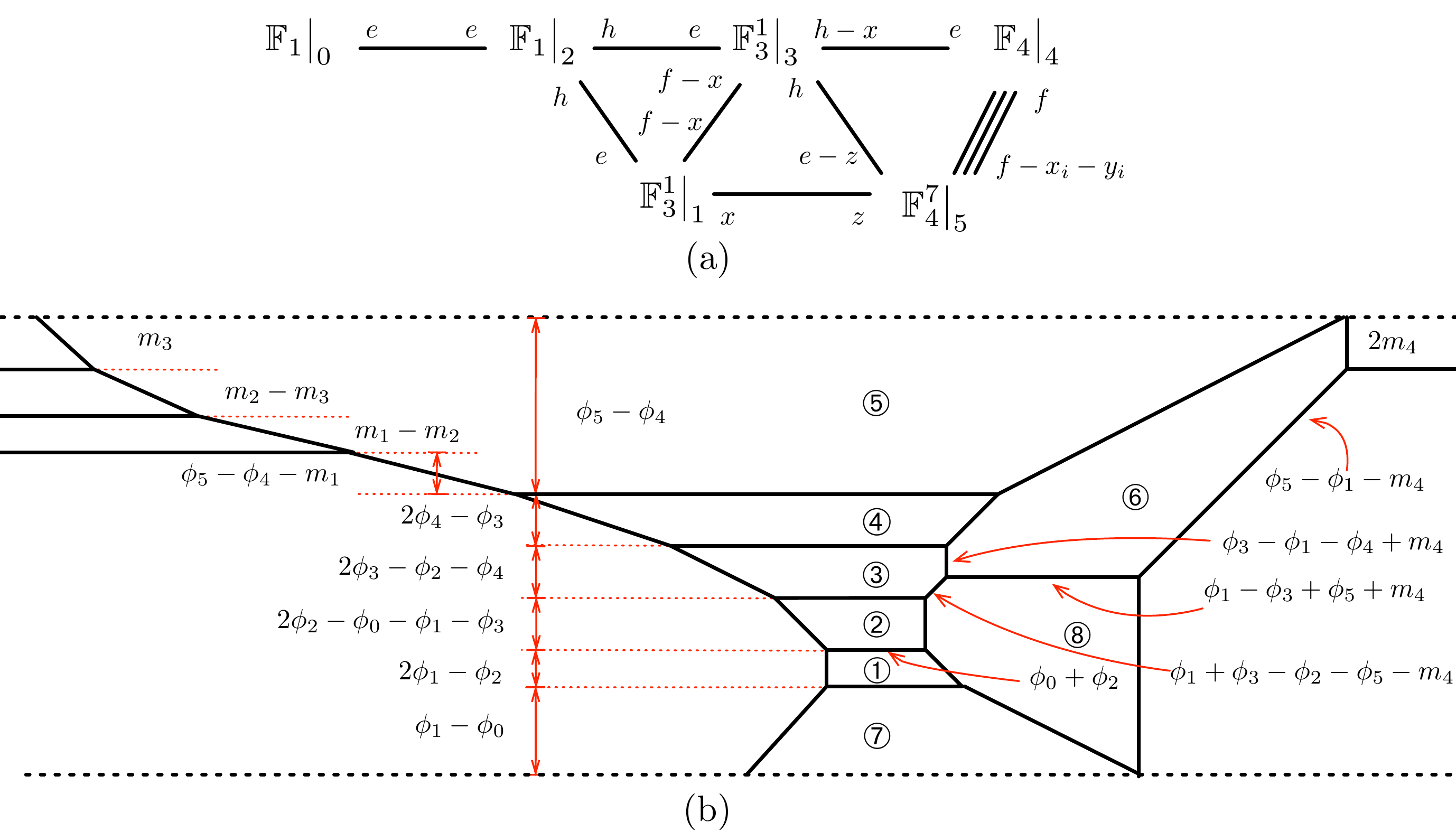}
  \caption{(a) The elliptic threefold for the 6d $SO(10)$ theory with 3 fundamentals and a spinor.
  (b) The dual brane web for (a).}
  \label{fig:SO10-3F-1S}
\end{figure}

\paragraph{\underline{$SO(10)$ gauge theory}}
Next, we will Higgs this 6d $SO(11)$ theory down to the 6d SCFT of $SO(10)$ gauge symmetry with 3 fundamentals and one spinor hyper. 
Similarly to the previous case, we give a Higgs vev to a fundamental scalar field and take the vev to infinity. In the brane web, this can be achieved by tuning the mass parameter $m_4$ to zero. The resulting brane web is given in Figure \ref{fig:SO10-3F-1S}(b) (up to an additional flop transition).

The resolved Calabi-Yau threefold for this theory \cite{Bhardwaj:2018yhy} is summarized in Figure \ref{fig:SO10-3F-1S}(a). The cubic prepotential of this threefold is
\begin{eqnarray}
  6\mathcal{F}_{SO(10)} &=& 8\phi_0^3+7\phi_1^3+8\phi_2^3+7\phi_3^3+8\phi_4^3+\phi_5^3 -3\phi_0^2\phi_2-3\phi_0\phi_2^2-9\phi_1\phi_2^2+3\phi_1^2\phi_2\nonumber \\
  &&-3\phi_1^2\phi_3-3\phi_1\phi_3^2 -3\phi_1^2\phi_5-3\phi_1\phi_5^2-9\phi_2^2\phi_3+3\phi_2\phi_3^2-12\phi_3^2\phi_4+6\phi_3\phi_4^2 \nonumber\\
  &&-15\phi_3^2\phi_5+9\phi_3\phi_5^2-18\phi_4^2\phi_5+6\phi_1\phi_2\phi_3+6\phi_1\phi_3\phi_5+18\phi_3\phi_4\phi_5 \ .
\end{eqnarray}
The map between the compact four-cycles in the geometry and the internal faces in the brane web is
\begin{align}
  &\mathbb{F}_1|_{0} = \textcircled{\scriptsize 1} \ , \qquad
   \mathbb{F}_3^1|_{1} = \textcircled{\scriptsize 1} + 2\cdot\textcircled{\scriptsize 7}+\textcircled{\scriptsize 8} \ , \qquad 
   \mathbb{F}_1|_{2} = \textcircled{\scriptsize 2} \ , \qquad 
   \mathbb{F}_3^1|_{3} = \textcircled{\scriptsize 3} \ ,  \nonumber \\
  &    F_4|_{4} = \textcircled{\scriptsize 4}
\ , \qquad 
   \mathbb{F}_4^{7}|_{5} = \textcircled{\scriptsize 4} + 2\cdot\textcircled{\scriptsize 5}+\textcircled{\scriptsize 6} \ .
\end{align}
We can compute the monopole string tensions from the brane web and the results are
\begin{align}
  &T_{\footnotesize\textcircled{\scriptsize 1}} = \frac{1}{2}(4\phi_0+\phi_2)(2\phi_0-\phi_2) \ , \cr 
  &T_{\footnotesize\textcircled{\scriptsize 1}+2\textcircled{\scriptsize 7}+\textcircled{\scriptsize 8}} = \frac{1}{2}(7\phi_1^2\!-\!3\phi_2^2\!-\!\phi_3^2\!-\!\phi_5^2\!+\!2\phi_1(\phi_2\!-\!\phi_3\!-\!\phi_5)\!+\!2\phi_3(\phi_2\!+\!\phi_5)) \ , \nonumber \\
  &T_{\footnotesize\textcircled{\scriptsize 2}} = \frac{1}{2}(4\phi_2\!+\!\phi_0\!-\!\phi_1\!-\!\phi_3)(2\phi_2\!-\!\phi_0\!-\!\phi_1\!-\!\phi_3) \ , \nonumber \\
  &T_{\footnotesize\textcircled{\scriptsize 3}} = \frac{1}{2}(-\phi_1^2\!-\!3\phi_2^2\!+\!7\phi_3^2\!+\!2\phi_4^2\!+\!3\phi_5^2\!+\!2\phi_1(\phi_2\!-\!\phi_3\!+\!\phi_5)\!+\!2\phi_3(\phi_2\!-\!4\phi_4\!-\!5\phi_5)\!+\!6\phi_4\phi_5) \ , \nonumber \\
  &T_{\footnotesize\textcircled{\scriptsize 4}} = (2\phi_4\!-\!\phi_3)(2\phi_4\!+\!2\phi_3\!-\!3\phi_5) \ , \nonumber \\
  &T_{\footnotesize\textcircled{\scriptsize 4}+2\footnotesize\textcircled{\scriptsize 5}+\footnotesize\textcircled{\scriptsize 6}} = \frac{1}{2}(-\phi_1^2\!-\!5\phi_3^2\!-\!6\phi_4^2\!+\!\phi_5^2\!-\!2\phi_1\phi_5\!+\!2\phi_3(\phi_1\!+\!3\phi_4\!+\!3\phi_5)) \ ,
\end{align}
which are in perfect agreement with the volumes of compact K\"ahler surfaces in the geometry obtained from the prepotential above.

\paragraph{\underline{$SO(9)$ gauge theory}}
\begin{figure}
  \centering
  \includegraphics[width=.9\linewidth]{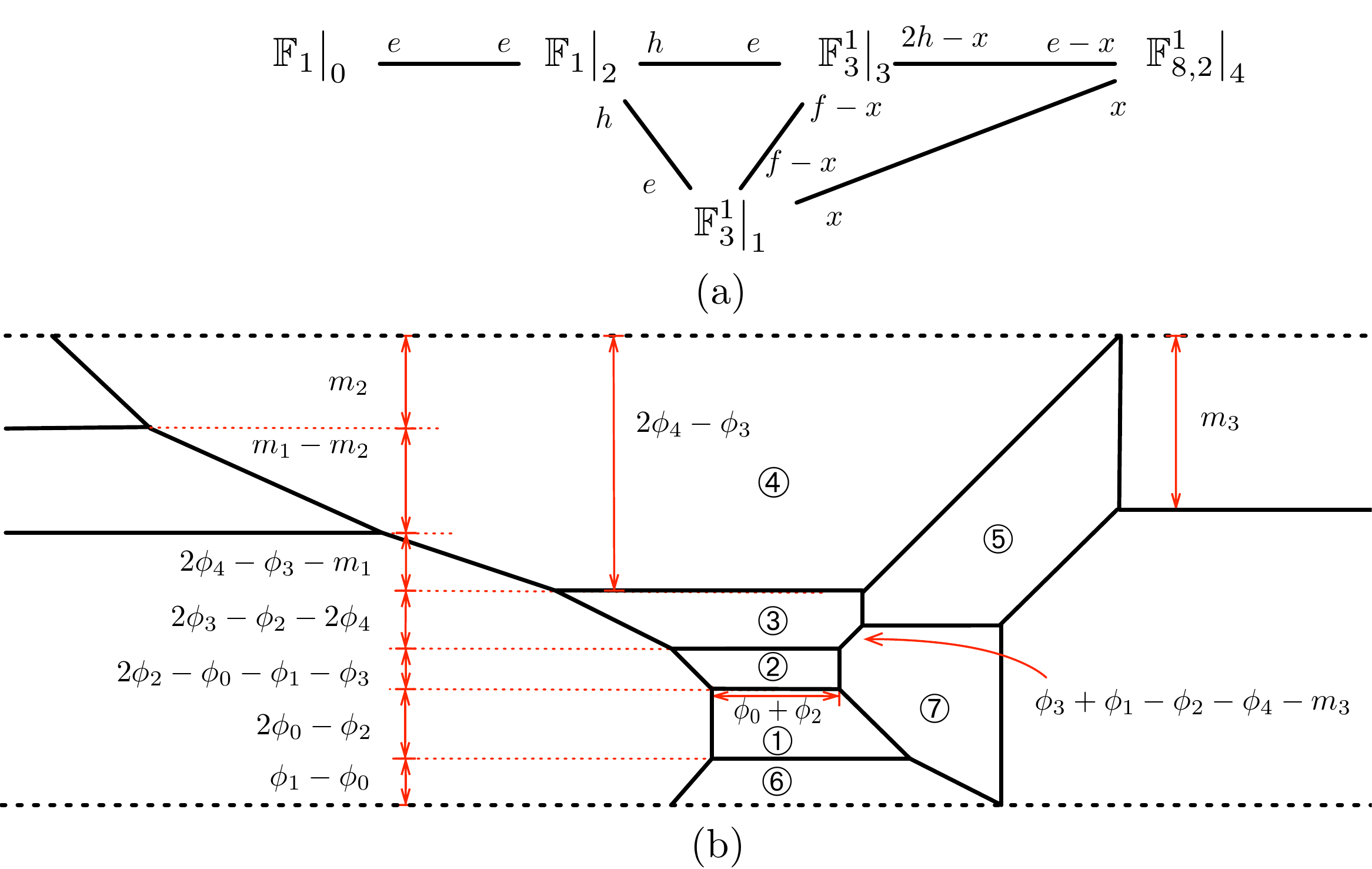}
  \caption{(a) The elliptic threefold for the 6d $SO(9)$ theory with 2 fundamentals and a spinor.
  (b) A dual brane web to the geometry (a).}
  \label{fig:SO9-2F-1S}
\end{figure}

By the same mechanism we can obtain all other 6d SCFTs on a $-3$ curve in the RG flow hierarchy.
It is hence straightforward to find the brane diagram of the 6d SCFT on $-3$ curve with $SO(9)$ gauge symmetry. We pull the top internal and the top external D5-branes on the left side toward the top orientifold plane. This corresponds to tuning the distances as $m_3=0$ and $\phi_5-\phi_4=0$ followed by a series of flop transitions. The final diagram is illustrated in Figure \ref{fig:SO9-2F-1S}(b). This is the brane web for the 6d $SO(9)$ gauge theory on $-3$ curve.

The corresponding CY$_3$ geometry is summarized in Figure \ref{fig:SO9-2F-1S}(a) and 
the prepotential of this geometry can be obtained from the geometry as
\begin{eqnarray}
	6\mathcal{F}_{SO(9)} &=& 8\phi_0^3 + 7\phi_1^3 + 8\phi_2^3 + 7\phi_3^3 -9\phi_4^3 -3\phi_0^2\phi_2 -3\phi_0\phi_2^2 -9\phi_1\phi_2^2+3\phi_1^2\phi_2  \nonumber \\
	&&-9\phi_2^2\phi_3+3\phi_2\phi_3^2-3\phi_1^2\phi_3-3\phi_1\phi_3^2-3\phi_1^2\phi_4-3\phi_1\phi_4^2 \nonumber \\
	&&-27\phi_3^2\phi_4+33\phi_3\phi_4^2+6\phi_1\phi_2\phi_3+6\phi_1\phi_3\phi_4 \ .
\end{eqnarray}
We can associate the geometry and the brane configuration by the map
\begin{align}
	&\mathbb{F}_1|_{0} = \textcircled{\scriptsize 1} \ , \qquad 
	\mathbb{F}^1_3|_{1} = \textcircled{\scriptsize 1}+2\cdot\textcircled{\scriptsize 6}+\textcircled{\scriptsize 7} \ , \qquad
	\mathbb{F}_1|_{2} = \textcircled{\scriptsize 2} \ , 
 \nonumber \\
	&	\mathbb{F}^1_3|_{3} = \textcircled{\scriptsize 3} \ ,\qquad \mathbb{F}_{8,2}^1|_{4} = 2\cdot \textcircled{\scriptsize 4} + \textcircled{\scriptsize 5} \ .
\end{align}
Note that the geometry in Figure \ref{fig:SO9-2F-1S}(a) does not look the same as the geometry for the 6d $SO(9)$ theory on a $-3$ curve given in \cite{Bhardwaj:2018yhy}. However it turns out that our geometry is equivalent to the geometry given in \cite{Bhardwaj:2018yhy} up to a flop on the exceptional curve $x$ in $F^1_{8,2}$. 

The monopole string tensions from the brane configuration are
\begin{eqnarray}
	&&T_{\footnotesize\textcircled{\scriptsize 1}} = \frac{1}{2}(4\phi_0+\phi_2)(2\phi_0-\phi_2) \ , \cr
	&&T_{\footnotesize\textcircled{\scriptsize 1}+2\footnotesize\textcircled{\scriptsize 6}+\footnotesize\textcircled{\scriptsize 7}} = \frac{1}{2}(7\phi_1^2\!-\!3\phi_2^{2}\!-\!\phi_3^2\!-\!\phi_4^2\!+\!2\phi_1(\phi_2\!-\!\phi_3\!-\!\phi_4)\!+\!2\phi_3(\phi_2\!+\!\phi_4)) \ , \nonumber \\
	&&T_{\footnotesize\textcircled{\scriptsize 2}} = \frac{1}{2}(4\phi_2\!+\!\phi_0\!-\!\phi_1\!-\!\phi_3)(2\phi_2\!-\!\phi_0\!-\!\phi_1\!-\!\phi_3) \ , \nonumber \\
	&&T_{\footnotesize\textcircled{\scriptsize 3}} = \frac{1}{2}(-\phi_1^2\!-\!3\phi_2^2\!+\!7\phi_3^2\!+\!11\phi_4^2\!+\!2\phi_1(\phi_2\!-\!\phi_3\!+\!\phi_4)\!+\!2\phi_3(\phi_2\!-\!9\phi_4)) \ , \nonumber \\
	&&T_{2\footnotesize\textcircled{\scriptsize 4}+\footnotesize\textcircled{\scriptsize 5}} = \frac{1}{2}(-\phi_1^2\!-\!9\phi_2^2\!-\!9\phi_4^2\!+\!2\phi_1(\phi_3\!-\!\phi_4)+22\phi_3\phi_4) \ .
\end{eqnarray}
This result also agrees with the volumes of four-cycles in the geometry obtained from the cubic prepotential above.




\paragraph{\underline{$SO(8)$ gauge theory}}
\begin{figure}
  \centering
  \includegraphics[width=.8\linewidth]{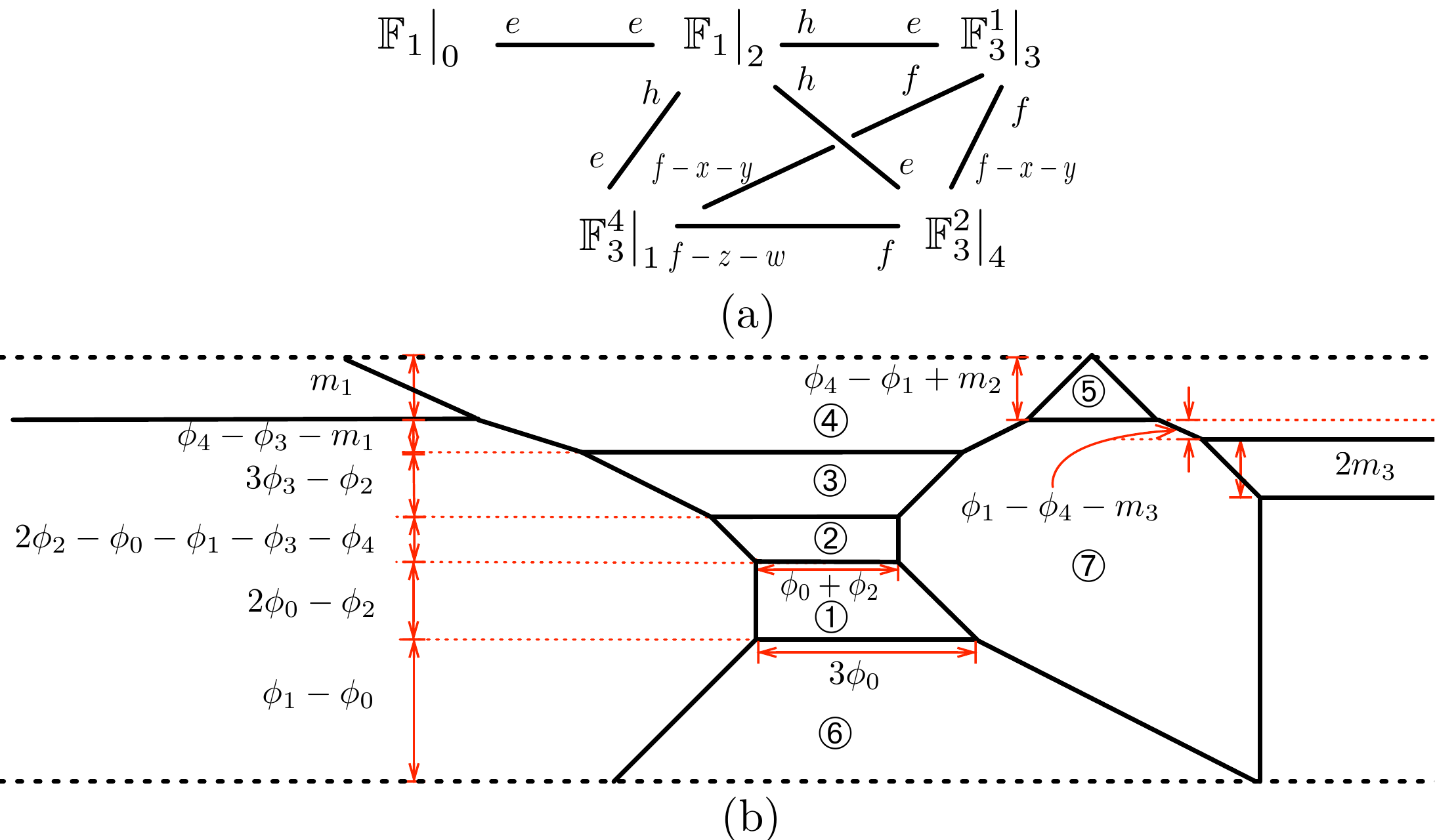}
  \caption{(a) The elliptic threefold for the 6d $SO(8)$ theory with a fundamental and a spinor and a conjugate spinor hyper.
  (b) The dual brane web for the 6d $SO(8)$ theory on a $-3$ curve. }
  \label{fig:SO8-1F-1S-1C}
\end{figure}

We can further Higgs the $SO(9)$ theory down to the 6d SCFT with $SO(8)$ gauge symmetry on a $-3$ curve.
Following the procedure discussed above, we pull the top external D5-brane toward the orientifold plane by turning off $m_2$. The resulting brane web after a series of flops is given in Figure \ref{fig:SO8-1F-1S-1C}(b). The geometry of the 6d $SO(8)$ gauge theory on a $-3$ curve  \cite{Bhardwaj:2018yhy} given in Figure \ref{fig:SO8-1F-1S-1C}(a). The threefold has the prepotential
\begin{eqnarray}
  6\mathcal{F}_{SO(8)} &=& 8\phi_0^3+4\phi_1^3+8\phi_2^3+8\phi_3^3+6\phi_4^3-3\phi_0^2\phi_2-3\phi_0\phi_2^{2}+3\phi_1^2\phi_2-9\phi_1\phi_2^2 \nonumber \\
  &&-6\phi_1\phi_3^2-6\phi_1\phi_4^2-9\phi_2^2\phi_3+3\phi_2\phi_3^2-9\phi_2^2\phi_4 +3\phi_2\phi_4^2 -6\phi_3^2\phi_4 \nonumber \\
  && +6\phi_1\phi_2\phi_3+6\phi_1\phi_2\phi_4+6\phi_2\phi_3\phi_4 \ .
\end{eqnarray}
We can easily identify this geometry with the brane web configuration through the map given as
\begin{align}
		&\mathbb{F}_1|_{0} = \textcircled{\scriptsize 1} \ , \qquad 
		\mathbb{F}^4_3|_{1} = \textcircled{\scriptsize 1} + 2\cdot\textcircled{\scriptsize 6}+\textcircled{\scriptsize 7} \ , \qquad 
		\mathbb{F}_1|_{2} = \textcircled{\scriptsize 2} \ , \nonumber \\
	&
\mathbb{F}_3|_{3} = \textcircled{\scriptsize 3} \ , \qquad 	
	\mathbb{F}_3^2|_{4} = \textcircled{\scriptsize 3}+2\cdot\textcircled{\scriptsize 4} +\textcircled{\scriptsize 5}
\ .
\end{align}
One can compute the monopole string tensions from the brane web and the results are
\begin{align}
&T_{\footnotesize\textcircled{\scriptsize 1}} = \frac{1}{2}(4\phi_0\!+\!\phi_2)(2\phi_0\!-\!\phi_2)  , \crcr
&T_{\footnotesize\textcircled{\scriptsize 1}+2\footnotesize\textcircled{\scriptsize 6}+\footnotesize\textcircled{\scriptsize 7}} = \frac{1}{2}\big(4\phi_1^2\!-\!3\phi_2^2\!-\!2\phi_3^2\!-\!2\phi_4^2\!+\!2\phi_2(\phi_1\!+\!\phi_3\!+\!\phi_4)\big) \ , \nonumber \\
&T_{\footnotesize\textcircled{\scriptsize 2}} = \frac{1}{2}(4\phi_2\!+\!\phi_0\!-\!\phi_1\!-\!\phi_3\!-\!\phi_4)(2\phi_2\!-\!\phi_0\!-\!\phi_1\!-\!\phi_3\!-\!\phi_4) \ , \nonumber \\
&T_{\footnotesize\textcircled{\scriptsize 3}} = \frac{1}{2}(2\phi_3\!-\!\phi_2)(4\phi_3\!+\!3\phi_2\!-\!2\phi_1\!-\!2\phi_4) \ , \nonumber \\
&T_{\footnotesize\textcircled{\scriptsize 3}+2\footnotesize\textcircled{\scriptsize 4}+\footnotesize\textcircled{\scriptsize 5}} = \frac{1}{2}\big(-3\phi_2^2\!-\!2\phi_3^2\!+\!6\phi_4^2\!+\!2\phi_2(\phi_1\!+\!\phi_3\!+\!\phi_4)\!-\!4\phi_1\phi_4\big) \ .
\end{align}
These tensions are the same as the volumes of surfaces in the geometry.

\paragraph{\underline{$SO(7)$ gauge theory}}
\begin{figure}
  \centering
  \includegraphics[width=.6\linewidth]{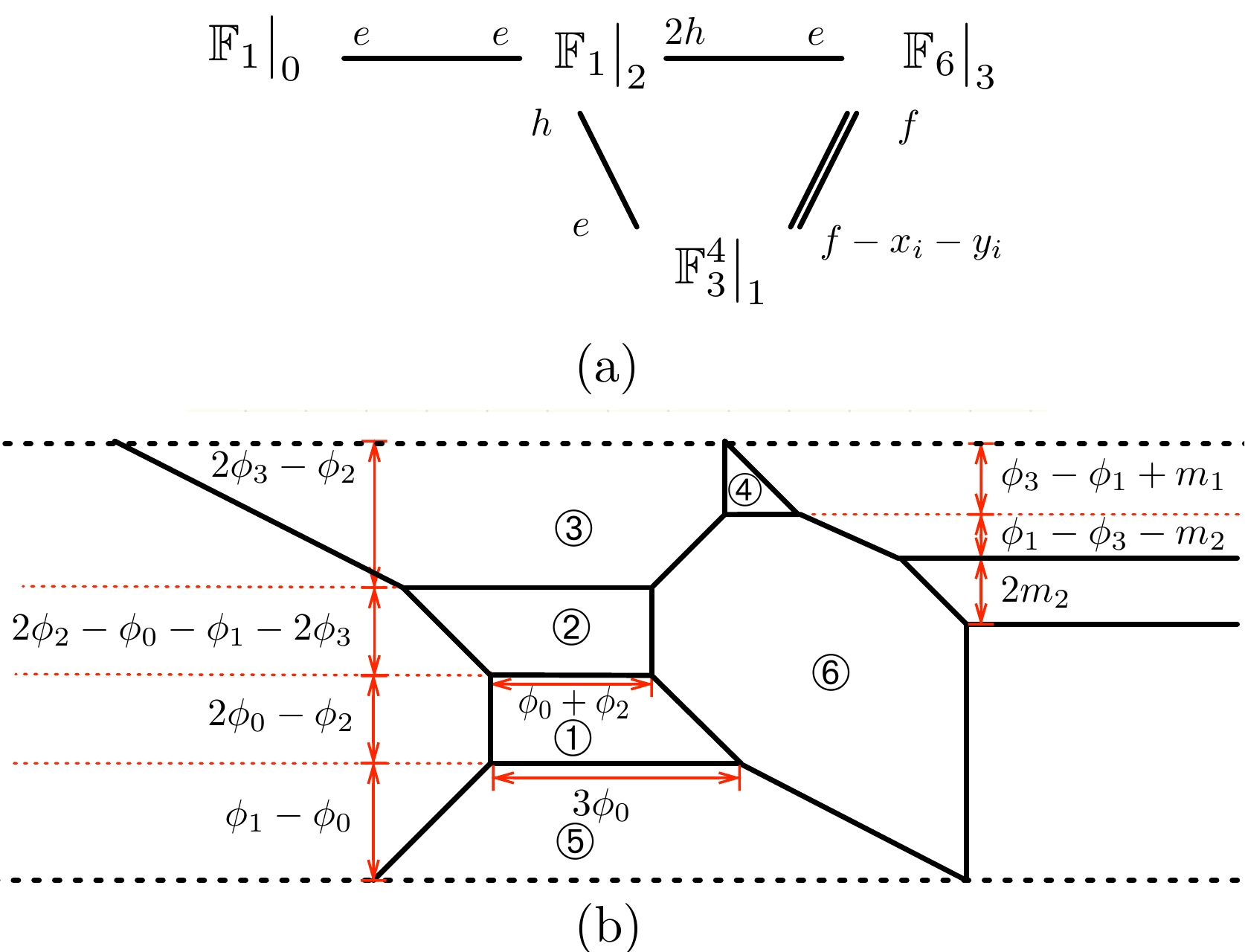}
  \caption{(a) Elliptic threefold for the 6d $SO(7)$ theory with two spinor hypers.
  (b) Brane web for the 6d $SO(7)$ theory on a $-3$ curve. }
  \label{fig:SO7-2S}
\end{figure}

The brane web for the 6d $SO(7)$ gauge theory on $-3$ curve can be obtained from the $SO(8)$ diagram by another standard Higgsing that uses an $SO(8)$ fundamental scalar vev. Then we obtain the web diagram in Figure \ref{fig:SO7-2S}(b) and this brane configuration amounts to the 6d SCFT with $SO(7)$ gauge symmetry coupled to two spinor hypermultiplets. The associated CY$_3$ geometry is drawn in Figure \ref{fig:SO7-2S}(a) \cite{Bhardwaj:2018yhy}. One computes the prepotential of this geometry as
\begin{eqnarray}
	6\mathcal{F}_{SO(7)} &=& 8\phi_0^3 + 4\phi_1^3+ 8\phi_2^3 + 8\phi_3^3 -3\phi_0^2\phi_2 - 3\phi_0 \phi_2^2 -9\phi_1\phi_2^2 + 3\phi_1^2\phi_2 \nonumber \\
	&& -12\phi_1\phi_3^2-18\phi_2^2\phi_3+12\phi_2\phi_3^2 +12 \phi_1\phi_2\phi_3 \ .
\end{eqnarray}
The brane diagram can also be compared to the geometry with the identification of compact surfaces as
\begin{align}
\mathbb{F}_1|_{0} = \textcircled{\scriptsize 1}\ , \quad \mathbb{F}_3^4|_{1} = \textcircled{\scriptsize 1}+2\cdot\textcircled{\scriptsize 5}+\textcircled{\scriptsize 6} \ , \quad \mathbb{F}_1|_{2} = \textcircled{\scriptsize 2} \ , \quad \mathbb{F}_6|_{3} = 2\cdot\textcircled{\scriptsize 3}+\textcircled{\scriptsize 4}
	\ .
\end{align}
The volume of these surfaces from the web diagram are
\begin{align}
&T_{\footnotesize\textcircled{\scriptsize 1}} = \frac{1}{2}(2\phi_0-\phi_2)(4\phi_0+\phi_2) \ , \nonumber\\
&T_{\footnotesize\textcircled{\scriptsize 1}+2\footnotesize\textcircled{\scriptsize 5}+\footnotesize\textcircled{\scriptsize 6}} = \frac{1}{2}(4\phi_1^2+2\phi_1\phi_2 -3\phi_2^2+4\phi_2\phi_3-4\phi_3^2) \ , \nonumber \\
&T_{\footnotesize\textcircled{\scriptsize 2}} = \frac{1}{2}(\phi_0+\phi_1-2\phi_2+2\phi_3)(-\phi_0+\phi_1-4\phi_2+2\phi_3) \ , \nonumber\\
&T_{2\footnotesize\textcircled{\scriptsize 3}+\footnotesize\textcircled{\scriptsize 4}} = (\phi_2-2\phi_3)(2\phi_1-3\phi_2-2\phi_3) \ .
\end{align}
This result is in perfect agreement with the geometric result.

\paragraph{\underline{$G_2$ gauge theory}}
\begin{figure}
  \centering
  \includegraphics[width=.5\linewidth]{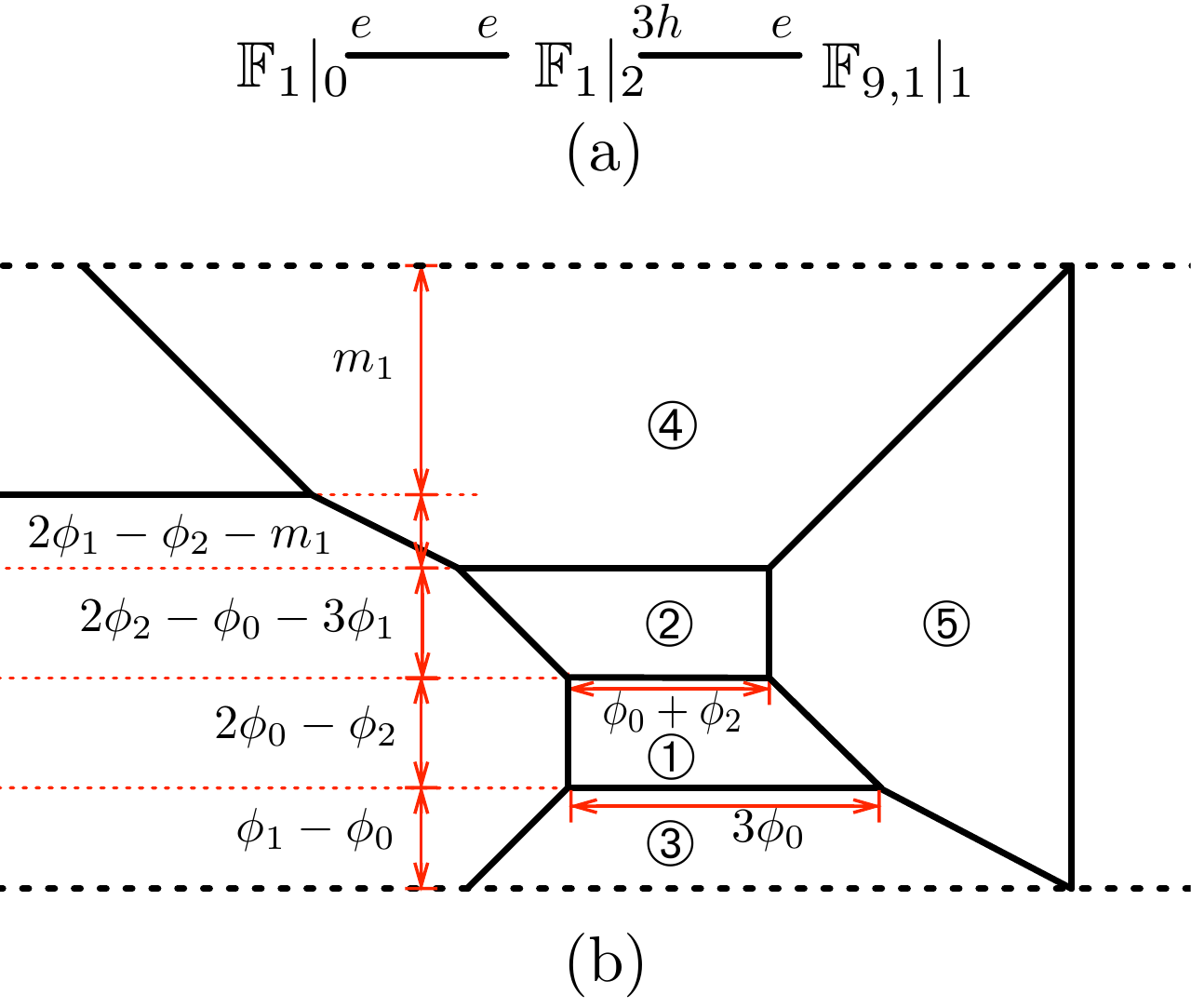}
  \caption{(a) The elliptic threefold for the 6d $G_2$ gauge theory with a fundamental hyper.
  (b) The brane web for this 6d $G_2$ theory. }
  \label{fig:G2-1F}
\end{figure}

We can now consider an interesting RG flow from an $SO(7)$ gauge theory to an exceptional $G_2$ gauge theory generated by a large Higgs vev of an $SO(7)$ spinor operator. This Higgsing has a natural realization in the $SO(7)$ brane web. Two spinor hypers in the $SO(7)$ gauge theory correspond to two external D5-branes on the right side in the brane web. So we need to adjust the positions of these two external D5-branes together so that a Higgs branch opens up. The Higgs branch diagram is drawn in  Figure \ref{fig:G2-1F}(b). This brane web at low energy implements the 6d SCFT of $G_2$ gauge symmetry with one fundamental hypermultiplet.

The corresponding threefold (studied in \cite{Bhardwaj:2018yhy}) is given in Figure \ref{fig:G2-1F}(a).
The cubic prepotential can be obtained from the threefold as
\begin{eqnarray}
	6\mathcal{F}_{G_2} &=& 8\phi_0^3 + 8\phi_2^3 -3\phi_0^3\phi_2 -3\phi_0\phi_2^2 -27\phi_1\phi_2^2+27\phi_1^2\phi_2 \ .
\end{eqnarray}
This geometry is associated with our brane web by the relation
\begin{align}
	\mathbb{F}_1|_{0} = \textcircled{\scriptsize 1} 
	\ , \quad \mathbb{F}_{9,1}|_{1}= \textcircled{\scriptsize 1}+2\cdot\textcircled{\scriptsize 3} + 2\cdot\textcircled{\scriptsize 4} + \textcircled{\scriptsize 5}\ , \quad \mathbb{F}_1|_{2} = \textcircled{\scriptsize 2}  \ .	
\end{align}
The volumes of these surfaces can be easily extracted from the brane diagram 
\begin{align}
&T_{\footnotesize\textcircled{\scriptsize 1}} = \frac{1}{2}(2\phi_0-\phi_2)(4\phi_0+\phi_2) \ , \nonumber \\
&T_{\footnotesize\textcircled{\scriptsize 1}+2\textcircled{\scriptsize 3}+2\textcircled{\scriptsize 4}+\textcircled{\scriptsize 5}} = \frac{9}{2}(2\phi_1-\phi_2)\phi_2 \ , \nonumber \\
&T_{\footnotesize\textcircled{\scriptsize 2}} = \frac{1}{2}(\phi_0 -3\phi_1 +4\phi_2)(-\phi_0-3\phi_1+2\phi_2) \ ,
\end{align}
and the result agrees with the volumes of K\"ahler surfaces in the geometry.

\paragraph{\underline{$SU(3)$ gauge theory}}
\begin{figure}
  \centering
  \includegraphics[width=.5\linewidth]{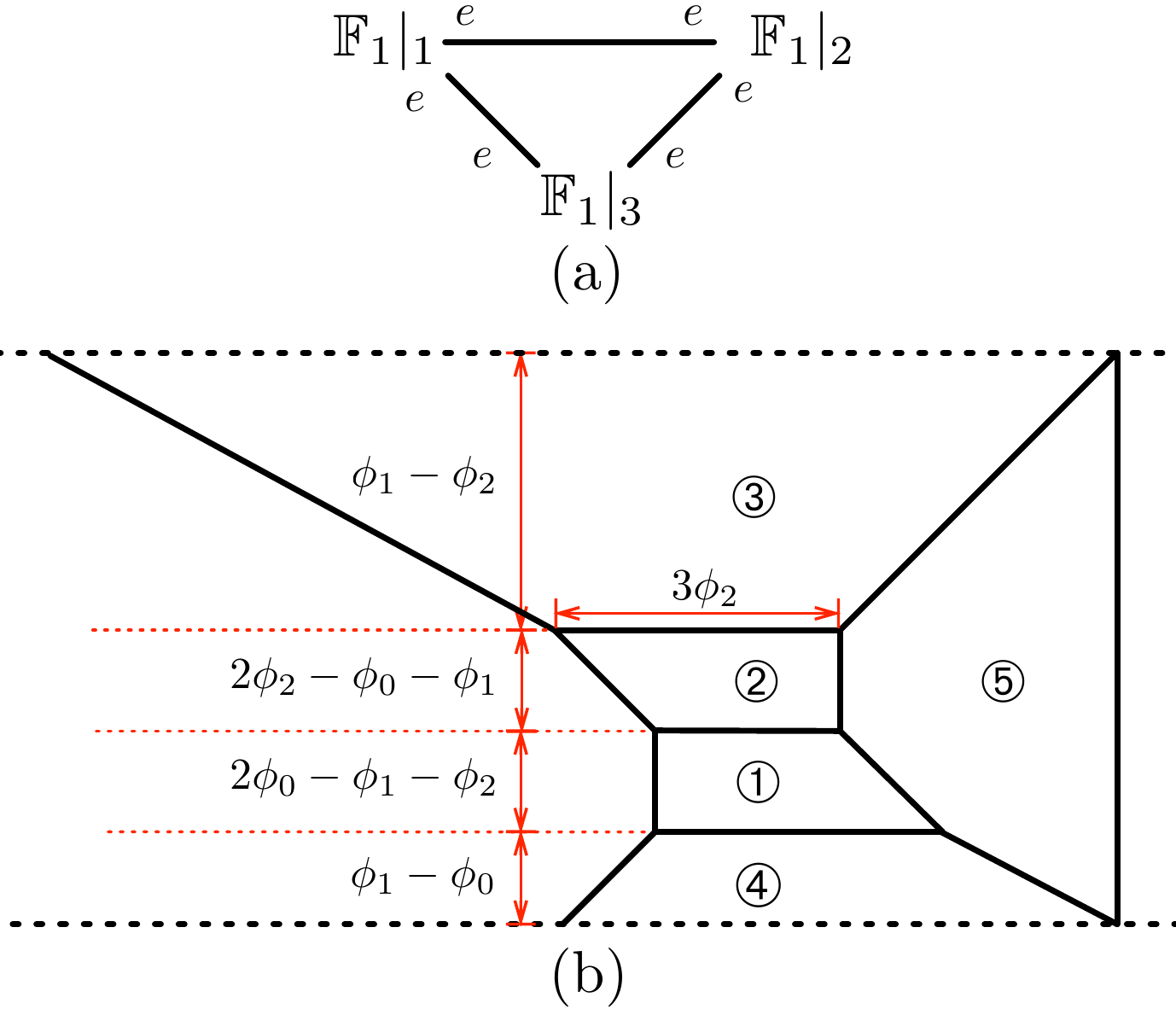}
  \caption{(a) The elliptic threefold for the 6d $SU(3)$ gauge theory on a $-3$ curve.
  (b) The brane web for the 6d $SU(3)$ theory on a $-3$ curve. }
  \label{fig:SU3}
\end{figure}

Lastly, we will Higgs this $G_2$ brane web down to the web diagram for the 6d $SU(3)$ gauge theory on a $-3$ curve. This can be done by simply tuning the mass parameter in the brane web as $m_1=0$. Then the resulting brane web is in Figure \ref{fig:SU3}(b).
The corresponding CY$_3$ consists of three surfaces glued as drawn in Figure \ref{fig:SU3}(a) and this threefold has the cubic prepotential
\begin{equation}
	6\mathcal{F}_{SU(3)} = 9\phi_0^3 + 9\phi_1^3 + 9\phi_2^3 - (\phi_0+\phi_1+\phi_2)^3 \ .
\end{equation}
The three surfaces in the geometry are identified with the faces in the brane web as
\begin{equation}
	\mathbb{F}_1|_{0} = \textcircled{\scriptsize 1} \ , \qquad \mathbb{F}_1|_{1} = \textcircled{\scriptsize 1} + \textcircled{\scriptsize 2} + 2\cdot\textcircled{\scriptsize 3} + 2\cdot\textcircled{\scriptsize 4} +  \textcircled{\scriptsize 5} \ , \qquad \mathbb{F}_1|_{2} = \textcircled{\scriptsize 2} \ .
\end{equation}
Moreover, the areas of the internal faces in the brane web
\begin{align}
T_{\footnotesize\textcircled{\scriptsize 1}} &= \frac{1}{2}(2\phi_0-\phi_1-\phi_2)(4\phi_0+\phi_1+\phi_2) \ , \nonumber \\
T_{\footnotesize\textcircled{\scriptsize 1}+\textcircled{\scriptsize 2}+2\textcircled{\scriptsize 3}+2\textcircled{\scriptsize 4}+\textcircled{\scriptsize 5}}&= \frac{1}{2}(-\phi_0+2\phi_1-\phi_2)(\phi_0+4\phi_1+\phi_2) \ , \nonumber \\
T_{\footnotesize\textcircled{\scriptsize 2}} &= \frac{1}{2}(-\phi_0-\phi_1+2\phi_2)(\phi_0+\phi_1+4\phi_2) \ 
\end{align}
perfectly match the volumes of the  three surfaces in the geometry.

Since the 6d $SU(3)$ gauge theory on $-3$ curve is a non-Higgsable theory \cite{Morrison:2012np}, we expect that there is no further Higgs branch flow from the $SU(3)$ brane web in Figure \ref{fig:SU3}(b). Indeed we find no other Higgs transition in this brane configuration.

\subsection{Twisted theories}
In the previous subsection, we discussed Higgsing procedures of 6d SCFTs induced by large vevs of scalar fields in hypermultiplets charged under ordinary gauge symmetries.
In this subsection, we consider a new type of RG flows from the 6d theories which opens up only after circle compactification at finite radius. When a 6d SCFT is compactified on a circle, the associated singular elliptic threefold can be resolved by introducing non-zero K\"ahler parameters for the compact four-cycles. From the gauge theory point of view, this corresponds to turning on non-zero holonomies both for gauge and global symmetries around the 6d circle. In the brane webs, all the compact faces and the internal edges have finite size by the holonomies. 

Now we can tune holonomies such that a charged scalar mode carrying non-zero Kaluza-Klein momentum along the 6d circle becomes light. We can then consider giving a non-zero vev to this light scalar mode. This will trigger a Higgs branch flow to a new 6d theory at low energy living on a circle. This is similar to the standard Higgs mechanism but now using a Kaluza-Klein momentum state.

Quite interestingly, we find that the RG flows by KK states in the brane systems we studied above lead to new brane configurations for twisted compactifications of the 6d $SO(2N)$ gauge theories on $-3$, which we shall discuss from now on.

\begin{figure}
\centering
{\includegraphics[width=.7\linewidth]{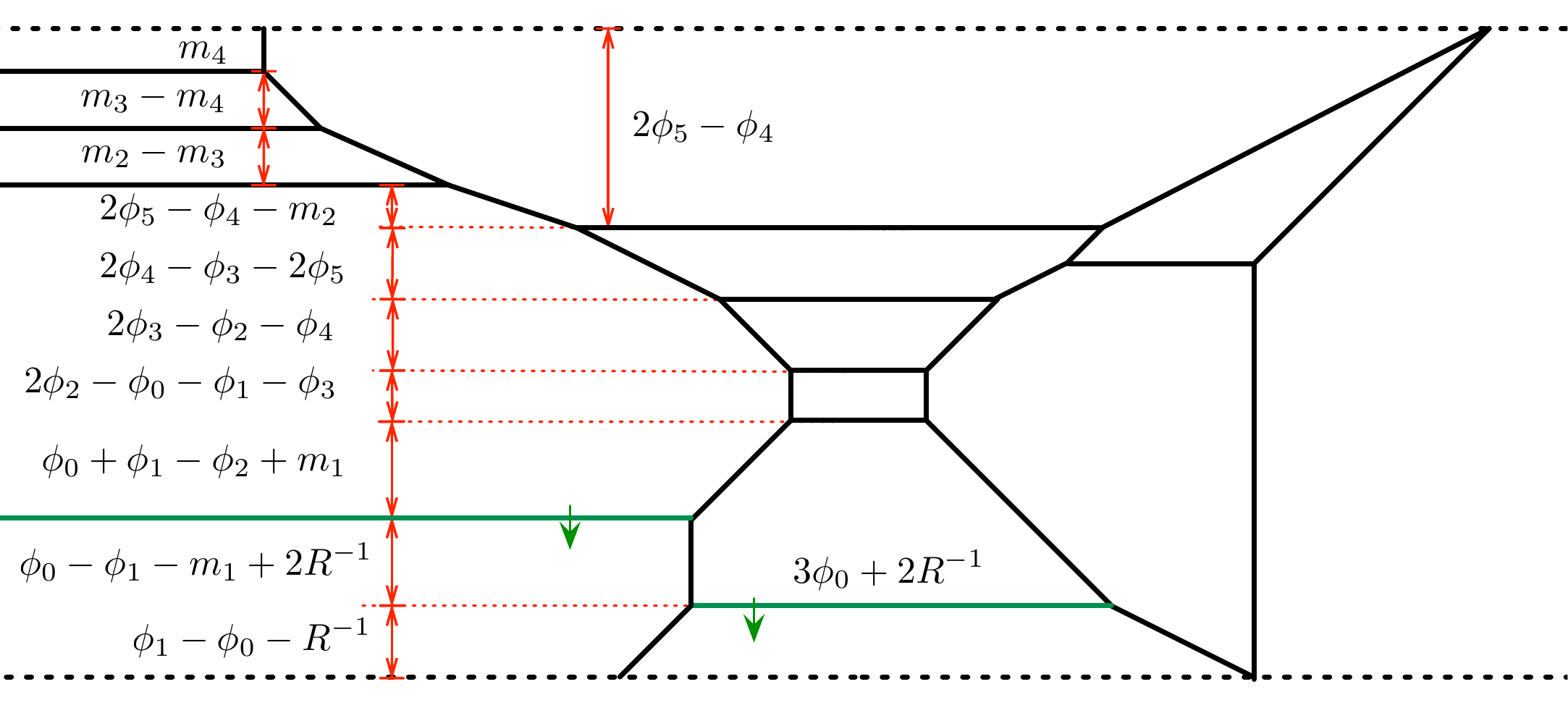}}\vspace{0.3cm}
{\includegraphics[width=.7\linewidth]{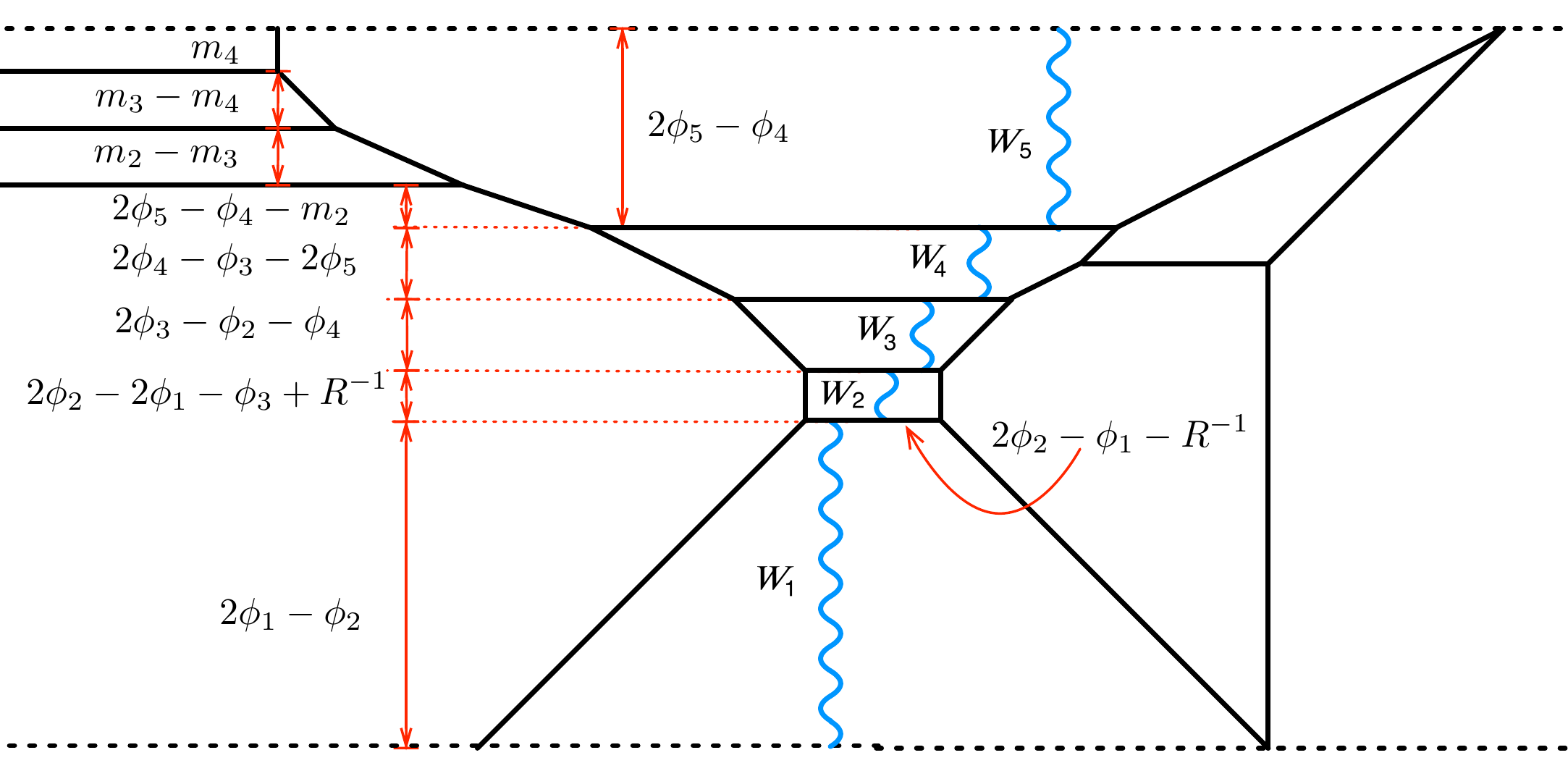}}
\caption{(a) The brane web for the 6d $SO(11)$ theory on $-3$ curve.
  (b) The Higgsed brane web realizes the 6d $SO(10)$ theory on a $-3$ curve with $Z_2$ twist. }
  \label{fig:SO10-twist}
\end{figure}

First, consider the brane web in Figure \ref{fig:SO11-4F}(b) for the $SO(11)$ gauge theory compactified on a circle. In fact, we need to introduce the 6d circle radius $R$ on this brane web diagram. Note that the height of the diagram is set by the circle radius. To restore the circle radius, we can redefine the K\"ahler parameters as
\begin{equation}
	2\phi_0 - \phi_2 \ \rightarrow \ 2\phi_0-\phi_2+ 2R^{-1}\ , \qquad \phi_1-\phi_0 \ \rightarrow \ \phi_1 - \phi_0 - R^{-1} \ ,
\end{equation}
while other parameters denoted in the diagram remain the same. Then we flop some two cycles to get the brane web in Figure \ref{fig:SO10-twist}(a). This brane diagram has another Higgs branch which opens when we pull the internal and the external D5-branes denoted by downward arrows toward the orientifold plane at the bottom.
The brane moves sets the mass parameters as
\begin{equation}\label{eq:mass-twist}
	\phi_0 = \phi_1 - R^{-1} \ , \qquad m_1 = R^{-1} \ .
\end{equation}
So the gauge and flavor holonomies are tuned to cancel the Kaluza-Klein mass proportional to $R^{-1}$. The Higgsed brane web at low energy is given in Figure \ref{fig:SO10-twist}(b).

We claim that this Higgsing turns the $SO(11)$ gauge theory into the $SO(10)$ gauge theory on a circle with $Z_2$ outer-automorphism twist. This Higgsing is in fact the same as the Higgsing of the $SO(12)$ brane web in Figure \ref{fig:SO12-5F}(b) to the $SO(11)$ brane web, but here we used a hypermultiplet scalar carrying KK momentum and affine gauge charge parameterized by $R^{-1}$ and $\phi_0$ respectively. So in the case at hand we move the D5-branes associated to this hypermultiplet to the bottom orientifold plane instead.
At the level of 6d gauge algebra after circle compactification, the previous Higgsing in the $SO(12)$ theory reduces affine $SO(12)$ gauge algebra to affine $SO(11)$ gauge algebra, which identifies two spinor nodes in the $SO(12)$ Dynkin diagram. On the other hand, the Higgsing of the $SO(11)$ brane web in this section identifies the affine node and the fundamental node in the other side of the affine $SO(11)$ Dynkin diagram. This implies the associated affine gauge algebra after Higgsing is the affine $D^{(2)}_{5}$ algebra which is the affine algebra of $SO(10)$ gauge symmetry with $Z_2$ outer-automorphism twist. The Higgsing thus leaves a brane configuration exhibiting $Z_2$ outer-automorphism twist of the $SO(10)$ gauge symmetry. We thus expect that the Higgsed brane web in Figure \ref{fig:SO10-twist}(b) is a brane realization of the 6d $SO(10)$ gauge symmetry on a $-3$ curve with $Z_2$ outer-automorphism twist.

This can also be checked from the spectrum of gauge bosons in our brane web. The gauge bosons come from strings suspending between D5-branes. The gauge bosons $W_i$ stretched between adjacent D5- and O5-branes have the following masses:
\begin{align}
	&m_{W_1} = 2\phi_1-\phi_2 \ , & m_{W_2}& = 2\phi_2-2\phi_1-\phi_3 + R^{-1} \ , && m_{W_3} = 2\phi_3-\phi_2-\phi_4 \ , \nonumber \\
	&m_{W_4}= 2\phi_4-\phi_3-2\phi_5 \ , & m_{W_5}& = 2\phi_5-\phi_4 \ .&&
\end{align}
Indeed, their gauge charges in terms of $\phi_i$ form the affine $D^{(2)}_5$ Cartan matrix. Note that three fundamental hypers in the $SO(10)$ theory are expected to remain as three fundamental hypers of $SO(9)$ subalgebra of the $D^{(2)}_5$, while one $SO(10)$ spinor hyper reduces to an half-hyper under the $SO(9)$ subalgebra under the $Z_2$ automorphism twist. The brane web in Figure \ref{fig:SO10-twist} shows only three masses $m_{i=2,3,4}$, as well as the KK mass parameter $R^{-1}$, for three fundamental hypers in IR. This satisfies our expectation for the $Z_2$ twisted theory. 
These results supports our brane web in Figure \ref{fig:SO10-twist}(b) for being the $Z_2$ twist of the 6d $SO(10)$ gauge theory.

\begin{figure}
\centering
{\includegraphics[width=.53\linewidth]{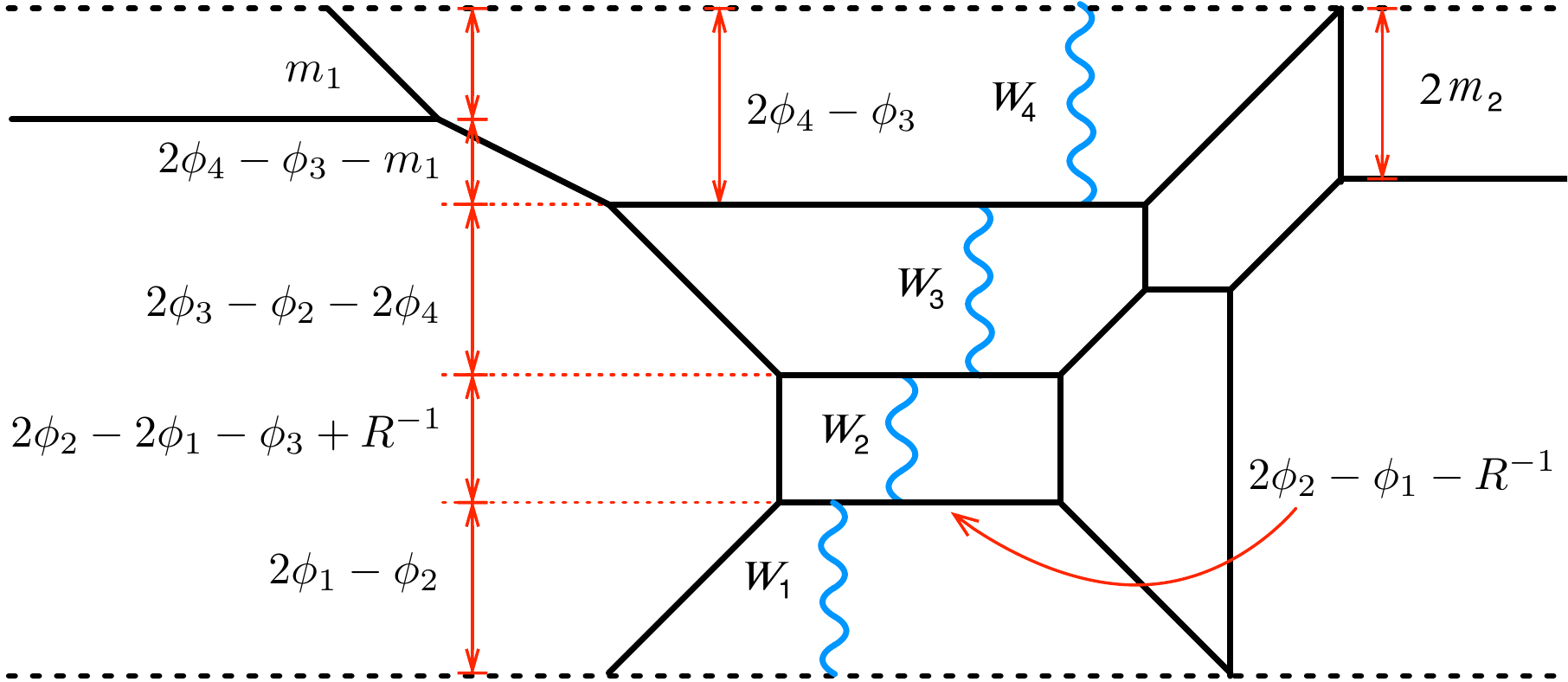}} \hspace{0.1cm} 
{\includegraphics[width=.44\linewidth]{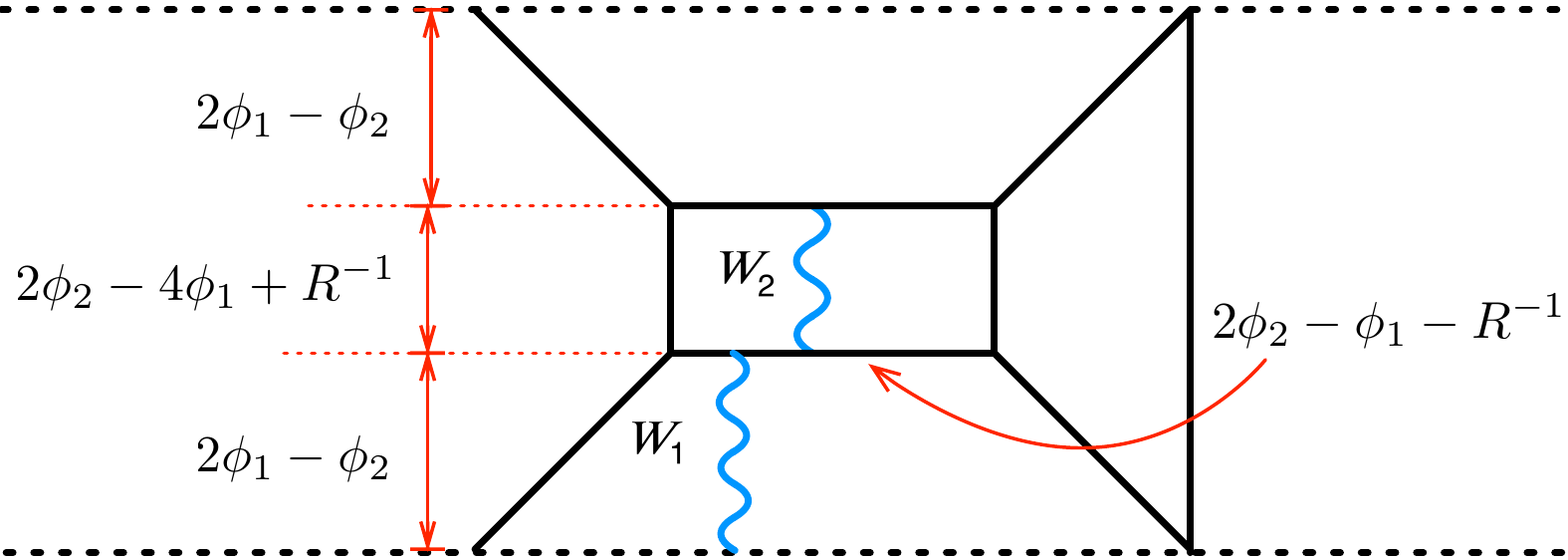}}
 \caption{(a) The brane web for the 6d $SO(8)$ theory on a $-3$ curve with $Z_2$ twist.
  (b) The brane web for the 6d $SU(3)$ theory on a  $-3$ curve with $Z_2$ twist. }
  \label{fig:SO8-SU3-twist}
\end{figure}


We can extend the new type of Higgsing to the other cases. The next example is the new type of Higgsing on the brane web for the $SO(9)$ gauge theory on a $-3$ curve depicted in Figure \ref{fig:SO9-2F-1S}(b).
We  move one of the $SO(9)$ D5-branes and one external D5-brane on the left-side toward the orientifold plane at the bottom by tuning the mass parameters as 
(\ref{eq:mass-twist}). We then obtain the brane web in Figure \ref{fig:SO8-SU3-twist}(a). We propose this brane web diagram for the $Z_2$ outer-automorphism twist of the 6d $SO(8)$ gauge theory on $-3$ curve.

This brane web shows the structure of affine $D^{(2)}_4$ gauge algebra, which is expected as the $Z_2$ twist of the affine $SO(8)$ gauge algebra. The gauge bosons from the brane web have the following masses: 
\begin{align}
	m_{W_1} &= 2\phi_1-\phi_2 \ , & m_{W_2} &= 2\phi_2-2\phi_1-\phi_3 \ ,\cr
	m_{W_3} & = 2\phi_3-\phi_2-2\phi_4 \ , & m_{W_4} &= 2\phi_4-\phi_3 \ ,
\end{align}
which form the affine $D^{(2)}_4$ Cartan matrix as expected. Moreover, the Higgsed brane web has only two mass parameter $m_1,m_2$ and it manifests the fact that the $Z_2$ twist of the $SO(8)$ theory with a fundamental and a spinor and a conjugate spinor hypermultiplet has two mass parameters for its global symmetry.

Another example is the Higgsing of the $G_2$ gauge theory with one fundamental hyper. We can consider the new Higgs brane flow in the brane diagram for this theory in Figure \ref{fig:G2-1F}(b). We find that the diagram can be Higgsed to the brane web in Figure \ref{fig:SO8-SU3-twist}(b) by putting one internal and one external D5-branes on the orientifold plane at the bottom. The Higgsed brane web turns out to be 
the brane web for the 5d $SU(3)$ gauge theory at Chern-Simons level $k=9$ proposed in \cite{Jefferson:2017ahm,Hayashi:2018lyv}. Incidentally, this 5d theory is conjectured to be the 6d $SU(3)$ minimal SCFT 
 with $Z_2$ outer-automorphism twist on a circle and the dual Calabi-Yau geometry is proposed as $S=\mathbb{F}_0\cup \mathbb{F}_{10}$ in \cite{Jefferson:2018irk}. So this example provides another Higgs branch flow leading to 6d theories with twisted compactification.
\begin{figure}
  \centering
  \includegraphics[width=.6\linewidth]{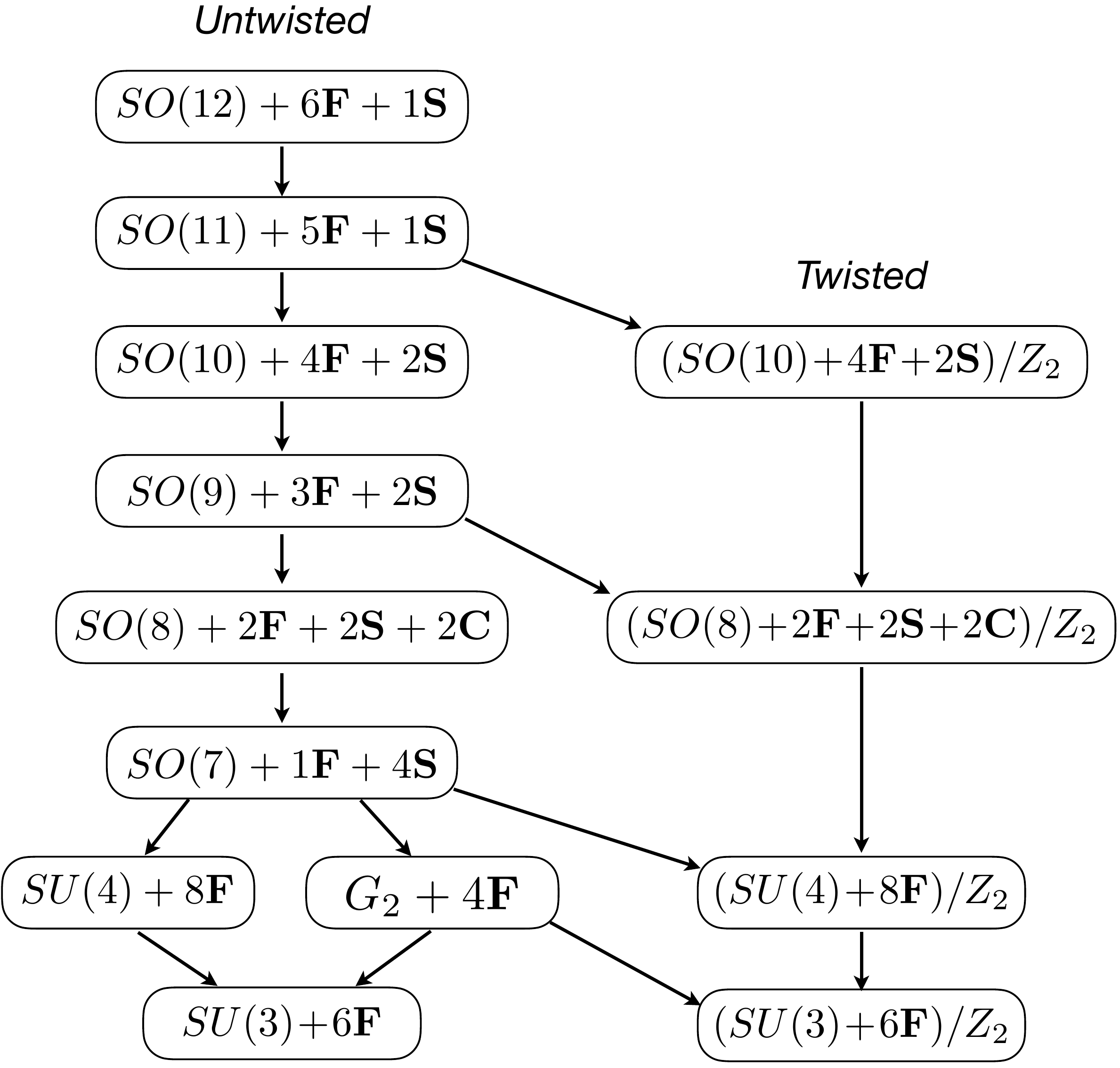}
  \caption{Higgsing hierarchy on a $-2$ curve.}
  \label{fig:HiggsChainO-2}
\end{figure}

\begin{figure}
  \centering
  \includegraphics[width=1\linewidth]{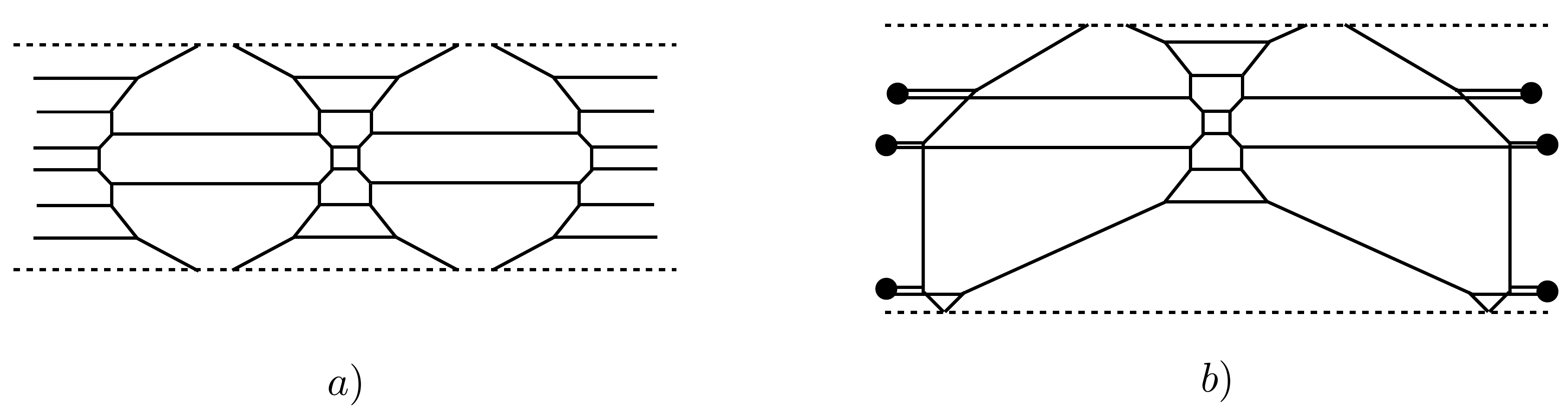}
  \caption{(a) The brane web for the 6d $SO(12)\times Sp(2)\times Sp(2)$ gauge theory.
  (b) The brane web for the 6d $SO(12)$ gauge theory on a $-2$ curve by Higgsing of (a). }
  \label{fig:SO12-Sp2-Sp2}
\end{figure}
 
\section{$SO(N)$ theories on $-2$ curve}\label{sec:SON02}
The discussion in the previous section shows that a family of $(p,q)$-brane webs for circle compactifications of 6d SCFTs on a $-3$ curve can be systematically obtained by RG-flows from a known brane configuration. In this section, we shall consider a natural generalization of this discussion now on the rank $2 \  (D_6,D_6)$ conformal matter theory \cite{DelZotto:2014hpa,Heckman:2015bfa}. This will allows us to engineer new interesting $(p,q)$-brane webs for 6d SCFTs now on a $-2$ curve. The hierarchy of RG fixed points we shall discuss in this section is drawn in Figure \ref{fig:HiggsChainO-2}.

\subsection{Untwisted theories}
\paragraph{\underline{$SO(12)$ gauge theory}}
We start with the brane web in Figure \ref{fig:SO12-Sp2-Sp2}(a). This web is for the 6d SCFT of 
$Sp(2)\times SO(12)\times Sp(2)$ gauge symmetry coupled to three tensor multiplets denoted as \cite{DelZotto:2014hpa,Heckman:2015bfa}
\begin{equation}
    [SO(12)] \ \stackrel{\mathfrak{sp}_2}{\bf1} \ \stackrel{\mathfrak{so}_{12}}{\bf4} \ \stackrel{\mathfrak{sp}_2}{\bf1} \ [SO(12)] \ ,
\end{equation}
which is also called as the rank 2 $(D_6,D_6)$ conformal matter theory.  We can Higgs this brane web by using external D5-branes of $SO(12)\times SO(12)$ global symmetry.  In the previous section we saw that the $Sp(2)$ gauge symmetry on the right side can be fully Higgsed by tuning 6 external D5-branes on the right side. We can consider the same Higgsing RG-flow for the $Sp(2)$ gauge symmetry on the left side. 
This brings the brane web in Figure \ref{fig:SO12-Sp2-Sp2}(a) to  the brane web in Figure \ref{fig:SO12-Sp2-Sp2}(b).
In F-theory, this blows down two $-1$ curves of $Sp(2)$ gauge symmetries in the base and the $-4$ curve in the middle reduces to a $-2$ curve.
After all, the 6d theory will be Higgsed to the 6d SCFT on a $-2$ curve with $SO(12)$ gauge symmetry coupled to a single tensor:
\begin{equation}
    [SO(12)] \ \stackrel{\mathfrak{sp}_2}{\bf1} \ \stackrel{\mathfrak{so}_{12}}{\bf4} \ \stackrel{\mathfrak{sp}_2}{\bf1} \ [SO(12)] \quad \rightarrow \quad \ \stackrel{\mathfrak{so}_{12}}{\bf2} [Sp(6)]
\end{equation}
We thus claim that the brane web in Figure \ref{fig:SO12-Sp2-Sp2}(b) is a brane construction of the 6d $SO(12)$ gauge theory on a $-2$ curve compactified on a circle.

\begin{figure}
  \centering
  \includegraphics[width=1\linewidth]{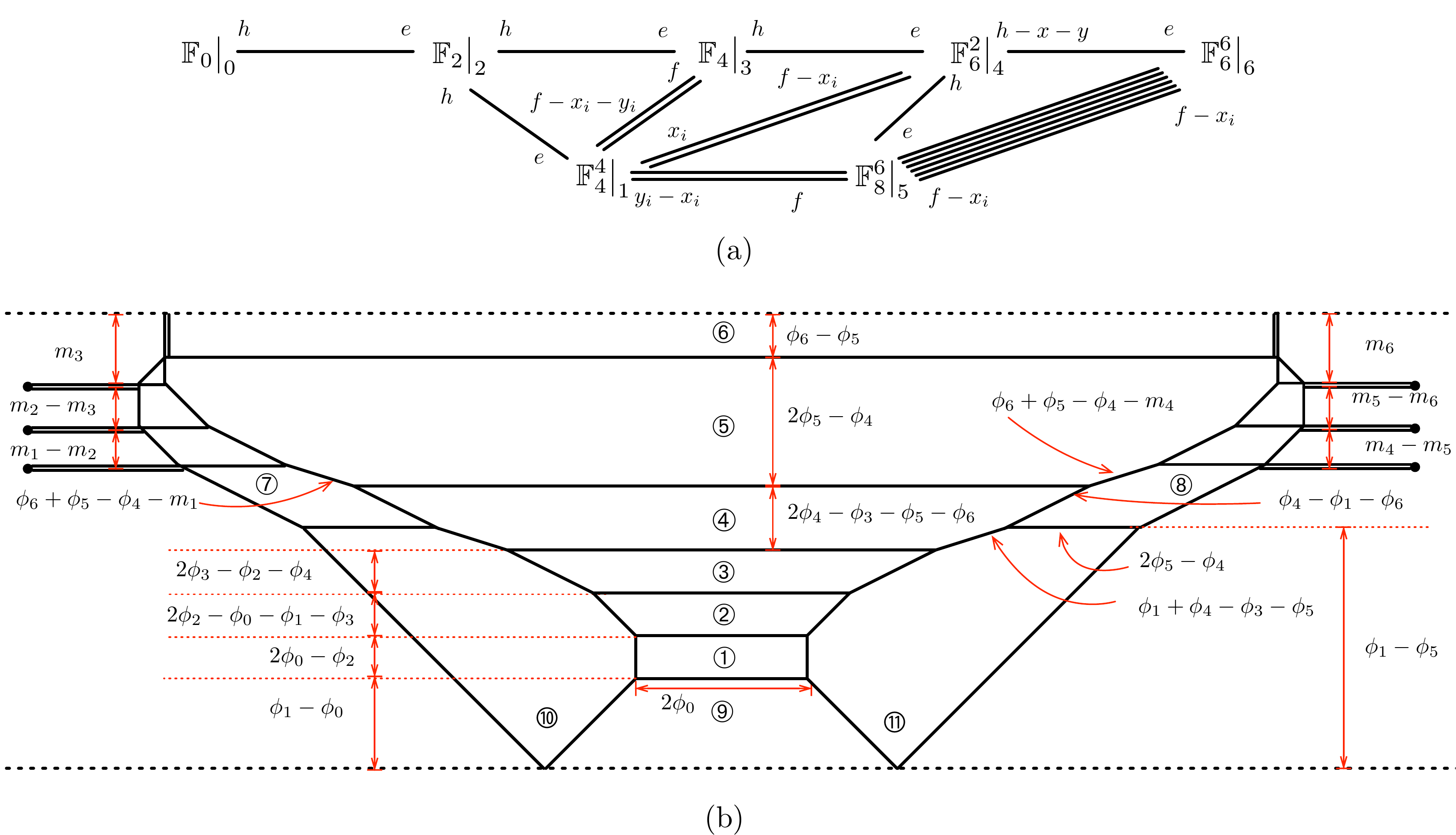}
  \caption{The elliptic threefold (a) and the brane web (b) for the 6d $SO(12)$ gauge theory on a $-2$ curve.}
  \label{fig:SO12-6F}
\end{figure}

Let us compare this brane construction against the geometric construction of the 6d $SO(12)$ theory given in \cite{Bhardwaj:2018yhy}.
For convenience, we shall consider the geometry in Figure \ref{fig:SO12-6F}(a) which is related to the geometry in \cite{Bhardwaj:2018yhy} by flop transitions of six exceptional curves denoted by $x_i$ in the surface $\mathbb{F}^6_8$. This threefold has the prepotential
\begin{align}
	6\mathcal{F}_{SO(12)} =& \ 8\phi_0^3 + 4\phi_1^3 + 8\phi_2^3 +8\phi_3^3+6\phi_4^3+2\phi_5^3+2\phi_6^3 -6\phi_0^2\phi_2 -12\phi_2^2\phi_3+6\phi_2 \phi_3^2 \nonumber \\
	& -12 \phi_1\phi_2^2+6\phi_1^2\phi_2-12\phi_1\phi_3^2-18\phi_3^2\phi_4+12\phi_3\phi_4^2-6\phi_1^2\phi_4 -6\phi_1\phi_4^2-12\phi_1\phi_5^2 \nonumber \\
	& -24\phi_4^2\phi_5+18\phi_4\phi_5^2 -18\phi_4^2\phi_6 + 12\phi_4\phi_6^2-18\phi_5^2\phi_6 -18\phi_5\phi_6^2 \nonumber\\
	&+12\phi_1\phi_2\phi_3 + 12\phi_1\phi_3\phi_4 + 12\phi_1\phi_4\phi_5 + 36\phi_4\phi_5\phi_6 \ .
\end{align}
To compare with this geometry we first need to move on to the chamber in Figure \ref{fig:SO12-6F}(b) by flopping some cycles in the brane web.
Then the compact four-cycles in the geometry can be identified with the internal faces in the brane web as
\begin{align}
  &\mathbb{F}_0|_{0} = \textcircled{\scriptsize 1} \ , \quad
  \mathbb{F}_4^4\big|_{1} = \textcircled{\scriptsize 1}+2\cdot \textcircled{\scriptsize9}+\textcircled{\scriptsize10}+\textcircled{\scriptsize11} \ , \quad 
   \mathbb{F}_2|_{2} = \textcircled{\scriptsize 2} \ , \quad \mathbb{F}_4|_{3} = \textcircled{\scriptsize 3} \ ,  \nonumber\\
 &\mathbb{F}_6^2\big|_{4} = \textcircled{\scriptsize 4} \ , \quad\mathbb{F}_8^6\big|_{5} = \textcircled{\scriptsize 5}+\textcircled{\scriptsize 7}+\textcircled{\scriptsize 8} \ , \quad \mathbb{F}_6^6\big|_{6} = \textcircled{\scriptsize 5} + 2\cdot \textcircled{\scriptsize 6} \ .
\end{align}
The monopole string tensions with respect to $\phi_i$ can be extracted from the brane web. The results are
\begin{align}
&T_{\footnotesize\textcircled{\scriptsize 1}} = 2\phi_0(2\phi_0\!-\!\phi_2) \ , \cr 
  &T_{\footnotesize\textcircled{\scriptsize 1}+2\textcircled{\scriptsize9}+\textcircled{\tiny10}+\textcircled{\tiny11}} = 2\phi_1^2\!-\!2\phi_2^2\!-\!2\phi_3^2\!-\!\phi_4^2\!-\!2\phi_5^2\!+\!2\phi_2(\phi_1\!+\!\phi_3)\!+\!2\phi_4(-\phi_1\!+\!\phi_3\!+\!\phi_5) \ , \nonumber \\
&T_{\footnotesize\textcircled{\scriptsize 2}} = (2\phi_2\!+\!\phi_0\!-\!\phi_1\!-\!\phi_3)(2\phi_2\!-\!\phi_0\!-\!\phi_1\!-\!\phi_3) \ , \nonumber \\
&T_{\footnotesize\textcircled{\scriptsize 3}} = 2(2\phi_3\!-\!\phi_2\!-\!\phi_4)(\phi_3\!+\!\phi_2\!-\!\phi_1\!-\!\phi_4) \ , \nonumber \\
&T_{\footnotesize\textcircled{\scriptsize 4}} = -\phi_1^2\!-\!3\phi_3^2\!+\!3\phi_4^2\!+\!3\phi_5^2\!+\!2\phi_6^2\!+\!2\phi_1(\phi_3\!-\!\phi_4\!+\!\phi_5)\!+\!2\phi_4(2\phi_3\!-\!4\phi_5\!-\!3\phi_6)\!+\!6\phi_5\phi_6 \ , \nonumber \\
&T_{\footnotesize\textcircled{\scriptsize 5}+\textcircled{\scriptsize 7}+\textcircled{\scriptsize 8}} =-4\phi_4^2\!+\!\phi_5^2\!-\!3\phi_6^2\!+\!2\phi_4(\phi_1\!+\!3\phi_5\!+\!3\phi_6)\!-\!2\phi_5(2\phi_1\!+\!3\phi_6) \ , \nonumber \\
 &T_{\footnotesize\textcircled{\scriptsize 5}+2\textcircled{\scriptsize 6}} = -3\phi_4^2\!-\!3\phi_5^2\!+\!\phi_6^2\!+\!2\phi_4(3\phi_5\!+\!2\phi_6)\!-\!6\phi_5\phi_6 \ ,
\end{align}
which perfectly match the volumes of surfaces in the geometry. More brane webs for the 6d SCFTs on a $-2$ curve can be obtained by a sequence of Higgs branch flows from this brane web.

\paragraph{\underline{$SO(11)$ gauge theory}}
\begin{figure}
  \centering
  \includegraphics[width=1\linewidth]{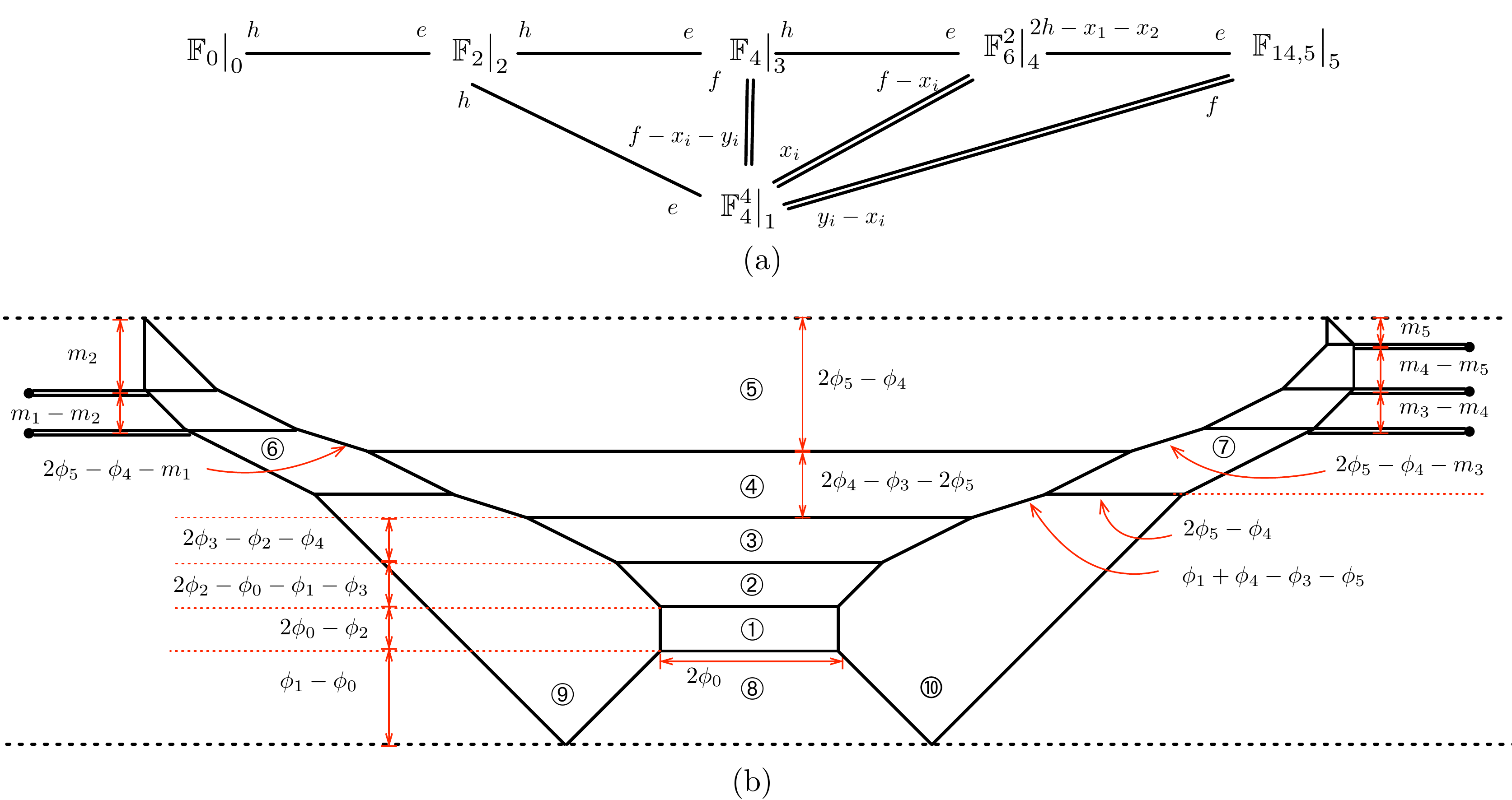}
  \caption{The elliptic threefold (a) and the brane web (b) for the 6d $SO(11)$ gauge theory on a $-2$ curve.}
  \label{fig:SO11-5F-1S}
\end{figure}

As we have shown, the RG-flow from the $SO(12)$ gauge theory down to the $SO(11)$ gauge theory can be generated by moving a set of D5-branes corresponding to an $SO(12)$ fundamental hyper toward the top orientifold plane in the brane system. In Figure \ref{fig:SO12-6F}(b), this amounts to setting mass parameters as
\begin{equation}
  \phi_6-\phi_5 = m_3 = 0 \ .
\end{equation} 
The Higgsed brane web is given in Figure \ref{fig:SO11-5F-1S}(b) and this realizes the 6d $SO(11)$ gauge theory on a $-2$ curve.

The corresponding CY$_3$ threefold consists of six compact 4-cycles intersecting each other as drawn in Figure \ref{fig:SO11-5F-1S}(a). The prepotential of this geometry is
\begin{align}
	6\mathcal{F}_{SO(11)} = & \ 8\phi_0^3 + 4\phi_1^3+8\phi_2^3+8\phi_3^3+6\phi_4^3-32\phi_5^3 -6\phi_0^2\phi_2 -12\phi_2^2\phi_3+6\phi_2\phi_3^2 \nonumber \\
	& -12\phi_1\phi_2^2+6\phi_1^2\phi_2 -12\phi_1\phi_3^2 -18\phi_3^2\phi_4 +12\phi_3\phi_4^2-6\phi_1^2\phi_4 -6\phi_1\phi_4^2-12\phi_1\phi_5^2 \nonumber \\
	& -42\phi_4^2\phi_5+66\phi_4\phi_5^2 +12\phi_1\phi_2\phi_3 + 12\phi_1\phi_3\phi_4 +12\phi_1 \phi_4\phi_5 \ .
\end{align}
We identify the 6 surfaces in the geometry with the internal faces in the brane web as
\begin{align}
  &\mathbb{F}_0|_{0} = \textcircled{\scriptsize 1} \ ,   \quad 
  \mathbb{F}_4^4|_{1} = \textcircled{\scriptsize 1}+2\cdot\textcircled{\scriptsize 8}+\textcircled{\scriptsize9}+\textcircled{\scriptsize10} \ , \quad
\mathbb{F}_2|_2 = \textcircled{\scriptsize 2} \ , \quad \mathbb{F}_4|_{3} = \textcircled{\scriptsize 3} \ ,  \nonumber \\
  & \mathbb{F}_6^2|_{4} = \textcircled{\scriptsize 4} \ , \quad\mathbb{F}_{14,5}|_{5} = 2\cdot\textcircled{\scriptsize 5}+\textcircled{\scriptsize 6}+\textcircled{\scriptsize 7} \ .
\end{align}
\begin{figure}
  \centering
  \includegraphics[width=1\linewidth]{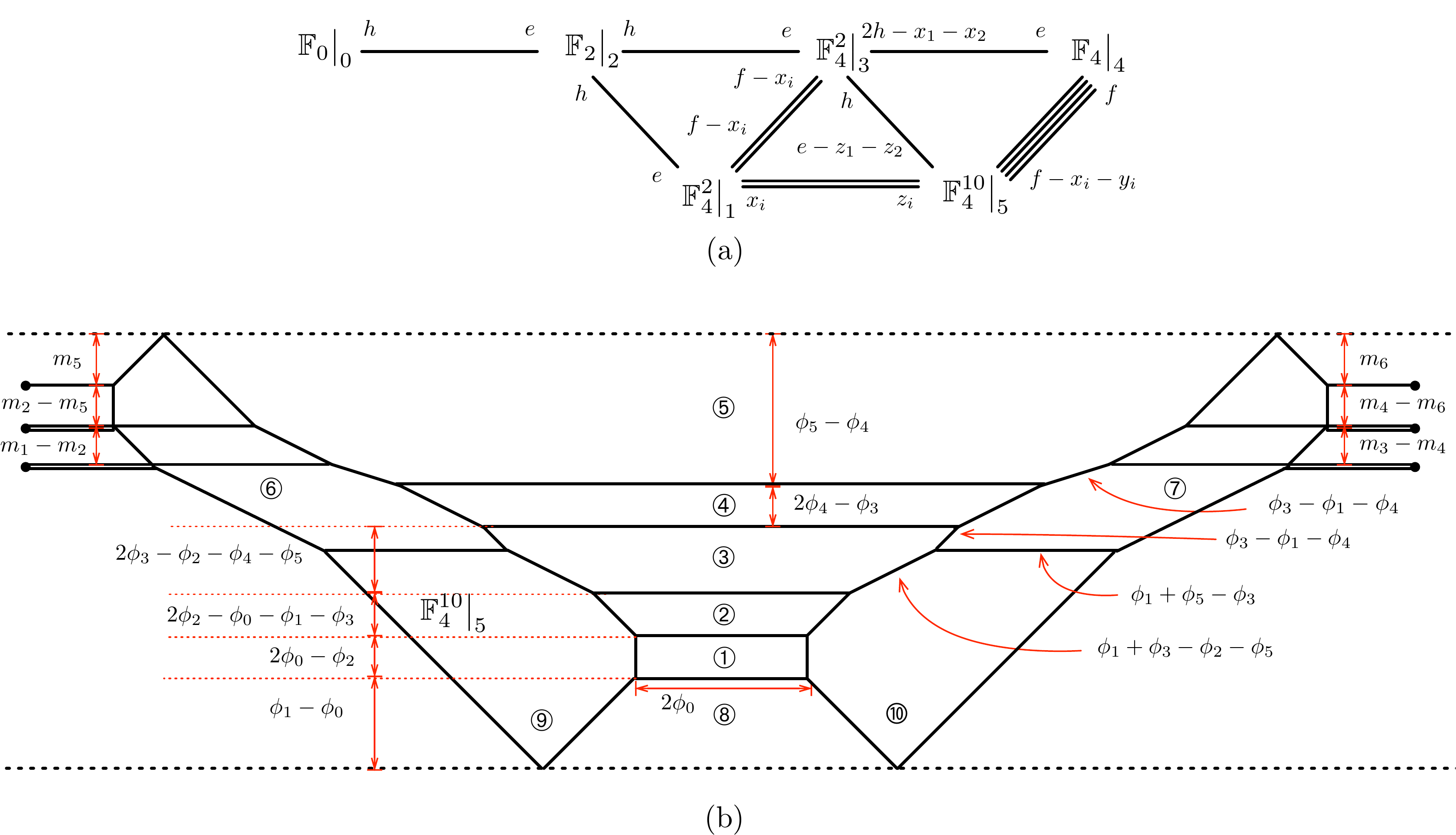}
  \caption{The elliptic threefold (a) and the brane web (b) for the 6d $SO(10)$ gauge theory on a $-2$ curve.}
  \label{fig:SO10-4F-2S}
\end{figure}
The volumes of these internal faces extracted from the branes are
\begin{align}
  &T_{\footnotesize\textcircled{\scriptsize 1}} = 2\phi_0(2\phi_0\!-\!\phi_2) \ , \cr 
  &T_{\footnotesize\textcircled{\scriptsize 1}+2\textcircled{\scriptsize 8}+\textcircled{\tiny9}+\textcircled{\tiny10}} = 2\phi_1^2\!-\!2\phi_2^2\!-\!2\phi_3^2\!-\!\phi_4^2\!-\!2\phi_5^2\!+\!2\phi_2(\phi_1\!+\!\phi_3)\!+\!2\phi_4(-\phi_1\!+\!\phi_3\!+\!\phi_5) \ ,\nonumber \\
  &T_{\footnotesize\textcircled{\scriptsize 2}} = (2\phi_2\!+\!\phi_0\!-\!\phi_1\!-\!\phi_3)(2\phi_2\!-\!\phi_0\!-\!\phi_1\!-\!\phi_3) \ , \nonumber \\
  &T_{\footnotesize\textcircled{\scriptsize 3}} = 2(2\phi_3\!-\!\phi_2\!-\!\phi_4)(\phi_3\!+\!\phi_2\!-\!\phi_1\!-\!\phi_4) \ , \nonumber \\
  &T_{\footnotesize\textcircled{\scriptsize 4}} = -\phi_1^2\!-\!3\phi_3^2\!+\!3\phi_4^2\!+\!11\phi_5^2\!+\!2\phi_1(\phi_3\!-\!\phi_4\!+\!\phi_5)\!+\!2\phi_4(2\phi_3\!-\!7\phi_5) \ , \nonumber \\
 &T_{\footnotesize2\textcircled{\scriptsize 5}+\textcircled{\scriptsize 6}} = (\phi_4\!-\!2\phi_5)(2\phi_1\!-\!7\phi_4\!+\!8\phi_5) \ .
\end{align}
This is in agreement with the volumes of compact four-cycles in the geometry.

\paragraph{\underline{$SO(10)$ gauge theory}}
Other SCFTs with $SO(N)$ gauge symmetry can be obtained by a series of Higgs branch flows. First, the brane realization of the $SO(10)$ gauge theory with 4 fundamentals and 2 spinor hypers is drawn in Figure \ref{fig:SO10-4F-2S}(b). The corresponding geometry is given in Figure \ref{fig:SO10-4F-2S}(a) and this geometry has the prepotential
\begin{align}
	6\mathcal{F}_{SO(10)} = & \ 8\phi_0^3 +6\phi_1^3 + 8\phi_2^3 + 6\phi_3^3 + 8\phi_4^3 -2\phi_5^3 -6\phi_0^2\phi_2 -12\phi_2^2\phi_3 +6\phi_2\phi_3^2 \\
	&-12\phi_1\phi_2^2 +6\phi_1^2\phi_2 -6\phi_1^2\phi_3 -6\phi_1\phi_3^2 -6\phi_1^2\phi_5-6\phi_1\phi_5^2-12\phi_3^2\phi_4 + 6\phi_3\phi_4^2 \nonumber \\
	&-18\phi_3^2\phi_5 +12\phi_3\phi_5^2-24\phi_4^2\phi_5+12\phi_1\phi_2\phi_3+12\phi_1\phi_3\phi_5+24\phi_3\phi_4\phi_5 \ .\nonumber
\end{align}
The map between
the compact faces in the brane web and the compact four-cycles in the geometry is
\begin{align}
	&\mathbb{F}_0|_{0} = \textcircled{\scriptsize 1} \ , \quad 
	\mathbb{F}_4^2|_{1} = \textcircled{\scriptsize 1}+2\cdot\textcircled{\scriptsize 8}+\textcircled{\scriptsize9}+\textcircled{\scriptsize10} \ ,
\quad	\mathbb{F}_2|_{2} = \textcircled{\scriptsize 2} \ , \quad \mathbb{F}_4^2|_{3} = \textcircled{\scriptsize 3} \ , \qquad \nonumber \\
	&
	\mathbb{F}_4|_{4} = \textcircled{\scriptsize 4} \ ,  \quad \mathbb{F}_4^{10}|_{5} = \textcircled{\scriptsize 4}+2\cdot\textcircled{\scriptsize 5}+\textcircled{\scriptsize 6}+\textcircled{\scriptsize 7} \ .
\end{align}
The volumes of these internal faces extracted from the branes are
\begin{align}
  &T_{\footnotesize\textcircled{\scriptsize 1}} = 2\phi_0 (2\phi_0 -\phi_2) \ , \cr 
  &T_{\footnotesize\textcircled{\scriptsize 1}+2\textcircled{\scriptsize 8}+\textcircled{\tiny9}+\textcircled{\tiny10}} = 3\phi_1^2 - 2\phi_2^2 - \phi_3^2 - \phi_5^2 +2\phi_1(\phi_2 -\phi_3-\phi_5)+2\phi_3(\phi_2+\phi_5) \ ,\nonumber \\
  &T_{\footnotesize\textcircled{\scriptsize 2}} = (2\phi_2\!+\!\phi_0\!-\!\phi_1\!-\!\phi_3)(2\phi_2\!-\!\phi_0\!-\!\phi_1\!-\!\phi_3) \ , \nonumber \\
  &T_{\footnotesize\textcircled{\scriptsize 3}} = -\phi_1^2\!-\!2\phi_2^2\!+\!3\phi_3^2\!+\!\phi_4^2\!+\!2\phi_5^2\!+\!2\phi_1(\phi_2\!-\!\phi_3-\!\phi_5)\!+\!2\phi_3(\phi_2\!-\!2\phi_4\!-\!3\phi_5)\!+\!4\phi_4\phi_5  \ , \nonumber \\
  &T_{\footnotesize\textcircled{\scriptsize 4}} =  2(2\phi_4-\phi_3)(\phi_3+\phi_4-2\phi_5)\ , \nonumber \\
 &T_{\footnotesize\textcircled{\scriptsize 4}+\footnotesize2\textcircled{\scriptsize 5}+\textcircled{\scriptsize 6}+\textcircled{\scriptsize 7}} = -\phi_1^2 -3\phi_3^2-4\phi_4^2-\phi_5^2+2\phi_1(\phi_3-\phi_5) + 4\phi_3(\phi_4+\phi_5)\ ,
\end{align}
which agree with the volumes of compact four-cycles in the geometry.

\paragraph{\underline{$SO(9)$ gauge theory}}
\begin{figure}
  \centering
  \includegraphics[width=1\linewidth]{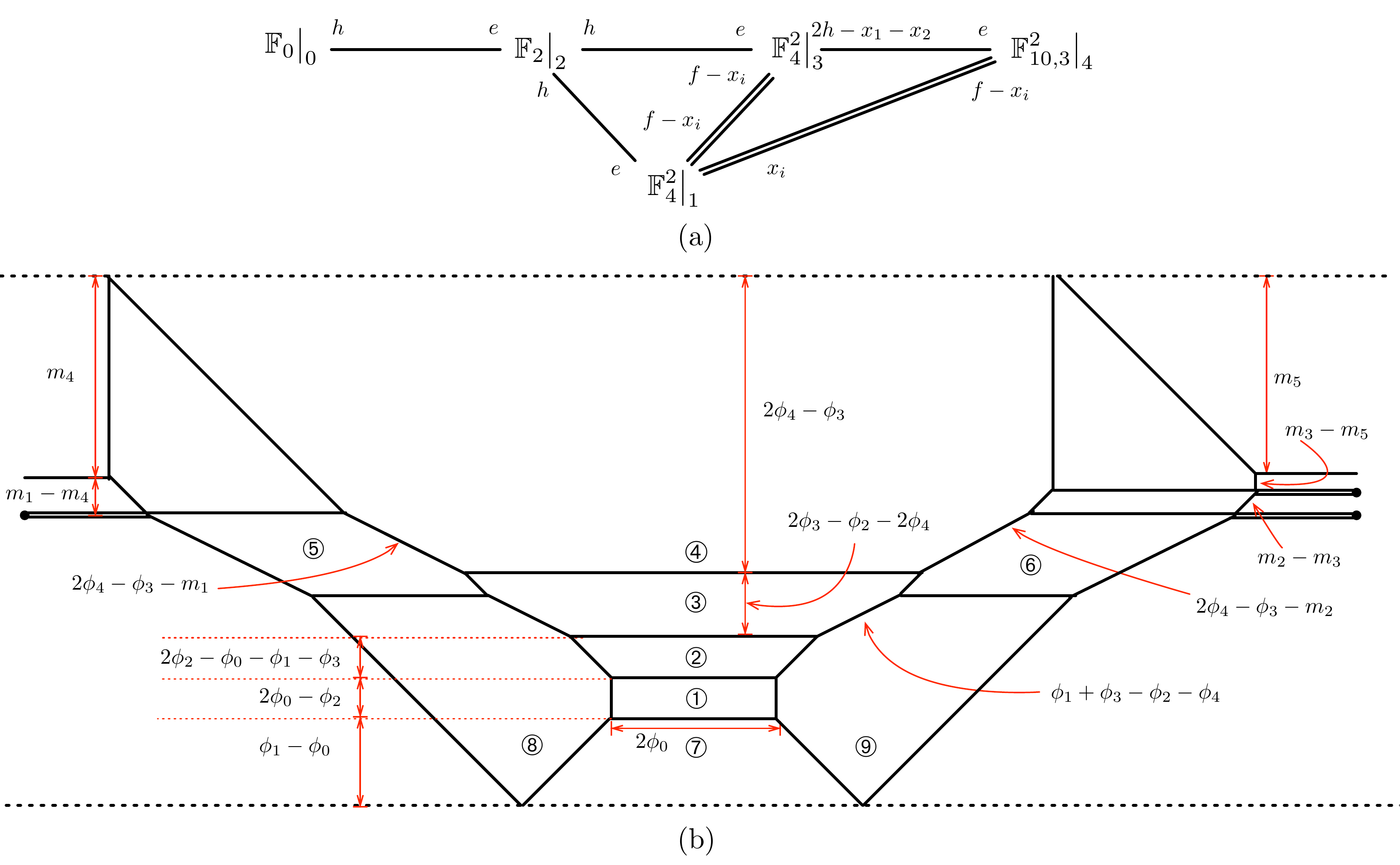}
  \caption{The elliptic threefold (a) and the brane web (b) for the 6d $SO(9)$ gauge theory on a $-2$ curve.}
  \label{fig:SO9-3F-2S}
\end{figure}

The brane web for the 6d SCFT on a $-2$ curve with $SO(9)$ gauge symmetry is depicted in Figure \ref{fig:SO9-3F-2S}(b). Its geometric counterpart is drawn in Figure \ref{fig:SO9-3F-2S}(a). This geometry is equivalent to the geometry for the $SO(9)$ gauge theory given in \cite{Bhardwaj:2018yhy} under flop transitions of two $-1$ curves $f-x_i$ in the surface $\mathbb{F}^2_{10,3}$.
The prepotential of this geometry is
\begin{align} 
	6\mathcal{F}_{SO(9)}=& \ 8\phi_0^3+6\phi_1^3 +8\phi_2^3+6\phi_3^3-18\phi_4^3 -6\phi_0^2\phi_2 -12\phi_1\phi_2^2+6\phi_1^2\phi_2\nonumber \\
	&
	-12\phi_2^2\phi_3+6\phi_2\phi_3^2 
	 -6\phi_1^2\phi_3 -6\phi_1\phi_3^2-6\phi_1^2\phi_4-6\phi_1\phi_4^2 -30\phi_3^2\phi_4 
	 \nonumber\\
	 &+42\phi_3\phi_4^2+12\phi_1\phi_2\phi_3 +12\phi_1\phi_3\phi_4 \ .
\end{align}
The map between the brane system and the geometry in Figure \ref{fig:SO9-3F-2S}(a) is
\begin{align}
	&\mathbb{F}_0|_{0} = \textcircled{\scriptsize 1} \ , \qquad 
	\mathbb{F}_4^2|_{1} = \textcircled{\scriptsize 1}+2\cdot\textcircled{\scriptsize 7}+\textcircled{\scriptsize 8}+\textcircled{\scriptsize9}\ , \qquad
	\mathbb{F}_2|_{2} = \textcircled{\scriptsize 2} \ , \nonumber \\
	& \mathbb{F}_4^2|_{3} = \textcircled{\scriptsize 3} \ ,   \qquad\mathbb{F}_{10,3}^2|_{4} = 2\cdot\textcircled{\scriptsize 4}+\textcircled{\scriptsize 5}+\textcircled{\scriptsize 6} \ .
\end{align}
The volumes of these internal faces extracted from the branes are
\begin{align}
  &T_{\footnotesize\textcircled{\scriptsize 1}} = 2\phi_0 (2\phi_0 -\phi_2) \ ,\cr  
  &T_{\footnotesize\textcircled{\scriptsize 1}+2\textcircled{\scriptsize 7}+\textcircled{\scriptsize8}+\textcircled{\scriptsize9}} = 3\phi_1^2 - 2\phi_2^2 - \phi_3^2 - \phi_4^2 +2\phi_1(\phi_2 -\phi_3-\phi_4)+2\phi_3(\phi_2+\phi_4) \ ,\nonumber \\
  &T_{\footnotesize\textcircled{\scriptsize 2}} = (2\phi_2\!+\!\phi_0\!-\!\phi_1\!-\!\phi_3)(2\phi_2\!-\!\phi_0\!-\!\phi_1\!-\!\phi_3) \ , \nonumber \\
  &T_{\footnotesize\textcircled{\scriptsize 3}} = -\phi_1^2\!-\!2\phi_2^2\!+\!3\phi_3^2\!+\!7\phi_4^2\!+\!2\phi_1(\phi_2\!-\!\phi_3-\!\phi_5)\!+\!2\phi_3(\phi_2\!-\!5\phi_4) \ , \nonumber \\
 &T_{\footnotesize2\textcircled{\scriptsize 4}+\textcircled{\scriptsize 5}+\textcircled{\scriptsize 6}} = -\phi_1^2 -5\phi_3^2-9\phi_4^2+2\phi_1(\phi_3-\phi_4) + 14\phi_3\phi_4\ ,
\end{align}
which agree with the volumes of compact four-cycles in the geometry.

\paragraph{\underline{$SO(8)$ gauge theory}}
\begin{figure}
  \centering
  \includegraphics[width=1\linewidth]{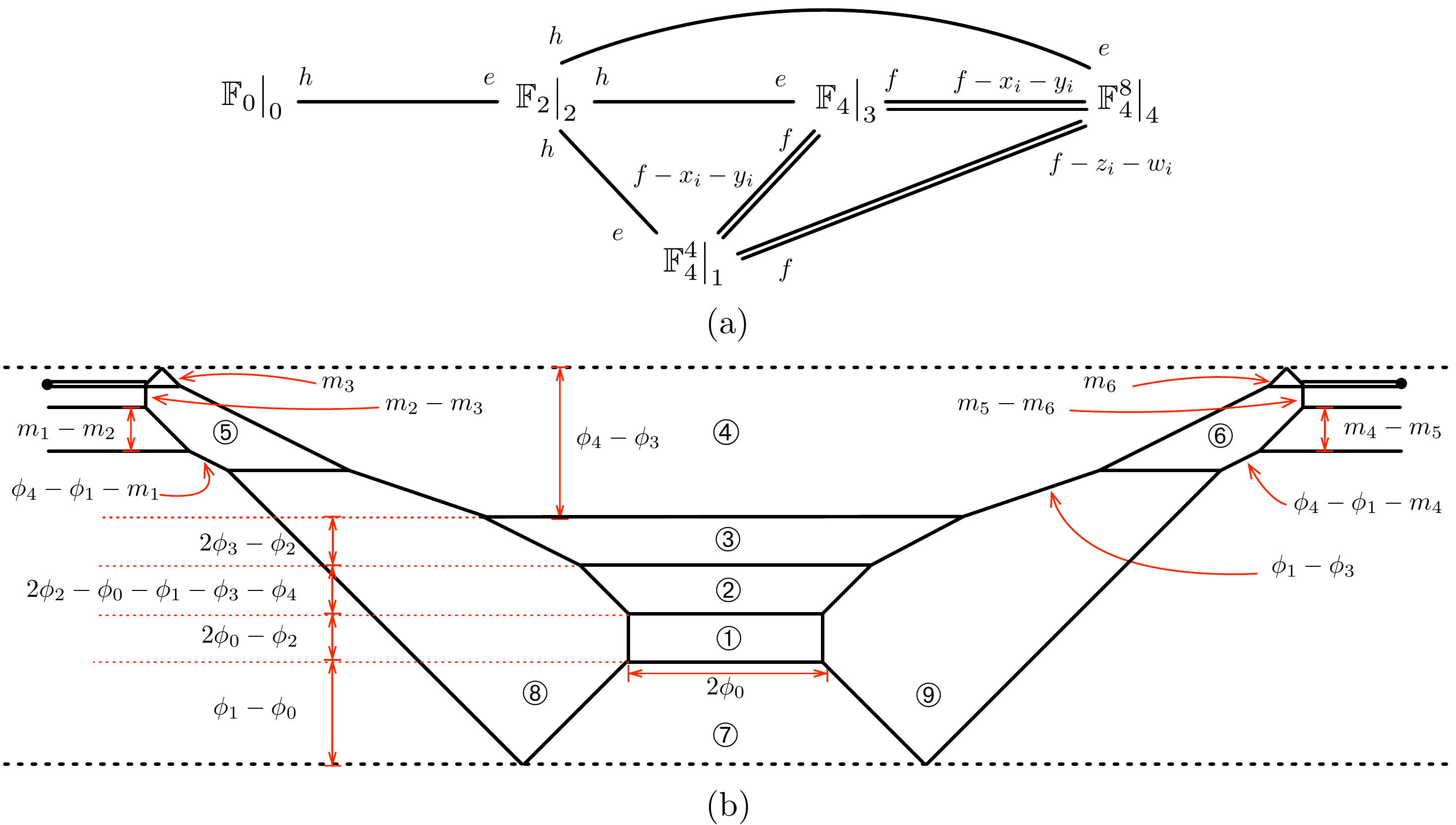}
  \caption{The elliptic threefold (a) and the brane web (b) for the 6d $SO(8)$ gauge theory on a $-2$ curve.}
  \label{fig:SO8-2F-2S-2C}
\end{figure}

The brane web for the 6d SCFT on a $-2$ curve with $SO(8)$ gauge symmetry is drawn in Figure \ref{fig:SO8-2F-2S-2C}(b). The geometry of this theory is summarized in Figure \ref{fig:SO8-2F-2S-2C}(a). We compute the cubic prepotential for this threefold as
\begin{align}
	6\mathcal{F}_{SO(8)} = & \ 8\phi_0^3 + 4\phi_1^3 + 8\phi_2^3 + 8\phi_3^3 -6\phi_0^2\phi_2 - 12\phi_1\phi_2^2 +6\phi_1^2\phi_2-12\phi_2^2\phi_3 + 6\phi_2 \phi_3^2  \nonumber \\
	&-12\phi_2^2\phi_4 + 6\phi_2\phi_4^2-12\phi_1\phi_3^2-12\phi_1^2\phi_4-12\phi_3^2\phi_4+12\phi_1\phi_2\phi_3\nonumber \\
	& + 12\phi_2\phi_3\phi_4 + 12\phi_1\phi_2\phi_4  \ .
\end{align}
The surfaces in the geometry is related to the internal faces in the brane web by the map given as
\begin{align}
	&\mathbb{F}_0|_{0} = \textcircled{\scriptsize 1} \ , \qquad 
	\mathbb{F}_4^4|_{1} = \textcircled{\scriptsize 1}+2\cdot\textcircled{\scriptsize 7}+\textcircled{\scriptsize 8}+\textcircled{\scriptsize9}\ , \qquad
	\mathbb{F}_2|_{2} = \textcircled{\scriptsize 2} \ , \nonumber \\
	& \mathbb{F}_4|_{3} = \textcircled{\scriptsize 3} \ , \qquad 
	\mathbb{F}_{4}^8|_{4} = \textcircled{\scriptsize 3}+2\cdot\textcircled{\scriptsize 4}+\textcircled{\scriptsize 5}+\textcircled{\scriptsize 6}\ .
\end{align}
The volumes of these internal faces extracted from the branes are
\begin{align}
  &T_{\footnotesize\textcircled{\scriptsize 1}} = 2\phi_0 (2\phi_0 -\phi_2) \ ,\cr  
  &T_{\footnotesize\textcircled{\scriptsize 1}+2\textcircled{\scriptsize 7}+\textcircled{\scriptsize8}+\textcircled{\scriptsize9}} = 2\big(\phi_1^2 - \phi_2^2 - \phi_3^2 +\phi_1(\phi_2 -2\phi_4)+\phi_2(\phi_3+\phi_4) \big)\ ,\nonumber \\
  &T_{\footnotesize\textcircled{\scriptsize 2}} = (-\phi_0\!+\!\phi_1\!-\!2\phi_2\!+\!\phi_3\!+\!\phi_4)(\phi_0\!+\!\phi_1\!-\!2\phi_2\!+\!\phi_3+\!\phi_4) \ , \nonumber \\
  &T_{\footnotesize\textcircled{\scriptsize 3}} = 2(\phi_2-2\phi_3)(\phi_1-\phi_2-\phi_3+\phi_4) \ , \nonumber \\
 &T_{\footnotesize2\textcircled{\scriptsize 3}+\footnotesize2\textcircled{\scriptsize 4}+\textcircled{\scriptsize 5}+\textcircled{\scriptsize 6}} = 2\big(\!-\!\phi_1^2 -\phi_2^2-\phi_3^2+\phi_2(\phi_1+\phi_3+\phi_4) \big)\ ,
\end{align}
which agree with the volumes of compact four-cycles in the geometry.


\paragraph{\underline{$SO(7)$ gauge theory}}
\begin{figure}
  \centering
  \includegraphics[width=.8\linewidth]{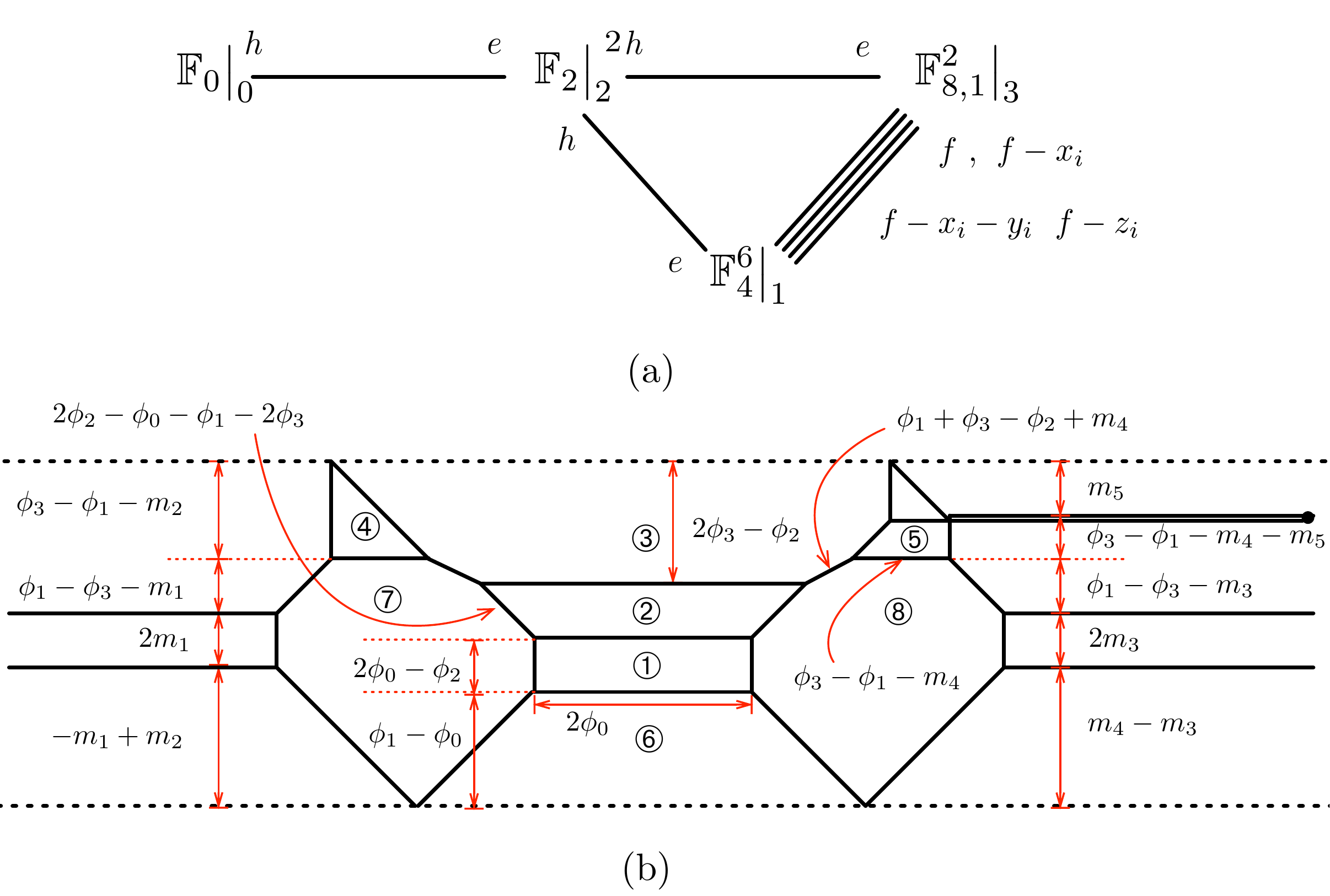}
  \caption{The elliptic threefold (a) and the brane web (b) for the 6d $SO(7)$ gauge theory on a $-2$ curve.}
  \label{fig:SO7-1F-4S}
\end{figure}
A further Higgs branch flow leads to the brane web for the 6d SCFT on a $-2$ curve with $SO(7)$ gauge symmetry which is drawn in Figure \ref{fig:SO7-1F-4S}(b). The geometry of this theory is in Figure \ref{fig:SO7-1F-4S}(a) and it has the prepotential
\begin{align}
	6\mathcal{F}_{SO(7)} = & \ 8\phi_0^3 +2\phi_1^3+ 8\phi_2^3 -2\phi_3^3-6\phi_0^2\phi_2 -12\phi_1\phi_2^2+6\phi_1^2\phi_2 -24\phi_2^2\phi_3 \nonumber \\
	& +24\phi_2\phi_3^2-18\phi_1\phi_3^2 -6\phi_1^2\phi_3+24\phi_1\phi_2\phi_3\ .
\end{align}
The 5-brane system can be mapped to the geometry by the following map:
\begin{align}
	&\mathbb{F}_0|_{0} = \textcircled{\scriptsize 1} \ , \qquad 
	\mathbb{F}_{4}^{6}|_{1} = \textcircled{\scriptsize 1}+2\cdot\textcircled{\scriptsize 6}+\textcircled{\scriptsize 7}+\textcircled{\scriptsize 8} \ , \qquad \mathbb{F}_2|_{2} = \textcircled{\scriptsize 2} \ , \nonumber \\
	& \mathbb{F}_{8,1}^2|_{3} = 2\cdot\textcircled{\scriptsize 3}+\textcircled{\scriptsize 4}+\textcircled{\scriptsize 5}  \ .
\end{align}
The geometry in Figure \ref{fig:SO7-1F-4S} is not quite equivalent to the geometry for this theory in \cite{Bhardwaj:2018yhy}. They are, in fact, related each other by flop transitions on two exceptional curves $x_i$ in $\mathbb{F}^2_{8,1}$.

We have also checked that the monopole string tensions obtained both from the brane webs and the geometries  are in perfect agreement. For example, the monopole string tensions from the brane web in Figure \ref{fig:SO8-2F-2S-2C}(b) are
\begin{align}
&T_{\footnotesize\textcircled{\scriptsize 1}} = 2\phi_0(2\phi_0\!-\!\phi_2) \ , \cr
&T_{\footnotesize\textcircled{\scriptsize 1}+2\textcircled{\scriptsize 6}+\textcircled{\scriptsize 7}+\textcircled{\scriptsize 8}} = \phi_1^2\!-2\phi_2^2\!-3\phi_3^2\!+\!2\phi_1\phi_2-2\phi_1\phi_3+4\,\phi_2\phi_3\, \crcr	
	&T_{\footnotesize\textcircled{\scriptsize 2}} = (-\phi_0\!+\!\phi_1\!-2\phi_2\!+2\phi_3)(\phi_0\!+\!\phi_1\!-2\phi_2\!+2\phi_3) \ , \nonumber \\
	&T_{2\textcircled{\scriptsize 3}+\textcircled{\scriptsize 4}+\textcircled{\scriptsize 5}} = -\phi_1^2\!-\!4\phi_2^2\!-\!\phi_3^2\!+\!4\phi_1\phi_2\!-\!6\phi_1\phi_3\!+\!8\phi_2\phi_3\ , 
\
\end{align}
which indeed match the volumes of the compact surfaces in the geometry.

\paragraph{\underline{$G_2$ gauge theory}}
\begin{figure}
  \centering
  \includegraphics[width=.9\linewidth]{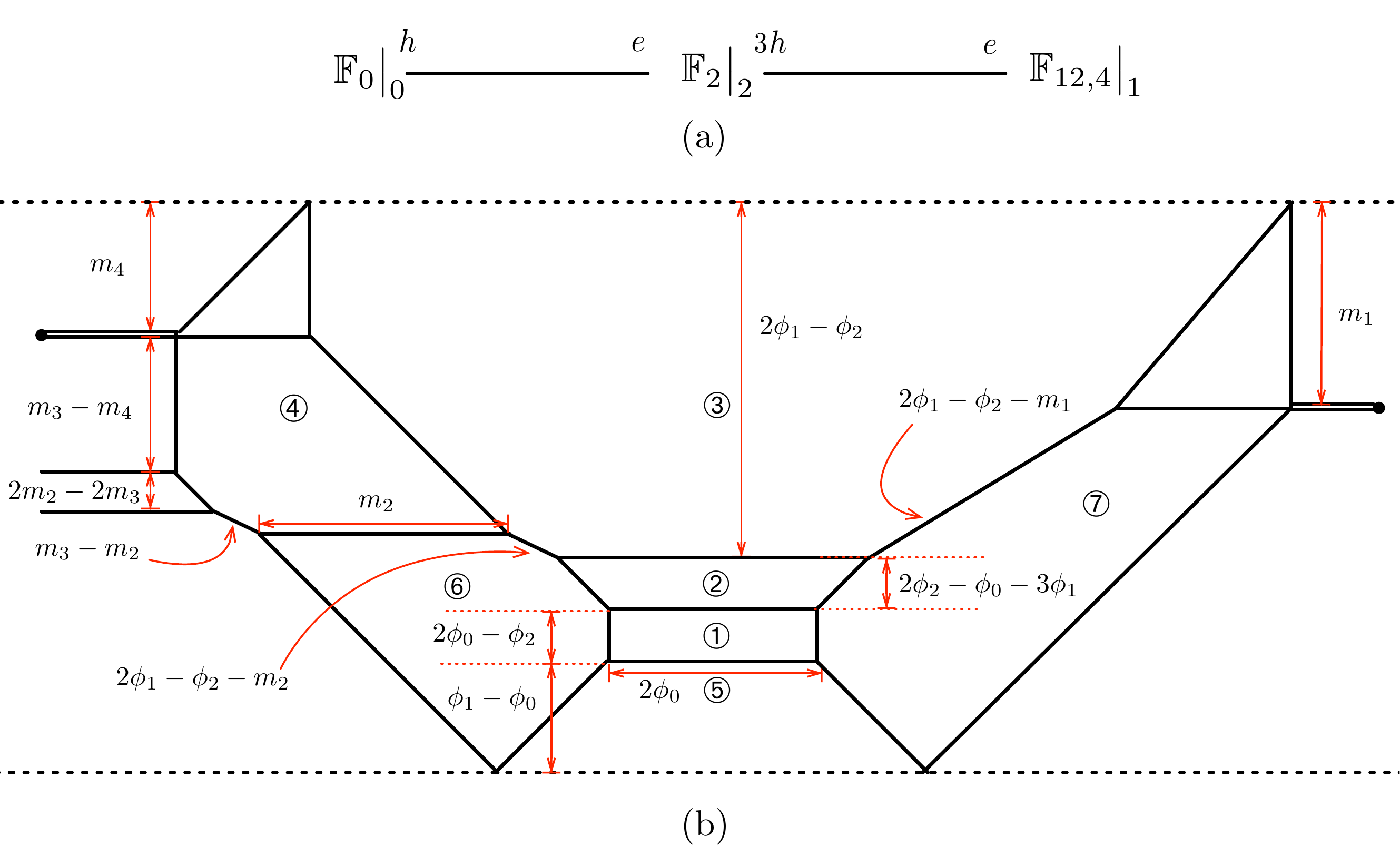}
  \caption{The elliptic threefold (a) and the brane web (b) for the 6d $G_2$ gauge theory on a $-2$ curve.}
  \label{fig:G2-4F}
\end{figure}

There are two distinct Higgs branch flows from the $SO(7)$ gauge theory and the corresponding brane web. One is the Higgsing to the $G_2$ gauge theory with four fundamentals. The brane web for this theory is illustrated in Figure \ref{fig:G2-4F}(b). This brane web corresponds to the CY$_3$ geometry in Figure \ref{fig:G2-4F}(a).
This threefold yields the prepotential 
\begin{align}
	6\mathcal{F}_{G_2} = & \ 8\phi_0^3 -24\phi_1^3 + 8\phi_2^3 -6\phi_0^2\phi_2 -36\phi_1\phi_2^2 +54\phi_1^2\phi_2 \ .
\end{align}
The map between the brane web and the geometry is given as
\begin{eqnarray}
	\mathbb{F}_0|_{0} = \textcircled{\scriptsize 1}  \ , \quad 	
	\mathbb{F}_{12,4}|_{1} = \textcircled{\scriptsize 1}+2\cdot\textcircled{\scriptsize 3}+\textcircled{\scriptsize 4}+2\cdot\textcircled{\scriptsize 5}+\textcircled{\scriptsize 6}+\textcircled{\scriptsize 7} \ , \quad 
		\mathbb{F}_2|_{2} = \textcircled{\scriptsize 2}\ .
\end{eqnarray}
As expected, the areas of the internal faces agree with the volumes of the compact surfaces in the geometry under this map
\begin{align}
&T_{\footnotesize\textcircled{\scriptsize 1}} = 2\phi_0(2\phi_0-\phi_2) \ , \cr
&T_{\footnotesize\textcircled{\scriptsize 1}+2\footnotesize\textcircled{\scriptsize 3}+\footnotesize\textcircled{\scriptsize 4}+2\footnotesize\textcircled{\scriptsize 5}+\footnotesize\textcircled{\scriptsize 6}+\footnotesize\textcircled{\scriptsize 7}} = 6 (-\phi_1+\phi_2) (2\phi_1-\phi_2)\ ,\cr
&T_{\footnotesize\textcircled{\scriptsize 2}} = (-2\phi_0-3\phi_1+2\phi_2)(\phi_0-3\phi_1+2\phi_2) \ .
\end{align}
\paragraph{\underline{$SO(6)$ gauge theory}}
\begin{figure}
  \centering
  \includegraphics[width=15.5cm]{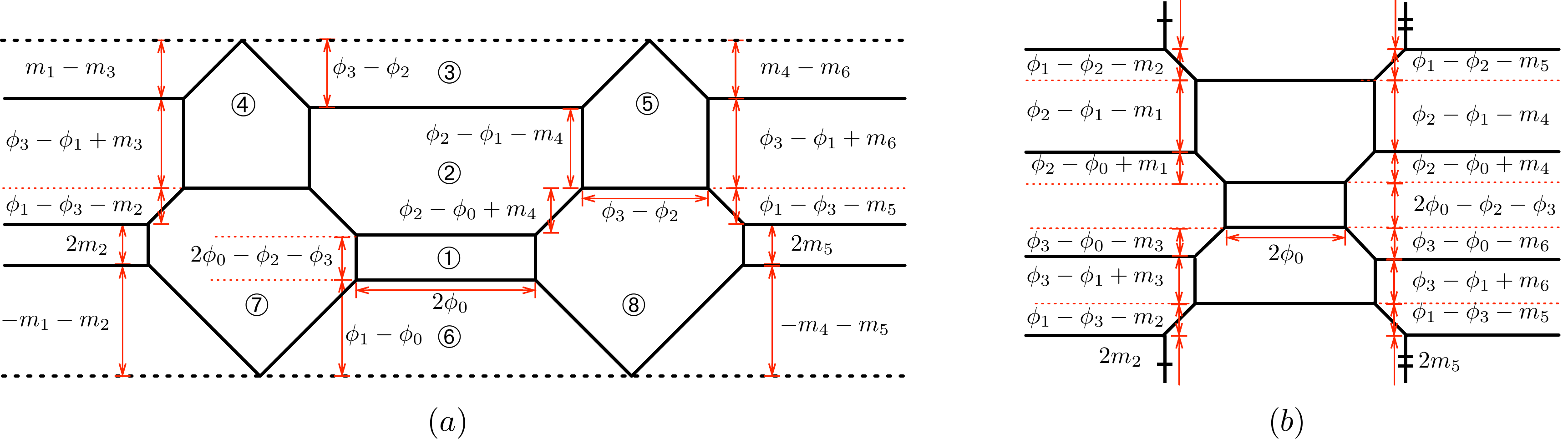}
  \caption{The elliptic threefold (a) and the brane web (b) for the 6d $SU(4)$ gauge theory with 8 fundamentals.}
  \label{fig:SO6-6S}
\end{figure}

Another Higgs branch limit is to the 6d SCFT with $SO(6)$ (or $SU(4)$) gauge symmetry and 8 spinor hypermultiplets. This theory is called as the $(A_3,A_3)$ conformal matter theory coupled to one tensor multiplet. The RG-flow from our $SO(7)$ brane web which Higgs the $SO(7)$ gauge symmetry to $SO(6)$ symmetry gives rise to the brane web in Figure \ref{fig:SO6-6S}(a). On the other hand, the brane realization of this theory is already known as Figure \ref{fig:SO6-6S}(b). We find that, though these two brane webs look quite different, they both engineer the $SO(6)$ gauge theory with 8 spinor hypers at low energy. We have also checked that the monopole strings in two brane realizations have the same tensions as their counterparts.

\subsection{Twisted theories}
In the previous section, we  obtained brane realizations of the $SO(2N)$ gauge theories with twisted compactification by Higgsing the brane webs of $SO(2N+1)$ gauge theories on a $-3$ curve. We can employ the same Higgsing procedure on the brane webs of the $SO(2N+1)$ theories we discussed in this section to obtain the brane realizations of the $SO(2N)$ gauge theories on a $-2$ curve with $Z_2$ outer-automorphism twist.

\begin{figure}
  \centering
  \includegraphics[width=0.85\linewidth]{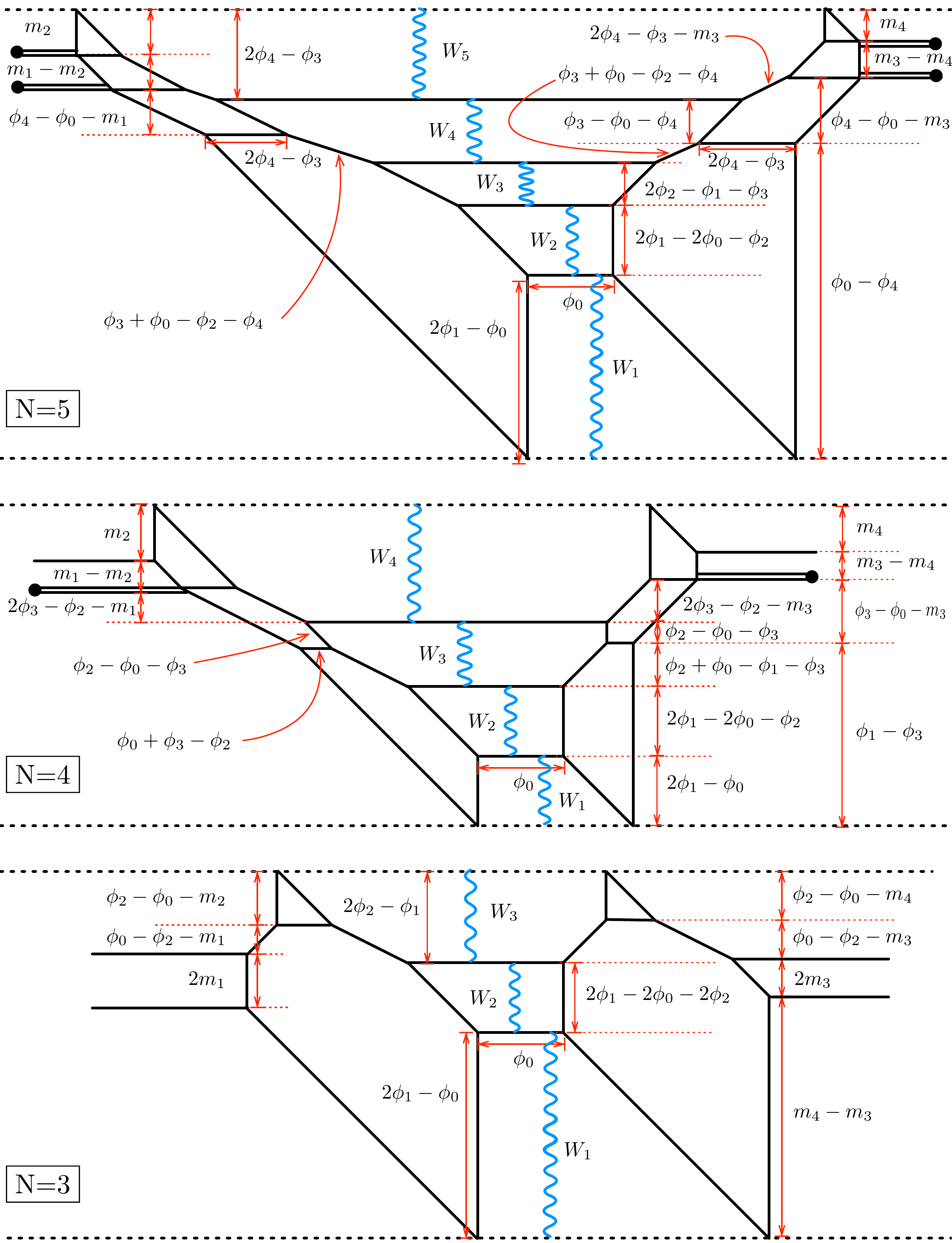}
  \caption{The brane webs for the 6d $SO(2N)$ gauge theories on a $-2$ curve with $Z_2$ twist, $N=5,4,3$.}
  \label{fig:SO-2-twist}
\end{figure}

We can first Higgs the brane web in Figure \ref{fig:SO11-5F-1S}(b) by moving one of the D7-branes toward the bottom orientifold plane. The resulting brane web is drawn in Figure \ref{fig:SO-2-twist}(a) and it realizes the SCFT of $SO(10)$ gauge symmetry on $-2$ curve with $Z_2$ outer-automorphism twist.
This insists that the strings $W_i$ connecting adjacent D5- or O5-branes must carry gauge charges of the affine $D_5^{(2)}$ algebra.
We can confirm this by look at the string masses. The charged string masses from the brane web are
\begin{eqnarray}
 &&m_{W_1} = 2\phi_0-\phi_1 \ , \quad m_{W_2} = 2\phi_1-2\phi_0-\phi_2 \ , \quad m_{W_3} = 2\phi_2-\phi_1-\phi_3 \ , \nonumber \\
&&m_{W_4} = 2\phi_3-\phi_2-2\phi_4 \ , \quad m_{W_5} = 2\phi_4-\phi_3 \ ,
\end{eqnarray}
which indeed exhibit the affine $D_5^{(2)}$ Cartan structure.

Similarly, the brane web for the $SO(9)$ gauge theory in Figure \ref{fig:SO9-3F-2S}(b) and that for the $SO(7)$ gauge theories in Figure \ref{fig:SO7-1F-4S}(b) can be Higgsed down to the brane configurations in Figure \ref{fig:SO-2-twist}(b) and Figure \ref{fig:SO-2-twist}(c) respectively. 
We conjecture that these brane webs in Figure \ref{fig:SO-2-twist}(b) and Figure \ref{fig:SO-2-twist}(c) are the $Z_2$ twisted compactifications of the $SO(8)$ and the $SO(6)$ gauge theories respectively on a $-2$ curve. As one can see, the strings $W_i$ have masses forming the affine $D_4^{(2)}$ Cartan matrix and the affine $D_3^{(2)}$ Cartan matrix respectively.

\begin{figure}
  \centering
  \includegraphics[width=.8\linewidth]{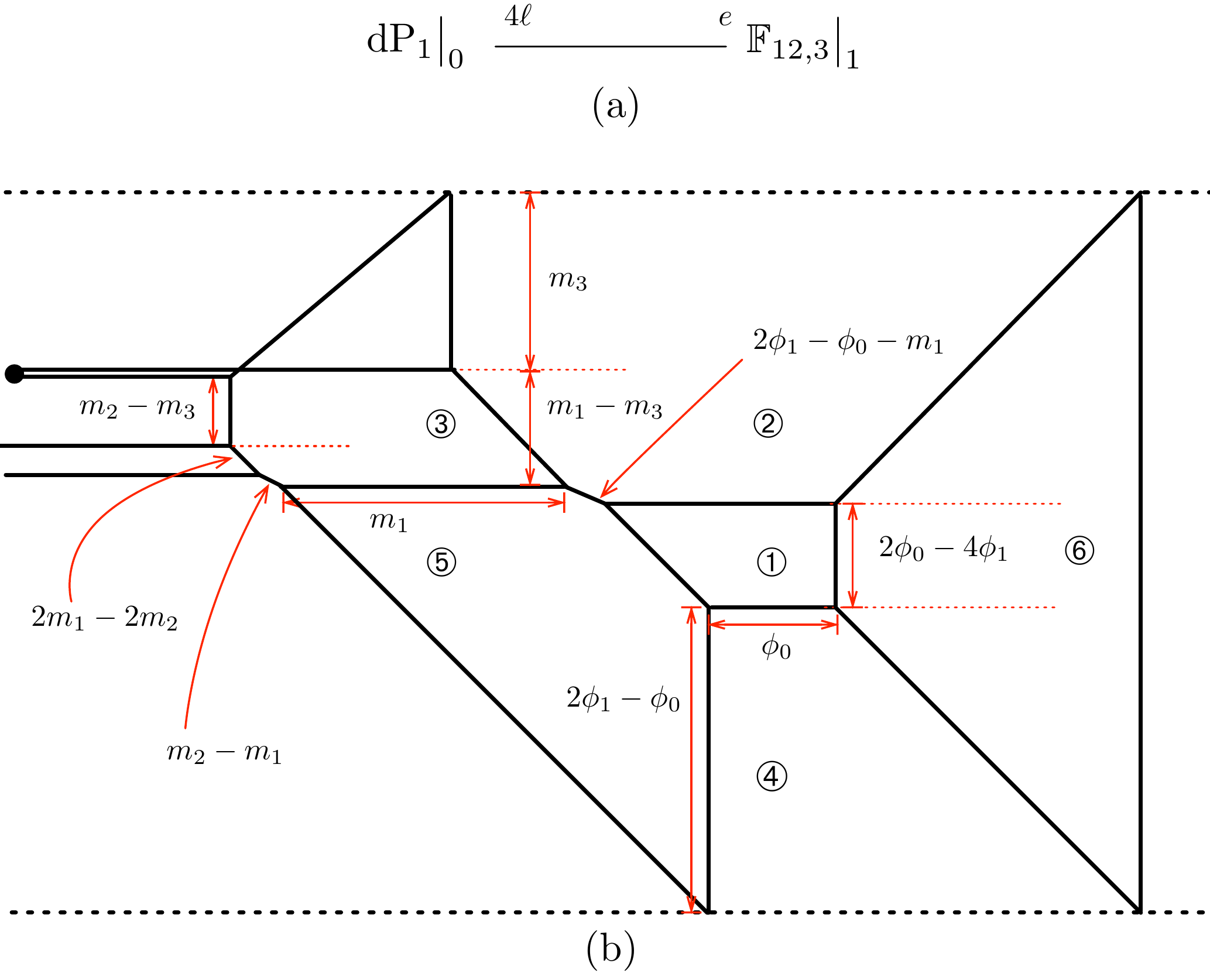}
  \caption{The elliptic threefold (a) and the brane web (b) for the 6d $SU(3)$ gauge theory with 6 fundamentals with $Z_2$ twist.}
  \label{fig:SU3-6F-twist}
\end{figure}

Lastly, we can Higgs the brane web in Figure \ref{fig:G2-4F}(b) and reduce it to the brane web for the $SU(3)$ gauge theory on a $-2$ curve with $Z_2$ outer-automorphism twist. The Higgsed diagram is illustrated in Figure \ref{fig:SU3-6F-twist}(b). We find that this brane web amounts to the geometry in Figure \ref{fig:SU3-6F-twist}(a) with the following map:
\begin{equation}
  {\rm dP}_1 = \textcircled{\scriptsize 1} \ , \qquad \mathbb{F}_{12,3} = 2\cdot\textcircled{\scriptsize 2}+\textcircled{\scriptsize 3}+2\cdot\textcircled{\scriptsize 4}+\textcircled{\scriptsize 5}+\textcircled{\scriptsize 6} \ .
\end{equation}
The monopole strings in the brane web have the tensions
\begin{equation}
  T_{\footnotesize\textcircled{\scriptsize 1}} = 4(\phi_0-2\phi_1)(\phi_0-\phi_1) \ , \quad T_{\footnotesize2\textcircled{\scriptsize 2}+\textcircled{\scriptsize 3}+2\textcircled{\scriptsize 4}+\textcircled{\scriptsize 5}+\textcircled{\scriptsize 6}} = -2(\phi_0-2\phi_1)(3\phi_0-2\phi_1) \ .
\end{equation}
This again agrees with the volumes of the surfaces in the geometry.

It was conjectured in \cite{Jefferson:2018irk}  that the $Z_2$ twisted compactification of the $SU(3)$ gauge theory on a $-2$ curve will reduce to the 5d $\mathcal{N}=1$ $Sp(2)_0$ gauge theory with 3 hypermultiplets in the anti-symmetric representation of $Sp(2)$ and, in addition, the associated geometry was proposed. We find that the geometry, which is of the form $\mathbb{F}_6\!\stackrel{2\ell}{\cup}\!{\rm dP}_4$, proposed in \cite{Jefferson:2018irk} can be deformed to the geometry in Figure \ref{fig:SU3-6F-twist}(a) by flopping three exceptional curves $\ell-x_1-x_i$ with $i=2,3,4$ in the dP$_4$.
This also support our brane proposal for this twisted theory in Figure \ref{fig:SU3-6F-twist}(b). 
We note also that another brane realization for this theory was proposed in \cite{Hayashi:2018lyv} which would be a dual description of our brane realization in Figure \ref{fig:SU3-6F-twist}(b).

\section{Twisting of $SO(2N)$ theories on $-4$ curve}
\label{sec:SON04}
\begin{figure}
  \centering
  \includegraphics[width=.9\linewidth]{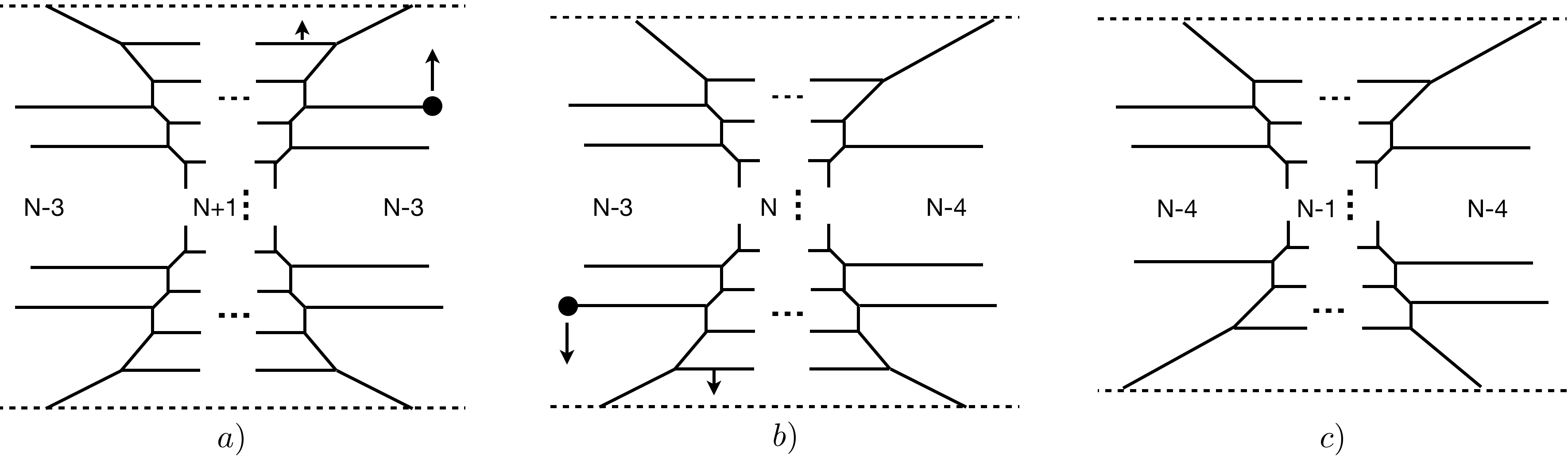}
  \caption{(a) The brane web for the 6d $SO(2N\!+\!2)$ gauge theory with $2N\!-\!6$ fundamental hypers. (b) The brane web for the 6d $SO(2N\!+\!1)$ gauge theory with $2N\!-\!7$ fundamental hypers. (c) The brane web for the $SO(2N)$ gauge theory with $2N\!-\!8$ fundamentals on $Z_2$ twisted compactification.}
  \label{fig:SO2N}
\end{figure}

One may wonder what if we consider a Higgsing of the $SO(2N\!+\!1)$ gauge theories with $2N-7$ fundamental hypers on a $-4$ curve by giving a vev to a fundamental scalar carrying unit Kaluza-Klein momentum along 6d circle. Similarly to the cases we discussed above, we expect that this type of Higgsing leads to the twisted compactifications of the $SO(2N)$ gauge theories on a $-4$ curve. This follows that we can engineer the brane systems for these twisted theories by a Higgsing of known brane webs for the $SO(2N\!+\!1)$ gauge theories with $2N-7$ fundamentals.

Let us start from the $(p,q)$-brane system in Figure \ref{fig:SO2N}(a) for the $SO(2N\!+\!2)$ gauge theory with $2N-6$ fundamental hypers. One can Higgs this diagram to the brane web for the $SO(2N\!+\!1)$ gauge theory. For this, we move one of the internal D5-branes and also one of the external D5-branes to the location of the orientifold plane at the top. Then we end up with the brane web in Figure \ref{fig:SO2N}(b). This web is for the $SO(2N\!+\!1)$ gauge theory with $2N-7$ fundamentals.
We now consider moving another set of D5-branes denoted by downward arrows to the bottom orientifold plane. The resulting brane web in Figure \ref{fig:SO2N}(c) is the brane web corresponding to the Higgsing triggered by a Higgs vev of a $SO(2N\!+\!1)$ fundamental scalar mode carrying KK momentum. This brane web is thus conjectured to realize the twisted compactification of the $SO(2N)$ gauge theory with $2N-8$ fundamental hypermultiplets.
The brane web now has $2N-8$ external D5-branes which is the same as those in the original $SO(2N)$ gauge theory before twisting. This respects the fact that fundamental hypermultiplets remain intact under $Z_2$ twisting of $SO(2N)$ gauge algebra.

Another simple check for this proposal is to compare the string charges between the internal D5-branes. The $i$-th string suspending between $i$-th and $(i+1)$-th D5-branes have the masses as
\begin{align}
	&m_1 = 2\phi_1 - \phi_2 \ , \nonumber\\
	&m_2 = 2\phi_2 - 2\phi_1 - \phi_3 \ , \nonumber \\
	&m_{i} = 2\phi_i - \phi_{i-1} - \phi_{i+1} \quad {\rm for} \ 2< i <N-1 \ , \nonumber \\
	&m_{N-1} = 2\phi_{N-1} - \phi_{N-2} - 2\phi_{N} \ ,\nonumber \\
	&m_{N} = 2\phi_{N} - \phi_{N-1} \ ,
\end{align}
after taking into account orientifold projections at the top and the bottom of the diagram. This result shows that the strings connecting internal branes form the affine $D^{(2)}_{N}$ gauge algebra. This signals that our brane web in Figure \ref{fig:SO2N}(c) would be the $Z_2$ twisting of the $SO(2N)$ gauge theory on a $-4$ curve. 
\section{Conclusion}\label{sec:conclusion}
In this paper, we constructed Type IIB 5-brane configurations for a family of 6d $\mathcal{N}=(0,1)$ SCFTs of $D_N$ gauge symmetry on a $-2$ or $-3$ curve in the base engineered in F-theory \cite{Heckman:2015bfa} when compactified on a circle. These brane webs are obtained by considering RG flows developed in certain Higgs branches in the brane webs associated to the 6d $(D_N,D_N)$ conformal matter theories introduced in \cite{DelZotto:2014hpa}. We have also explored new RG flows triggered by Higgs vevs of scalar operators carrying non-zero KK momentum along the compactification circle. Surprisingly, these RG flows realize twisted compactifications of 6d SCFTs with $SO(2N)$ or $SU(3)$ gauge symmetry on a $-2$ or $-3$ curve.

Let us make some comments about our results and future researches. First, we have considered Higgs branch flows by scalar vevs with unit KK momentum in 6d theories compactified on a circle and found that these RG flows give rise to consistent low energy theories. The low energy theories discussed in this paper turn out to be circle compactifications of 6d theories with outer-automorphism twist which are already known to exist. One may wonder if similar RG flows by using Higgs vevs of KK modes can provide us a new family of consistent 6d theories living on a circle.

Another important point is that ordinary Type IIB branes in exotic configurations appear to support new realizations of 6d theories and also 5d theories which either have exceptional $G_2$ gauge symmetry or share some unusual properties with exceptional gauge theories. For example, the 6d minimal $SU(3)$ gauge theory was previous engineered by $(p,q)$ string junctions between mutually non-perturbative 7-branes admitting no weakly coupled Type IIB description. However we could find a $(p,q)$ 5-brane web for this theory where the non-abelian gauge states come from perturbative strings stretched between D5-branes. It would be very interesting to further investigate those exotic 5-brane configurations and see if we can construct $(p,q)$ 5-brane systems for other exceptional gauge theories.

Lastly, brane webs found in this paper can be used to compute the partition functions of the corresponding 6d theories  on $\Omega$-background. For this computation, we can employ topological vertex formalism introduced in \cite{Aganagic:2003db,Iqbal:2007ii} and also additional techniques in \cite{Kim:2017jqn,Hayashi:2018bkd} to take into account the orientifold planes. The results for the theories on a $-3$ curve can be compared with some known results in \cite{Kim:2018gjo,Kim:2016foj}, which would exhibit more concrete checks for our proposal \cite{KKL}.

\section*{Acknowledgements}

We would like to thank Hirotaka Hayashi, Songling He, Seok Kim, Noppadol Mekareeya, Alessandro Tomasiello, Xing-Yue Wei, Futoshi Yagi
and Gabi Zafrir for useful discussions. This work was motivated from discussions at the Aspen Center for Physics, which is supported by National Science Foundation grant PHY-1607611.
We acknowledge APCTP for hosting the Focus program ``Strings, Branes and Gauge Theories 2019'', and YMSC for hosting the Workshop on Mirror Symmetry and Related Topics at Tsinghua Sanya International Mathematics Forum (TSIMF) where part of this work is done.  This work has been benefited from  the 2019 Pollica summer workshop, which was supported in part by the Simons Foundation (Simons Collaboration on the Non-perturbative Bootstrap) and in part by the INFN.  H.K. and K.L. would like to hank the SCGP workshop, Simons Summer Workshop: Cosmology and String Theory where the part of the work is done. 
S.K. thanks KIAS and POSTECH for hospitality for his visit.
The research of H.K. is supported by the POSCO Science Fellowship of POSCO TJ Park Foundation and the National Research Foundation of Korea (NRF) Grant 2018R1D1A1B07042934.
S.K. is supported by the UESTC Research Grant 
A03017023801317.
K.L. is supported in part by the National Research Foundation of Korea Grant  NRF-2017R1D1A1B06034369.

\appendix
 \begin{figure}
  \centering
  \includegraphics[width=.9\linewidth]{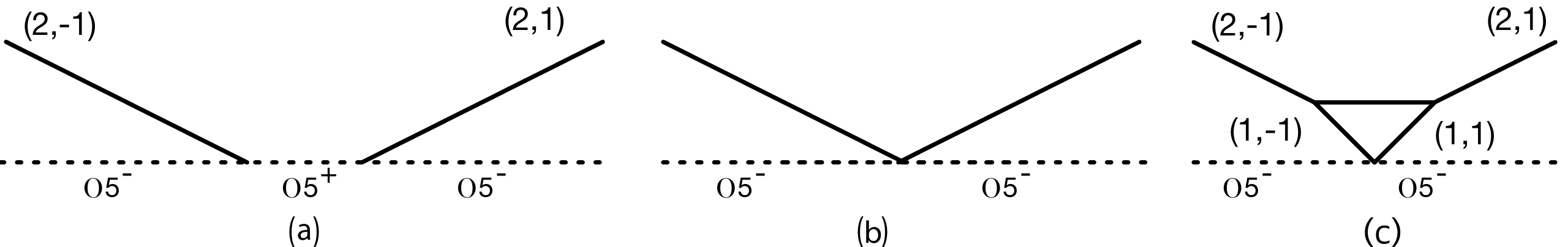}
  \caption{(a) A 5-brane configuration before a generalized transition where $(2,-1)$ and $(2,1)$ 5-branes are located with some distance. (b) Two 5-branes meeting at a point on an O5-plane. (c) A 5-brane configuration after a generalized transition where a D5-brane in the middle is generalized while $(1,-1)$ and $(1,1)$ 5-branes remain intact.}
  \label{fig:generalizedFlop}
\end{figure}
\section{Brane transitions on orientifold 5-branes}\label{sec:appA}
We briefly discuss possible brane transitions on an O5-plane. In Type IIB 5-brane webs, spinor matter of an $SO(N)$ gauge theory can be realized non-perturbatively \cite{Zafrir:2015ftn}. An $Sp(0)$ instanton of coupled to an $SO(N)$ gauge theory is a 5-brane realization of a hypermultiplet in the spinor representation of $SO(N)$ gauge theory where the ``coupling'' of  $Sp(0)$ naturally provides a mass of a spinor hypermultiplet which appears as the distance between two 5-branes ending on an O5-plane. In order to realize small mass spinor parameter, one needs to reduce the distance between two 5-branes on an O5-plane. Along the way, two such 5-branes meet at a point on a O5-plane. To realize a massless spinor hypermultiplet, one may bring them even more close. In doing so, 
it was observed that 5-branes undergo ``generalized flop transitions'' \cite{Hayashi:2017btw}, which looks similar to usual flop transitions in 5-brane webs without orientifold planes. An instructive example is given in Figure \ref{fig:generalizedFlop}. One can also think of a brane configuration in Figure \ref{fig:generalizedFlop} as a part of web diagram for an $Sp(N)$ gauge theory, as discussed in \cite{Kim:2017jqn}.  

With ``generalized  flop transitions'' one can realize various 5-brane configurations. Of course, the massless limit of a spinor hypermultiplet for an $SO(N)$ gauge theory is straightforward to construct. For instance, 5d $G_2$ gauge theory with flavors was  constructed as a Higgsing of the 5d $SO(7)$ gauge theory with spinor \cite{Hayashi:2018bkd}. Moreover, in a similar way, 5-brane webs for 5d gauge theory with hypermultiplet in the rank-3 antisymmetric representation was also constructed \cite{Hayashi:2019yxj}. 

 These brane transitions are allowed phases in the parameter space of Seiberg-Witten curves for 5d theories. 
Before and after the transitions,  the diagrams should be a charge conserving configuration. Figure \ref{fig:generalizedFlop} is a good example. 
 A 5-brane configuration in Figure \ref{fig:generalizedFlop}(a) changes successively to \ref{fig:generalizedFlop}(b) and to \ref{fig:generalizedFlop}(c), where the height of the newly generated D5-brane can be higher and higher as the distance between $(2,-1)$ and $(2,1)$ 5-branes getting closer and closer, keeping the 5-brane configuration on an O5-plane where  $(1,-1)$ and $(1,1)$ 5-branes met a point. The position of the newly generated D5-brane can be placed even higher by applying usual flop transitions, and then can be used for a non-perturbative Higgsing as discussed in the main text. Many more examples are also illustrated in \cite{Hayashi:2017btw, Hayashi:2018bkd, Hayashi:2018lyv, Hayashi:2019yxj}.

\section{Hanany-Witten transitions on a 5-brane web with O5-planes and 7-brane monodromy cuts}\label{sec:appB}
\begin{figure}
  \centering
  \includegraphics[width=.60\linewidth]{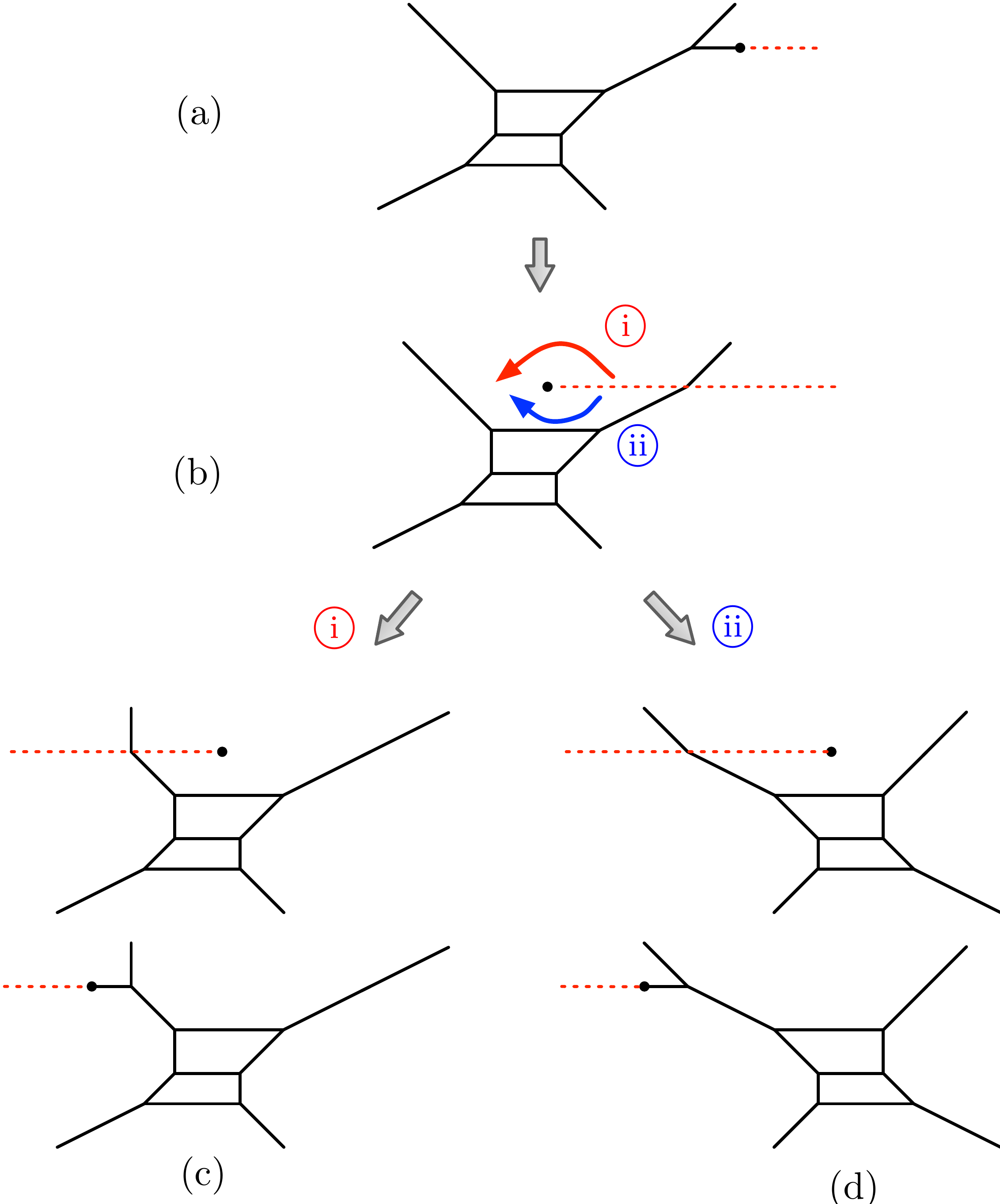}
  \caption{Hanany-Witten transitions and 7-brane monodromy cut. By rotating the 7-brane monodromy cut (counter-)clockwise, one obtains 5-brane configuration either (d) for (e), and two configurations are related by an $SL(2, \mathbb{Z})$ transformation.}
  \label{fig:HWusual}
\end{figure}
Hanany-Witten transitions on a 5-brane web are realized as moving a $(p,q)$ 7-brane on one side, split by a $(p', q')$ 5-brane, to the other side along the direction of the 7-brane charge. This procedure is repeated till there remain no further 5-branes along its way so that the 7-brane can be taken to infinity without crossing other 5-branes. A 7-brane has the monodromy which is given by the charge of the 7-brane, and the corresponding monodromy  cut can be put in an arbitrary direction on a 5-brane web but the charges of 5-branes crossing the monodromy cut should be carefully treated. Though the orientation of a $(p,q)$ 7-brane monodromy cut is usually chosen to be aligned along the orientation of its charge  $(p,q)$ on a 5-brane web, one can align a $(p,q)$ 7-brane cut arbitrary or even rotate it clockwise or counterclockwise. This is often understood when one performs Hanany-Witten transitions. 

\begin{figure}
  \centering
  \includegraphics[width=.65\linewidth]{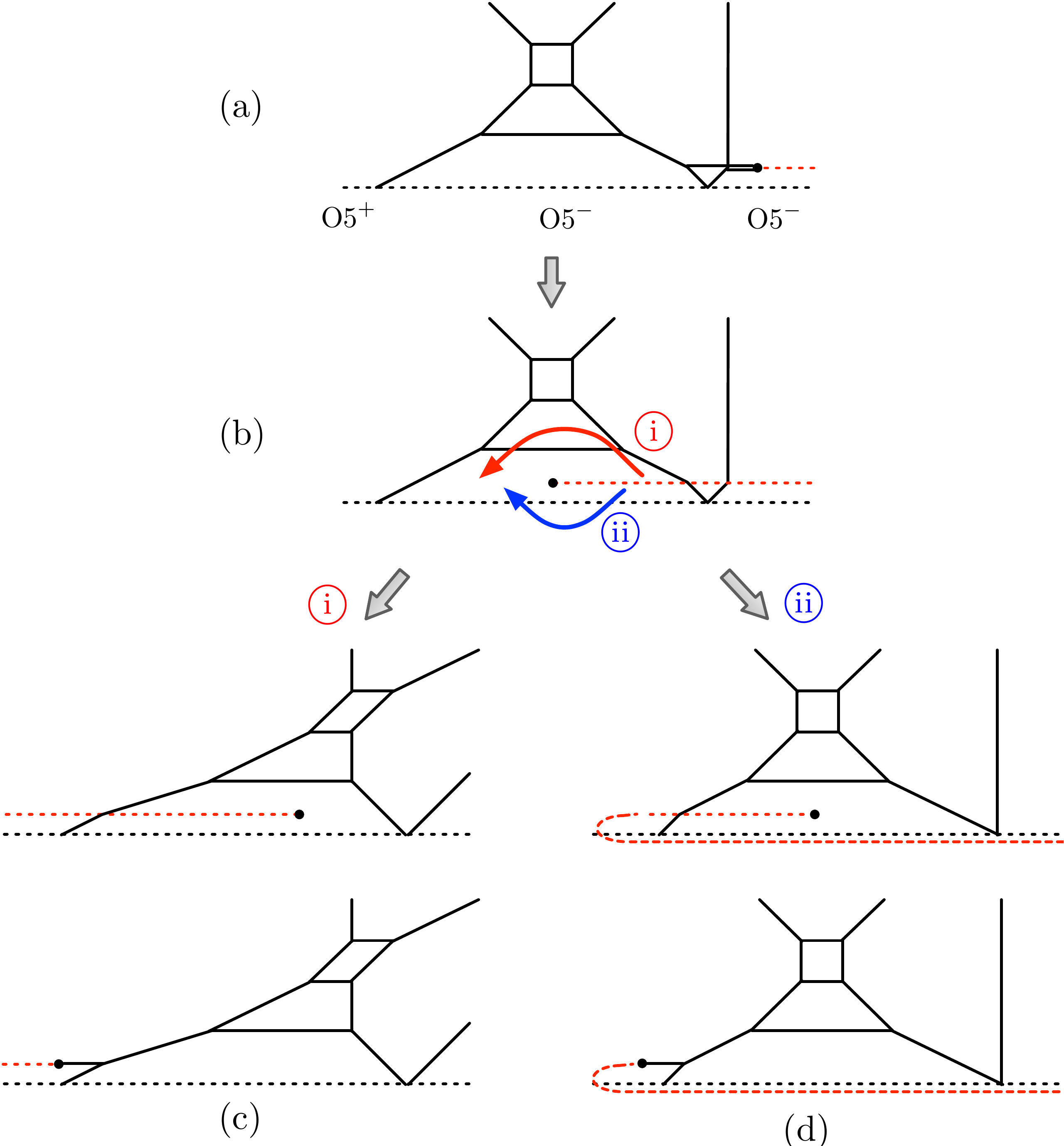}
\caption{Hanany-Witten transitions and 7-brane monodromy cut for a 5-brane configuration with an O5-plane. By rotating the 7-brane monodromy cut (counter-)clockwise, one obtains 5-brane configuration either (c) for (d).}
  \label{fig:HWwO5}
\end{figure}

Consider an example of Hanany-Witten transition on a 5-brane configuration, drawn in Figure \ref{fig:HWusual}, where the monodromy cuts are depicted as the red dotted lines.  Figure \ref{fig:HWusual}(b) is an intermediate step of the Hanany-Witten transition where the $(1,0)$ 7-brane is being moved to the middle from the right, where the $(1,0)$ 5-brane ending on this $(1,0)$ 7-brane is now annihilated as it crossed a 5-brane. Figure \ref{fig:HWusual}(b) is still a charge conserving configuration due to the orientation of the 7-brane cut. The next step is then to move this $(1,0)$ 7-brane to the left. Here one can rotate the monodromy cut clockwise or counterclockwise. Figure \ref{fig:HWusual}(c) is the resulting configuration when the monodromy cut is rotated counterclockwise \textcircled{\scriptsize i}, followed by pulling the 7-brane to the left. On the other hand, Figure \ref{fig:HWusual}(d) is the configuration after the monodromy cut is rotated clockwise \textcircled{\scriptsize ii}, followed by pulling the 7-brane to the left. We note that both configurations are a charge conserving configuration and they are related by an $SL(2, \mathbb{Z})$ $T$-transformation. 

\begin{figure}
  \centering
\includegraphics[width=.9\linewidth]{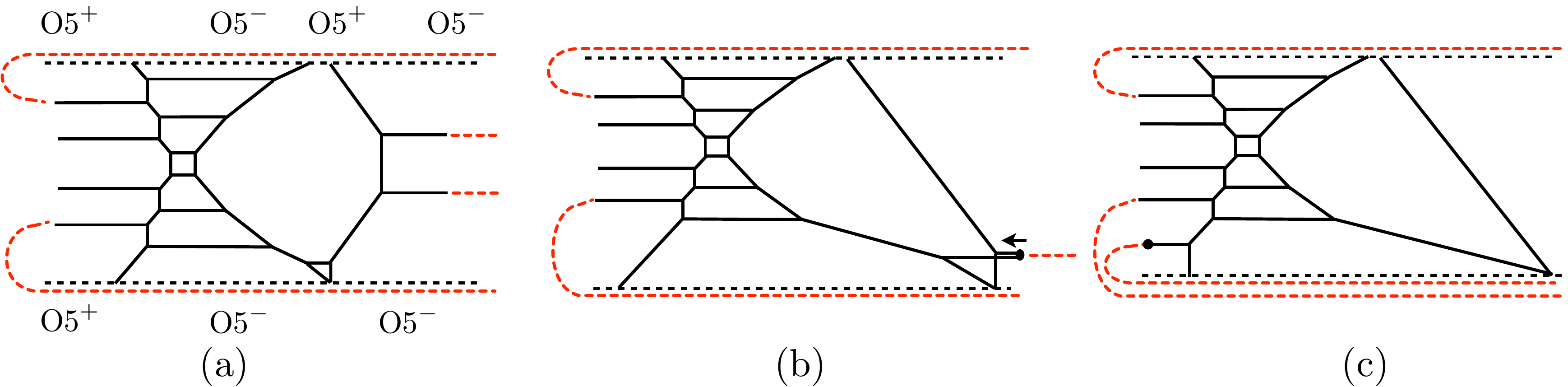}
  \caption{Monodromy cuts for Figure \ref{fig:SO12-Sp0}. The red dotted lines denote the monodromy cuts for 7-branes which are moved to the left via Hanany-Witten transitions.}
  \label{fig:SO12-Sp0-1}
\end{figure}

When a 5-brane configuration of interest has an O5-plane, Hanany-Witten transitions can also be understood in a similar way but with some caution. For instance, see Figure \ref{fig:HWwO5}(a). It is a part of 5-brane configuration discussed in the main text. 
Just as in the previous example of Hanany-Witten transition, one can put a $(1,0)$ 7-brane in the middle as depicted in Figure \ref{fig:HWwO5}(b), and there again one can rotate around the monodromy cut clockwise or counterclockwise. The counterclockwise rotation \textcircled{\scriptsize i} is straightforward and the configuration can be understood as the same way as the case without  an O5-plane, and the result configuration is given in Figure \ref{fig:HWwO5}(c). A clockwise rotation \textcircled{\scriptsize ii}, however, requires some caution. By Figure  \ref{fig:HWwO5}(c), we imagine a partial clockwise rotation of the 7-brane monodromy cut such that the monodromy cut is curved so that it still points its initial orientation along the O5-plane. This configuration is in particular useful as it gives rise to a charge conserving 5-brane configuration on an O5-plane, as depicted in Figure \ref{fig:HWusual}(d). Moreover, when we consider a 5-brane configuration with two O5-planes, rotations of monodromy cuts of flavor 7-brane can be understood in this way. As an example, recall Figure \ref{fig:SO12-Sp0} discussed in section \ref{sec:SON03}. If we were to denote the monodromy cuts of those $(1,0)$ 7-branes which were moved to the left by Hanany-Witten transitions, the direction of monodromy cuts should be understood as in Figure \ref{fig:SO12-Sp0-1}, where the charges are conserved on O5-planes.

\clearpage
\vspace{1cm}

\bibliographystyle{JHEP}

\bibliography{HiggsD}

\end{document}